\documentclass[rmp,twocolumn,amsmath,amssymb,showpacs,aps]{revtex4-1}

\usepackage{graphicx}
\usepackage[space]{grffile}
\usepackage{dcolumn}
\usepackage{bm}
\usepackage[ansinew]{inputenc}
\usepackage[T1]{fontenc}
\usepackage{color}
\usepackage{ae,aecompl}
\usepackage{textcomp}
\usepackage[colorlinks=true, linkcolor=black, citecolor=black, urlcolor=blue]{hyperref}

\newcommand{\bra}[1]{\langle#1|}
\newcommand{\ket}[1]{|#1\rangle}
\newcommand{\Rb}{^{87}\text{Rb}}

\begin{document}

\title{Cavity-based quantum networks with single atoms and optical photons}
\author{Andreas Reiserer}
\email{Electronic address: a.a.reiserer@tudelft.nl}
\affiliation{Kavli Institute of Nanoscience Delft, Delft University of Technology, Lorentzweg 1, 2628CJ Delft, The Netherlands}
\author{Gerhard Rempe}
\email{Electronic address: gerhard.rempe@mpq.mpg.de}
\affiliation{Max-Planck-Institut f{\"u}r Quantenoptik, Hans-Kopfermann-Stra{\ss}e 1, 85748 Garching, Germany}

\date{\today}

\begin{abstract}
Distributed quantum networks will allow users to perform tasks and to interact in ways which are not possible with present-day technology. Their implementation is a key challenge for quantum science and requires the development of stationary quantum nodes that can send and receive as well as store and process quantum information locally. The nodes are connected by quantum channels for flying information carriers, i.e. photons. These channels serve both to directly exchange quantum information between nodes as well as to distribute entanglement over the whole network. In order to scale such networks to many particles and long distances, an efficient interface between the nodes and the channels is required. This article describes the cavity-based approach to this goal, with an emphasis on experimental systems in which single atoms are trapped in and coupled to optical resonators. Besides being conceptually appealing, this approach is promising for quantum networks on larger scales, as it gives access to long qubit coherence times and high light-matter coupling efficiencies. Thus, it allows one to generate entangled photons on the push of a button, to reversibly map the quantum state of a photon onto an atom, to transfer and teleport quantum states between remote atoms, to entangle distant atoms, to detect optical photons nondestructively, to perform entangling quantum gates between an atom and one or several photons, and even provides a route towards efficient heralded quantum memories for future repeaters. The presented general protocols and the identification of key parameters are applicable to other experimental systems.
\end{abstract}



\maketitle

\tableofcontents

\section{Introduction} \label{sec:Introduction}

Quantum theory has originally been formulated as a statistical theory that describes ensembles of particles. Meanwhile, however, experiments are being carried out in many laboratories around the world with single particles, without finding any deviation from the predictions of quantum theory. In fact, quantum physics is now the best studied and most accurately tested theory of our time.

Beyond the desire to understand fundamental quantum phenomena occurring in small-scale systems of individual particles, the dream has therefore emerged to harness the strangeness and the power of the quantum to facilitate novel technologies that provide possibilities beyond those offered by any classical device \cite{dowling_quantum_2003}.

Quantum functionality has been successfully demonstrated in many different physical systems. Experimental platforms are as diverse as nuclear magnetic resonance systems \cite{vandersypen_nmr_2005}, single ions or neutral atoms trapped in the vacuum \cite{blatt_entangled_2008, bloch_quantum_2008, saffman_quantum_2010}, microwave \cite{raimond_manipulating_2001} and optical \cite{obrien_photonic_2009} photons, superconducting circuits \cite{schoelkopf_wiring_2008}, spins in quantum dots or other solid state host materials \cite{hanson_coherent_2008, warburton_single_2013, greve_ultrafast_2013, lodahl_interfacing_2015}, and even optomechanical systems \cite{aspelmeyer_cavity_2014}. Each of these systems has its own advantages, but also drawbacks which prevent a simple and direct scaling of current proof-of-concept experiments to larger systems. This, however, is required to exploit the full potential of quantum technology.

The main challenge is that the individual particles forming a quantum system have to be largely decoupled from the environment to prevent decoherence, while at the same time a controllable long-range interaction is required for up-scaling. It was realized early on \cite{cirac_quantum_1997, duan_long-distance_2001} that a hybrid system of light and matter qubits could tackle this problem: Single matter qubits can be well isolated from the environment, while single photons can be used to connect them over large distances to form a quantum network \cite{kimble_quantum_2008, duan_colloquium:_2010, meter_quantum_2014}.

A quantum network is a coherent few- or many-body system with genuine quantum properties such as entanglement. This makes it profoundly different from any classical network. For example, the information stored in the network is largely encoded \emph{between} the network nodes, not \emph{in} the network nodes. The amount of stored information can be huge as the accessible state space increases exponentially with the number of nodes, not linearly like in classical networks. Quantum networks can be large both in terms of number and distance, and they can display any topology and arbitrary connectivity, with entangling links between pairs of nodes being turned on and off at will.

These controlled long-range and potentially even infinite-range interactions form the basis for numerous possible applications. For example, distributed quantum networks will allow for the connection of small quantum processors to form larger computational units for scalable universal quantum information processing \cite{monroe_scaling_2013, awschalom_quantum_2013}. In this context, it is worthwhile to mention that a large physical separation between individual qubits might prevent correlated errors that preclude efficient quantum error correction \cite{lidar_quantum_2013}, which is a prerequisite for scalable quantum computation with realistic, i.e. imperfect, devices \cite{ladd_quantum_2010, nielsen_quantum_2010}.

Furthermore, entangled states that are distributed across a quantum network lead to correlations that are both nonlocal and nonclassical \cite{brunner_bell_2014}. Besides enabling fundamental tests of the predictions of quantum theory, these correlations can be the basis for communication with unbreakable encryption \cite{gisin_quantum_2007, scarani_security_2009, ekert_ultimate_2014}, possibly even on a global scale using quantum repeaters \cite{briegel_quantum_1998, van_enk_photonic_1998}.

Looking far into the future, to a time when today's small-scale quantum network prototypes might have grown into a full-scale quantum internet \cite{kimble_quantum_2008}, quantum computers might download quantum software \cite{preskill_plug-quantum_1999, gottesman_demonstrating_1999} just like classical computers download classical software via today's classical internet. Alternatively, individual users with limited quantum capabilities might place private quantum queries in a future quantum cloud or perform private quantum calculations on remote quantum supercomputers \cite{barz_demonstration_2012}. Finally, one might even envision the teleportation \cite{bennett_teleporting_1993} of macroscopic quantum objects between remote locations, a fascinating research perspective for the generations to come.

Outside the immediate realms of quantum computation and quantum communication, distributed quantum networks could also find applications in other fields of science and technology. To mention just a few examples, recent proposals consider telescopes with increased resolution \cite{gottesman_longer-baseline_2012} or timekeeping with unprecedented precision \cite{komar_quantum_2014}. Moreover, scaling to larger networks might enable the simulation of complex quantum many-body systems \cite{georgescu_quantum_2014} with the goal to understand and possibly design biologically and chemically relevant molecules, functional materials like high-temperature superconductors, and emerging nanotechnological devices.

The few examples mentioned above illuminate the foreseeable prospects of quantum networks, but the whole cornucopia of potential applications has not yet been imagined, nor will it be until large-scale networks actually become available. With this in mind, a realistic avenue towards scaling needs to be found and implemented. From today's perspective, it will critically rely on the following criteria:

First, the use of optical photons, ideally in the near infrared, as easily controllable flying qubits. This allows for the transmission of quantum information at room temperature over large distances using existing telecommunication fiber technology.

Second, the availability of versatile quantum nodes that can be initialized and manipulated individually, and that can process quantum information and store it for a time that is much longer than the time it takes to distribute quantum states over the network.

Third, the implementation of a protocol that is robust with respect to photon loss. This can be achieved using a heralded scheme that allows for the efficient correction of errors caused by photon loss. Clearly, this requires that quantum information is not encoded in the presence of a photon, but in its polarization, path, frequency, or arrival time.

Finally, a quantum interface that facilitates the connection of stationary and flying qubits with efficiency that is high enough to enable a deterministic transfer of quantum information between the network nodes well within their coherence time.

Clearly, due to the small interaction cross section of atoms and photons in free space, the last criterion is hard to fulfill. A possible solution is the use of atomic ensembles that consist of a large number of particles and therefore are optically thick. The optical nonlinearity necessary for the operation of a quantum-network node in a long-distance communication architecture can be established by suitable measurements \cite{duan_long-distance_2001, hammerer_quantum_2010, sangouard_quantum_2011}. However, the randomness of the measurement outcome makes the communication protocol intrinsically probabilistic. An alternative avenue towards deterministic quantum-nonlinear optics and quantum information processing at the single-photon level is provided by the Rydberg blockade mechanism \cite{lukin_dipole_2001} which is based on the giant interaction between highly excited atoms. Effectively, an interaction between photons is achieved by mapping the light quanta onto Rydberg atoms by means of slow and stopped light \cite{fleischhauer_electromagnetically_2005}. Remarkable experimental progress has been made with Rydberg blockaded ensembles during the last few years \cite{chang_quantum_2014}. More specifically, this includes the storage and the generation of nonclassical light fields \cite{dudin_strongly_2012, peyronel_quantum_2012, li_entanglement_2013, maxwell_storage_2013}, the demonstration of photon-photon interactions \cite{peyronel_quantum_2012, firstenberg_attractive_2013}, as well as single-photon switches and amplifiers \cite{baur_single-photon_2014, tiarks_single-photon_2014, gorniaczyk_single-photon_2014}.

This article, however, describes a different approach, where the stationary nodes consist of \emph{single} trapped atoms. This has the advantage that the quantum information is localized in individually addressable particles that can be manipulated and controlled in a straightforward way. Moreover, single atoms arguably constitute the most natural way to implement quantum-nonlinear media with the feature that their optical properties can be controlled by single photons, provided the light-matter coupling is large. To realize the required efficient interface between atomic and photonic qubits, the atoms are placed in optical resonators. The many mirror images of the atom, each with a stable phase relative to the atomic dipole, effectively boost the light-matter interaction strength and give access to the rich physics of cavity quantum electrodynamics (CQED).

Apart from the experiments in the optical domain which are described in this article, CQED has been pioneered with flying atoms that interact with trapped photons in microwave resonators \cite{raimond_manipulating_2001, walther_cavity_2006}. Many of the proposed concepts have been adapted and further developed using superconducting qubits in the electromagnetic field of microwave cavities and microwave stripline resonators, giving rise to a rapidly growing and progressing field which is very promising for future quantum technologies \cite{devoret_superconducting_2013}. However, in order to realize large-distance quantum networks with this platform, longer coherence times and the conversion to optical photons would be required. Therefore, microwave experiments are not covered in this article.

Apart from superconducting circuits, there are also other promising approaches to solid-state based quantum networks, e.g. quantum dots or crystalline defect centers \cite{hanson_coherent_2008, zwanenburg_silicon_2013, awschalom_quantum_2013, warburton_single_2013, greve_ultrafast_2013, lodahl_interfacing_2015}. A detailed discussion of these systems would go beyond the scope of this review. We only note that the encoding of quantum information in the spin of individual nuclei in an otherwise spin-free or spin-polarized crystal lattice can be achieved with remarkably long coherence times \cite{chekhovich_nuclear_2013, maurer_room-temperature_2012}, with a current record of several hours for the storage of strong light pulses in a spin ensemble \cite{zhong_optically_2015}. The major remaining challenge is to efficiently couple individual spins to photonic quantum channels. Typically, this requires exquisite control over of the involved spatio-temporal light mode, posing a formidable challenge to system design and nano-fabrication \cite{awschalom_quantum_2013, loncar_quantum_2013, lodahl_interfacing_2015}. Otherwise, these solid-state-based systems employ similar techniques and concepts as the atom-based approach described in this review.

CQED with atomic systems has been a very active field of research during the past decades. Overviews on earlier work can be found in \cite{kimble_strong_1998, mabuchi_cavity_2002, kuhn_optical_2002, vahala_optical_2003, kimble_quantum_2008, kuhn_cavity-based_2010, solomon_single_2013} and in many textbooks, e.g. in \cite{carmichael_statistical_2002, carmichael_statistical_2007, walls_quantum_2008, agarwal_quantum_2013, haroche_exploring_2013}. The focus of this article will lie on recent experiments aimed towards quantum networks. Before turning to this topic in detail, a basic introduction to the relevant principles of CQED will be given. Then, in Sec. \ref{sec:ExperimentalTechniques}, we will explain the different types of employed optical resonators and the basic experimental techniques to observe, capture, trap, cool, and control single atoms within optical resonators. Subsequently, experiments towards the generation of nonclassical light fields are summarized in Sec. \ref{sec:QuantumLightSources}. In Sec. \ref{sec:QuantumNetworks}, we will discuss the distribution and processing of quantum information in an elementary quantum network. To conclude, an outlook onto possible future directions of the field is given in Sec. \ref{sec:Outlook}.

\subsection{Quantum electrodynamics in a cavity} \label{sub:IntroductionToCQED}

For an intuitive introduction to the objectives of CQED, consider a two-level atom which is resonantly interacting with a faint light pulse that is focused to a beam waist of $w_0$ and that contains only one photon. To estimate the photon scattering probability, the area of the light beam, $A=\frac{\pi}{4}w_0^2$ is compared to the absorption cross section of the atom, $\sigma_\text{abs}=\frac{3 \lambda^2}{2 \pi}$, with $\lambda$ denoting the optical wavelength. Deterministic atom-photon interaction requires $\sigma_\text{abs}\gg A$. Clearly, this condition is not achievable in free space, even when using tight but diffraction limited focusing of a Gaussian input mode \cite{van_enk_single_2000, tey_strong_2008, wrigge_efficient_2008}.

The situation is different when the atom is located in a Fabry-Perot cavity, see Fig. \ref{fig:AtomInCavity}a. To this end, two mirrors with reflectivity $R_1$ and $R_2$ are placed in the beam on each side of the atom, such that the photon can bounce back and forth many times. Effectively, this increases the absorption cross section by the number of bounces, $\frac{\mathcal{F}}{\pi}$, which is determined by the quality of the resonator as characterized by its finesse $\mathcal{F}=\frac{\pi(R_1 R_2)^{1/4}}{1-\sqrt{R_1 R_2}}$. The condition to achieve a deterministic interaction then reads:

\begin{equation} \label{eq:AbsorptionCrossSectionInCavity}
\underbrace{\frac{3 \lambda^2}{2 \pi}}_{\sigma_{\text{abs}}} \times \underbrace{\frac{\mathcal{F}}{\pi}}_{\text{bounces}} \gg \underbrace{\frac{\pi}{4}w_0^2}_{\text{beam area}}
\end{equation}

This equation can be rewritten in terms of the cavity field decay rate, $\kappa$, the polarization decay rate of the atom, $\gamma$, and a parameter $g$ that can be identified as an atom-cavity coupling constant. It is determined by the electric dipole matrix element $\mu_{\text{ce}}$ of the atomic transition from the (coupled) ground state $\ket{c}$ to the excited state $\ket{e}$, and by the electric field $E$ of a single photon in the mode volume $V$ of the resonator:

\begin{equation} \label{eq:couplingStrength}
g=\frac{\mu_{\text{ce}}E}{\hbar}=\sqrt{\frac{\mu_{\text{ce}}^2 \omega}{2\epsilon_0 \hbar V}}u(\vec{x})
\end{equation}

Here, $\hbar$ is the reduced Planck constant, $\epsilon_{0}$ the permittivity of free space, $\omega$ the cavity frequency, and $V=\int u^2(\vec{x}) d^3x$ is the integral over the dimensionless electric-field mode function $u(\vec{x})$ of the resonator, normalized to one at the field maximum. With the speed of light $c$ and the resonator length $L$, and using the relations

\begin{subequations}
\begin{align}
\kappa &= \frac{\pi c}{2 L \mathcal{F}} \\
\gamma &= \frac{\mu_{ce}^2\omega^3}{6 \pi \epsilon_0 \hbar c^3}
\end{align}
\end{subequations}

one can rewrite Eq. (\ref{eq:AbsorptionCrossSectionInCavity}) as:

\begin{equation} \label{eq:cooperativity_Def}
C=\frac{g^2}{2\kappa\gamma} \gg 1
\end{equation}

Here, $C$ denotes the single-atom cooperativity parameter of the optical-bistability literature \cite{gibbs_optical_2012}, originally derived for many atoms. Clearly, with $N_a$ atoms in the mode, the cooperativity increases to $C_{N_a} = N_a C$. Therefore, the inverse of $C$ is often called the critical atom number, which gives the number of atoms required to strongly affect the transmission properties of the resonator. For $C \gg 1$, even a single atom is able to induce a big effect on a light beam transmitted through the cavity. From the above picture of a bouncing photon it is clear that $C$ should not directly depend on the cavity length. In fact, the length-dependence in the definition of $C$ drops out as $\kappa \propto \frac{1}{L}$ and $ g \propto \frac{1}{\sqrt{L}} $.

In a similar way, one can define a critical photon number, $n_c$, which determines the number of photons required to significantly change the radiation properties of the atom. For a two-level atom, $n_c$ can be identified as the number of photons needed to saturate the atomic transition. To derive this number, the rate of spontaneous emission, $2\gamma$, has to be compared to the rate of stimulated emission per photon, $\frac{3\lambda^2 c}{2\pi V}$. This leads to the condition:

\begin{equation} \label{eq:CriticalPhotonNumber}
n_c=\frac{\gamma^2}{2g^2} \ll 1
\end{equation}

The simple picture developed above already allows one to determine the most relevant experimental parameters. As can be seen from Eq. (\ref{eq:couplingStrength}), the coupling strength on a given atomic transition only depends on $V$. Therefore, the use of a cavity with a small mode volume is essential for the realization of a quantum interface where the atom-photon coupling rate is supposed to be larger than all dissipative loss rates. Before the influence of these losses will be discussed in Sec. \ref{sub:damped_AtomCavitySystems}, an idealized lossless situation is assumed in Sec. \ref{sub:JaynesCummings_Model}. This allows for a basic understanding of the dynamics of a coupled atom-cavity system within the canonical Jaynes-Cummings model that has first been introduced in \cite{jaynes_comparison_1963}.

\subsection{The Jaynes-Cummings model} \label{sub:JaynesCummings_Model}

\begin{figure}
\includegraphics[width=86mm]{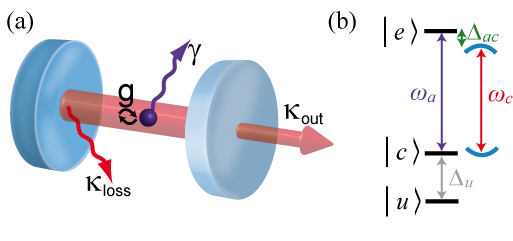}
\caption{ \label{fig:AtomInCavity}
(color online) Basic physical situation considered in CQED. A three-level emitter, e.g. a single atom, is located in an optical cavity, schematically represented by two facing mirrors. The atom exchanges energy with the optical field at a rate $2g$. The decay rate of photons out of the resonator is $2\kappa=2\kappa_\text{out}+2\kappa_{loss}$. The atomic polarization decay rate is $\gamma$.
(b) Atomic energy level scheme. The transition frequency between the coupled ground state $\ket{c}$ and the excited state $\ket{e}$ is $\omega_a$. The cavity resonance frequency is $\omega_c$. Atom and cavity are detuned by $\Delta_{ac}=\omega_a-\omega_c$. The third, uncoupled level $\ket{u}$ is detuned from $\ket{c}$ by $\Delta_u$.
}
\end{figure}

The basic physical situation described by the Jaynes-Cummings model is depicted in Fig. \ref{fig:AtomInCavity}(a). A single emitter is located in a resonator that supports a single optical mode at a frequency $\omega_c$. This frequency is in (or close to) resonance with an atomic transition from a coupled ground state $\ket{c}$ to an excited state $\ket{e}$ at a frequency $\omega_a$. In addition, the atom typically also exhibits one or several uncoupled energy states $\ket{u}$ that are far detuned from the cavity field, see the level scheme in Fig. \ref{fig:AtomInCavity}(b).

The coupling of the atomic states $\ket{c}$ and $\ket{e}$ to the electromagnetic field of the cavity is described in the dipole approximation. Thus, the Hamiltonian of the atom-photon interaction reads:

\begin{subequations}
\begin{align}
\mathcal{H}_{I} &= -\vec{d}\vec{E} \rightarrow d E (\sigma_{ce}^\dagger + \sigma_{ce}) (a^\dagger + a) = \\
&= \left( \sigma_{ce}^\dagger a^\dagger + \sigma_{ce}^\dagger a + \sigma_{ce} a^\dagger + \sigma_{ce} a \right )
\end{align}
\end{subequations}

In this equation, $\vec{d}$ denotes the atomic dipole moment operator and $\vec{E}$ the electric-field operator of a photon in the cavity mode, $\sigma_{ij}=\ket{i}\bra{j}$ and $\sigma_{ij}^\dagger=\ket{j}\bra{i}$ are the atomic lowering and raising operators, and $a^\dagger$ and $a$ are the creation and annihilation operators of a photon in the cavity mode, respectively. The term $d E / \hbar$ can be identified as the atom-cavity coupling strength $g$.

In the interaction picture with time dependent operators, the terms $\sigma_{ce}^\dagger a^\dagger$ and $\sigma_{ce} a$ exhibit a very fast time dependence $\propto e^{i (\omega_a + \omega_c) t}$. As long as $g \ll (\omega_a, \omega_c)$, these off-resonant terms can be neglected in comparison with the resonant terms $\sigma_{ce} a^\dagger$ and $\sigma_{ce}^\dagger a$ which evolve $\propto e^{i (\omega_a - \omega_c) t}$. This approximation is called the rotating wave or the secular approximation. In the following, we chose the energy origin such that the combined atom-cavity ground state $\ket{c,n=0}$ has zero energy. In this state, the photon number in the cavity, $n$, is zero and the atom is in the resonantly coupled state $\ket{c}$. Assuming that the atom exchanges energy with the cavity field at a frequency $2g$ then gives the Jaynes-Cummings Hamiltonian $\mathcal{H}_{JC}$:

\begin{equation}
\mathcal{H}_{JC}=\underbrace{\hbar \omega_{a} \sigma_{ce}^\dagger\sigma_{ce}}_{\text{bare atom}} + \underbrace{\hbar \omega_c a^\dagger a }_{\text{empty cavity}}+ \underbrace{\hbar g (\sigma_{ec}a+\sigma_{ce}a^\dagger)}_{\text{dipole coupling}}
\end{equation}

Note that $ \sigma_{ce}^\dagger\sigma_{ce}+a^\dagger a$ commutes with the Hamiltonian, which means that the total number of energy quanta in the system is conserved. Thus, for the diagonalization of the full Hamiltonian it is sufficient to consider separate two-level systems [see e.g. \cite{haroche_exploring_2013, agarwal_quantum_2013}]. This leads to the following energy eigenstates:

\begin{subequations}
\begin{align}
E_0&=0 \\
E_{N,\pm}&=N\hbar \omega_c + \hbar \frac{\Delta_{ac}}{2} \pm \frac{\hbar}{2} \sqrt{\Omega_N^2+\Delta_{ac}^2}
\end{align} \label{eq:EnergyEigenstatesJaynescumming}
\end{subequations}

\begin{figure}
\includegraphics[width=86mm]{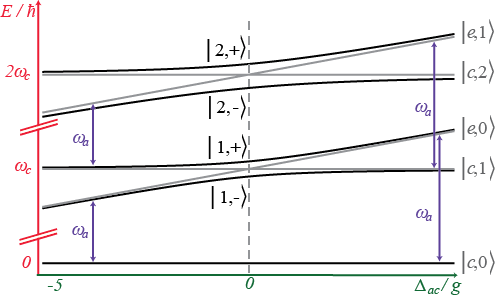}
\caption{ \label{fig:JaynesCummingsLadder}
(color online) The Jaynes-Cummings ladder of the dressed energy eigenstates $\ket{N,\pm}$ of a strongly coupled atom-cavity system, where the cavity frequency is kept fixed and the emitter frequency is scanned over the resonance. Compared to the uncoupled situation (grey), the coupled states (black) exhibit pronounced avoided level crossings. At zero detuning, the levels are separated by $2g\sqrt{N}$.
}
\end{figure}

Here, $N$ is a positive integer denoting the total number of excitation quanta in the system, and $\Omega_N=2g \sqrt{N}$ is the corresponding $N$-quanta Rabi frequency. The term $\Delta_{ac}=\omega_a-\omega_c$ denotes the detuning of the cavity frequency from the atomic transition. The resulting spectrum of energy eigenstates, often called the Jaynes-Cummings-ladder, is depicted in Fig. \ref{fig:JaynesCummingsLadder} as a function of the atomic detuning. The ground state of the coupled system, $\ket{c,0}$, remains unaffected. Similarly, when the detuning is large the coupled energy eigenstates almost coincide with those of an uncoupled system with $n$ photons in the cavity, $\ket{c,n}$ and $\ket{e,n}$. For $\Delta_{ac} \gg g$, the coupled states are shifted by $\Delta_{e,n} = (n+1)\frac{g^2}{\Delta_{ac}}$ and $\Delta_{c,n} = -n\frac{g^2}{\Delta_{ac}}$ \cite{haroche_exploring_2013}. Opposite shifts of equal magnitude occur for $-\Delta_{ac} \gg g$. Thus, in this regime the presence of the atom slightly shifts the cavity frequency.

The consequences of the coupling are much more pronounced on resonance. Here, the energies of the excited states $E_{N,+}$ and $E_{N,-}$ differ by $2\hbar g \sqrt{N}$. Note that the splitting increases nonlinearly with $N$. This single-photon nonlinearity can be exploited to engineer interactions between photons, which will be further discussed in Sec. \ref{subsub:PhotonPhoton_Gates}.

The eigenstates of the coupled system, $\ket{N,\pm}$, are linear combinations of $\ket{e,N-1}$ and $\ket{c,N}$. On resonance, $\Delta_{ac}=0$, these states are:

\begin{equation} \label{eq:EigenvectorsJaynesCummings}
\ket{N,\pm}=\frac{1}{\sqrt{2}}(\ket{e,N-1} \pm \ket{c,N})
\end{equation}

In order to explore the consequences of this superposition state, assume for example that a resonant atom is prepared in the excited state $\ket{e}$ inside a cavity containing $N-1$ photons. The initial state is thus $\ket{e,n=N-1}=\frac{1}{\sqrt{2}}(\ket{N,+}+\ket{N,-})$. Calculating the probability of the atom being in $\ket{e}$ as a function of time gives $P_{\ket{e}}(t)=\cos^2(\Omega_N t / 2)$, which means that the system undergoes a Rabi oscillation between the states $\ket{c,N}$ and $\ket{e,N-1}$ at a frequency $\Omega_N = 2 g \sqrt{N}$. The value $\Omega_1=2g$ that is obtained when the cavity is initially in the vacuum state, $N-1=0$, is therefore called the vacuum-Rabi frequency.

The prerequisite for an experimental observation of these vacuum-Rabi oscillations is that their rate is faster than all dissipative processes in the system, as will be explained in more detail in the following.

\subsection{Damped atom-cavity system} \label{sub:damped_AtomCavitySystems}

In contrast to the idealized Jaynes-Cummings model, any experimental atom-cavity system is an open system that is coupled to the environment via two loss channels:

First, the electric field of the cavity mode decays at a rate $\kappa=\kappa_{\text{out}}+\kappa_{\text{loss}}$. This can be caused by coupling to a propagating field mode (with rate $\kappa_{\text{out}}$) or by scattering and absorption losses of both mirrors (with a combined rate of $\kappa_{\text{loss}}$). Note that the energy in the cavity decays at a rate $2\kappa$.

Second, the rate $\gamma$ at which the atomic polarization decays. This can either be caused by atomic decay to an uncoupled state $\ket{u}$, which can be avoided using a closed transition, or caused by the emission of a photon at the resonant frequency, but into free space rather than into the cavity mode. Assuming purely radiative decay and a cavity mode which covers only a small solid angle like in typical Fabry-Perot resonators, $\gamma=\Gamma/2$, with $\Gamma$ being the free-space spontaneous-emission rate (modifications will be discussed in Sec. \ref{subsub:PurcellRegime}).

The standard approach to calculate the dynamics of the open atom-cavity system with density operator $\rho$ is to use a Master equation in the Lindblad form:

\begin{equation} \label{eq:MasterEquation}
\dot{\rho}=\frac{1}{i\hbar}[\mathcal{H}_{JC},\rho]+\sum_i [L_i \rho L_i^\dagger - \frac{1}{2}(L_i^\dagger L_i \rho + \rho L_i^\dagger L_i)]
\end{equation}

Here, $L_{1,2}$ are the jump operators that correspond to the atomic ($L_1=\sqrt{2\gamma}\sigma_{ce}$) and cavity ($L_2=\sqrt{2\kappa}a$) decay. Note that at optical frequencies, the incoupling of energy from a thermal reservoir is negligibly small. Thus, the system decays to the vacuum state $\ket{c,0}$. Depending on the ratio of the three rates $(g,\kappa,\gamma)$, different dynamics of this decay are expected from a solution of the master equation.

In case $C \ll 1$, the dynamics is almost the same that is observed with a free-space atom. The situation changes when $C$ approaches unity. But even if $C \gg 1$, there are still two experimental regimes with different dynamics: First, the "Purcell" or "fast cavity" regime, where the emission of a photon is an irreversible process since $\kappa > g$. Second, the regime of "strong coupling", where $g$ is the highest rate in the system and a reversible energy exchange between the atom and the resonator is possible. The basic physical effects encountered in the different regimes are explained in more detail in the following sections.

\subsubsection{The strong-coupling regime} \label{subsub:StrongCoupling}

In the regime of strong coupling, the atom-cavity coupling is the highest rate in the system:

\begin{equation}
g\gg(\gamma,\kappa)
\end{equation}

This also implies that $C \gg 1$ [compare Eq. (\eqref{eq:cooperativity_Def})]. Only in this regime the reversible atom-photon coupling is faster than all irreversible processes, such that the vacuum-Rabi oscillations predicted by the Jaynes-Cummings model can be observed experimentally.

\begin{figure}
\includegraphics[width=86mm]{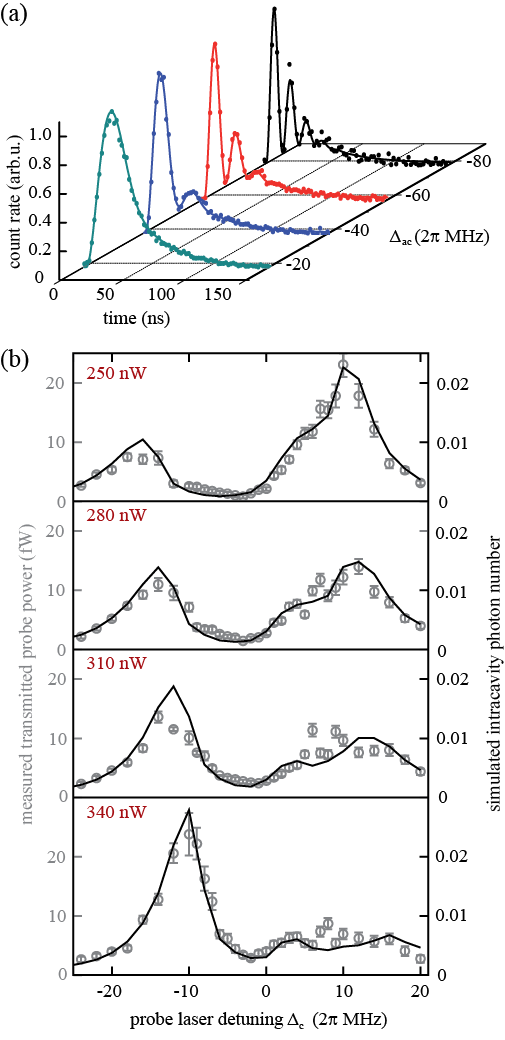}
\caption{ \label{fig:NormalModesSingleAtom}
(color online) (a) Vacuum-Rabi oscillations in the time domain. A short laser pulse excites a single atom to the excited state $\ket{e}$, from which it decays back to the ground state while emitting a single photon. In the cavity output mode, damped vacuum-Rabi oscillations are observed. With increasing detuning $|\Delta_{ac}|$ (front to back), the vacuum-Rabi frequency increases. The solid curves are fits to the data using a theoretical model. Adapted from \cite{bochmann_fast_2008}.
(b) Vacuum-Rabi splitting in the spectral domain, recorded with exactly one trapped atom. The detuning of the atom with respect to the cavity is varied by changing the power of the intracavity dipole-trap laser from $250\,\text{nW}$ (top) to $340\,\text{nW}$ (bottom). The data points (grey) clearly show a vacuum-Rabi splitting and are in good agreement with Monte-Carlo simulations of the system (black lines). Adapted from \cite{maunz_normal-mode_2005}.
}
\end{figure}

Due to the presence of the losses, one expects a damped oscillation in the time domain. Fig. \ref{fig:NormalModesSingleAtom}(a) shows such an experiment \cite{bochmann_fast_2008}. An atom in the resonator has been excited to the state $\ket{e}$ using a laser pulse of $3\,\text{ns}$ duration, which is short compared to the timescale of the atomic polarization decay ($\sim 50\,\text{ns}$). Subsequently, the atom can exchange energy with the resonator at a rate $\Omega_{N=1,\Delta_{ac}}=\sqrt{\Delta_{ac}^2+4g^2}$. The emission of photons from the cavity mode is measured with single-photon counting modules (SPCMs), giving the measured signal in Fig. \ref{fig:NormalModesSingleAtom}(a). With larger detuning $|\Delta_{ac}|$, the oscillation frequency increases. The fit curves (solid lines) calculated from a theoretical model are in good agreement with the data.

In the spectral domain, the damped vacuum-Rabi oscillations correspond to a double-peak structure, dubbed the vacuum-Rabi splitting. This splitting can be observed when the frequency dependence of the cavity transmission is measured using a faint laser (such that the system excitation probability remains low, i.e. the average number of photons in the cavity $\bar{n} \ll 1$). After the first demonstration in the optical domain \cite{thompson_observation_1992}, the vacuum-Rabi splitting has been observed in numerous experiments, thus proving the attainment of the strong-coupling regime. The first observations with a single trapped atom were reported in \cite{maunz_normal-mode_2005} and \cite{boca_observation_2004}.

Fig. \ref{fig:NormalModesSingleAtom}(b) shows the results of \cite{maunz_normal-mode_2005}. In the different panels, the atomic transition frequency is scanned by changing the atomic AC-Stark shift via the power (characterized by the transmitted intensity, red numbers) of the dipole-trap laser which holds the atom inside the cavity. One clearly observes the expected vacuum-Rabi splitting. The obtained spectra are in excellent agreement with Monte-Carlo simulations of the system (black lines). The simulation takes into account the probe-laser induced random motion of the atom in the intracavity dipole trap. A more recent spectrum obtained with a better localized atom will be discussed below in Sec. \ref{sub:CouplingToPropagatingFields}.

\subsubsection{The Purcell regime} \label{subsub:PurcellRegime}

An atom-cavity system in the Purcell regime is characterized by the following relations: $\kappa \gg g \gg \gamma$ and $C \gg 1$. In this regime, the vacuum-Rabi oscillations cannot be observed, as a photon that is emitted into the cavity mode is lost from the resonator before it can be reabsorbed by the atom. Nevertheless, the dynamics of the atomic decay is strongly different from observations in free space, since the density of photonic modes is considerably changed by the presence of the cavity. This can lead to an increased \cite{purcell_spontaneous_1946} or decreased \cite{kleppner_inhibited_1981} rate of atomic radiative decay, depending on the detuning of the cavity frequency from the atomic frequency. Solving the master equation of the coupled atom-cavity system, the atomic dipole decay rate into the cavity, $\gamma_c$, can be derived [see e.g. \cite{agarwal_quantum_2013, haroche_exploring_2013}]. This gives:

\begin{equation} \label{eq:Gamma_PurcellEffect}
\gamma_c=\frac{g^2\kappa}{\kappa^2+\Delta_{ac}^2}
\end{equation}

With increasing detuning $\Delta_{ac}$ between atom and cavity, the decay into the resonator mode is suppressed. When the cavity mode covers only a small fraction $\zeta$ of the solid angle, the atomic polarization decay rate $\gamma$ via emission into free space modes remains unchanged. Otherwise, $\gamma=\frac{\Gamma}{2}(1-\frac{\zeta}{4\pi})$, which can lead to an increased lifetime of the atomic excited state in the cavity.

It is remarkable that on resonance $\gamma_{c}=2C\gamma$ can be much larger than $\gamma$. The radiative interaction of the atom with the environment is then dominated by the cavity mode rather than the free-space modes. This leads to the notion of a "one-dimensional atom" \cite{kimble_strong_1998}.

\begin{figure}
\includegraphics[width=86mm]{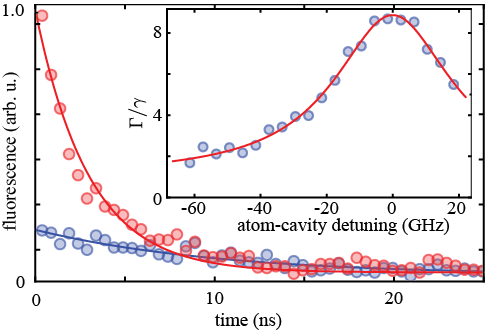}
\caption{ \label{fig:PurcellEffect}
(color online) Purcell-enhanced atomic emission. On resonance (red data and upper fit curve), the presence of a photonic crystal cavity with $\kappa=2\pi\times12.5\,\text{GHz}$ guides the fluorescence of a single atom into the cavity mode, and shortens the excited state lifetime from $26\,\text{ns}$ to $3\,\text{ns}$. Both effects are less pronounced at a detuning of $-41\,\text{GHz}$ (blue data and lower lying fit curve). The inset shows the dependence of the decay rate on the atom-cavity detuning. Adapted from \cite{tiecke_nanophotonic_2014}.
}
\end{figure}

Early observations of atomic lifetime reduction in optical resonators are reported in \cite{heinzen_enhanced_1987, kreuter_spontaneous_2004}. Fig. \ref{fig:PurcellEffect} shows the results obtained in a recent study \cite{tiecke_nanophotonic_2014} in the Purcell regime with $C=4$ and $\kappa=25 g$. An almost tenfold reduction of the atomic excited state lifetime is observed.

The best explored application of the modified decay rate is the implementation of an efficient source of single photons, which has been realized in many atomic and solid-state systems [for a review, see e.g. \cite{eisaman_invited_2011}]. To this end, the atom is excited, e.g. by a short resonant laser pulse. Subsequently, it decays back to the ground state via the resonant cavity mode, which is coupled to a single propagating mode at a rate $\kappa_{\mathrm{out}}$. The fraction of photons emitted into this mode is then

\begin{equation}
\frac{\kappa_{\text{out}}}{\kappa}\times \frac{\gamma_c}{\gamma+\gamma_c} = \frac{\kappa_{\text{out}}}{\kappa}\times \frac{2C}{1+2C}.
\end{equation}

Note that this formula has been derived under the assumption $\kappa \gg g$. For an efficient photon source, clearly a large cooperativity of the atom-cavity system is required. The topic of photon generation in atom-cavity systems will be revisited in Sec. \ref{sec:QuantumLightSources}.

\subsection{Coupling to propagating light fields} \label{sub:CouplingToPropagatingFields}

\begin{figure}
\includegraphics[width=86mm]{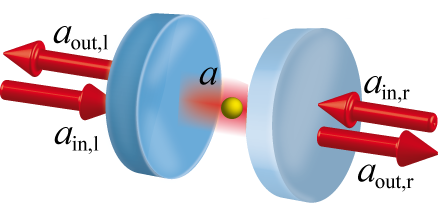}
\caption{ \label{fig:InputOutput_Operators}
(color online) Visualization of the field annihilation operators describing the input and output of the cavity field.
}
\end{figure}

In the treatment of Sec. \ref{sub:damped_AtomCavitySystems}, the coupling of the cavity mode to the outside world has been considered as a damping term. Now, this approach is extended using the framework of input-output theory \cite{gardiner_input_1985, carmichael_quantum_1993} to calculate the coherent response of the system to impinging light fields.

Consider an asymmetric Fabry-Perot cavity with field decay rates $\kappa_{r}$ and $\kappa_{l}$ through the right and left mirror, respectively, and losses $\kappa_{\text{loss}}$ as depicted in Fig. \ref{fig:AtomInCavity}. The operators $a_{\text{in,l}}$ ($a_{\text{in,r}}$) are the field operators of an electromagnetic mode impinging from the left (right) side, see Fig. \ref{fig:InputOutput_Operators}, while $a_{\text{out,l}}$ ($a_{\text{out,r}}$) describe the outgoing field to the left (right) side, respectively. $a$ is the annihilation operator for a photon in the cavity. Then the relation between the input and output modes (in the Heisenberg picture) is given by \cite{agarwal_quantum_2013}:

\begin{subequations} \label{eq:InputOutput}
\begin{align}
a_{\text{out,r}}\left(t\right)+a_{\text{in,r}}\left(t\right) &= \sqrt{2\kappa_{r}}a\left(t\right) \\
a_{\text{out,l}}\left(t\right)+a_{\text{in,l}}\left(t\right) &= \sqrt{2\kappa_{l}}a\left(t\right)
\end{align}
\end{subequations}

\begin{figure*}
\includegraphics[width=172mm]{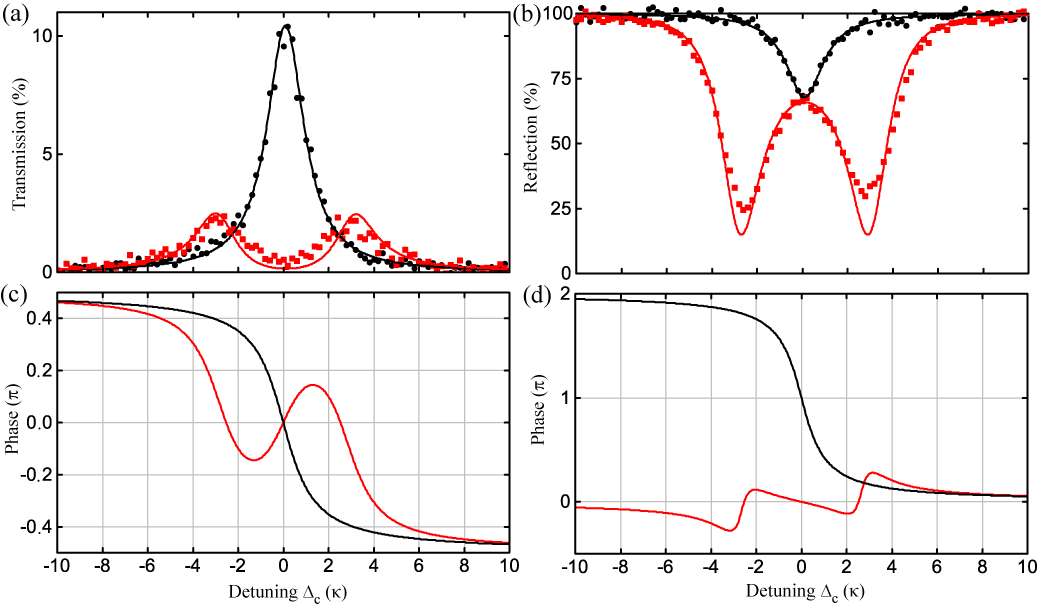} \caption{ \label{fig:Transmission_Reflection}
(color online) Transmission (a) and reflection (b) spectra of a single-sided, overcoupled cavity in the strong-coupling regime. Without the presence of an atom (black circles and black fit curves), the cavity exhibits a Lorentzian transmission resonance, which is also observed as a drop in reflection. The sum of transmission and reflection is less than $100\%$ due to mirror absorption and scattering. With an atom trapped in the mode (red squares and red theory curve), the vacuum-Rabi peaks are clearly resolved. The discrepancy of the data points from the theory curve is explained by inhomogeneous broadening of the atomic transition frequency. (c,d) Phase response in the coupled (red) and uncoupled (black) case in transmission (c) and reflection (d). On resonance, a phase shift of $\pi$ is observed in reflection for the case of an empty cavity.
}
\end{figure*}

For simplicity, assume the system is weakly driven from the left side only. Then the above relations allow one to calculate the reflection and transmission properties by solving the Heisenberg-Langevin equation of motion \cite{cirac_quantum_1997, carmichael_statistical_2002, duan_scalable_2004, walls_quantum_2008, agarwal_quantum_2013}. These equations follow from the master equation and the relation $\frac{d}{dt}\langle A \rangle=\text{tr}(A\dot{\rho})$, valid for the expectation value of any system operator $A$. For an atom-cavity-system, this gives:

\begin{subequations} \label{eq:Langevin}
\begin{align}
\langle \dot{a} \rangle &= -\frac{i}{\hbar} \langle \left[a,\mathcal{H}_{JC}\right]\rangle-\kappa \langle a \rangle+\sqrt{2\kappa_{l}}\langle a_{\text{in,l}}\rangle \left(t\right) \\
\langle \dot{\sigma}_{ce} \rangle &= -\frac{i}{\hbar} \langle \left[\sigma_{ce},\mathcal{H}_{JC}\right]\rangle-\gamma \langle \sigma_{ce} \rangle
\end{align}
\end{subequations}

Here, $\kappa=\kappa_{l}+\kappa_{r}+\kappa_{\text{loss}}$ is the total cavity field decay rate. The above equations can be integrated numerically to calculate the dynamics of the system.

In addition, an analytic result can also be obtained [see e.g. \cite{hu_giant_2008, tiecke_nanophotonic_2014}] when the atomic excitation is negligibly small and the photonic wavepacket envelope only varies on a timescale longer than the cavity decay time. This gives the following relations for the amplitude reflection $r(\omega)$ and transmission $t(\omega)$ coefficients, depending on the detuning $\Delta_{c(a)}\equiv\omega-\omega_{c(a)}$ of the driving laser with respect to the cavity (atomic transition) frequency $\omega_{c}$ ($\omega_{a}$), respectively:

\begin{subequations}
\begin{align}
r(\omega) &= 1-\frac{2\kappa_{l}(i\Delta_{a}+\gamma)}{(i\Delta_{c}+\kappa)(i\Delta_{a}+\gamma)+g^{2}} \\
t(\omega) &= \frac{2\sqrt{\kappa_{l}\kappa_{r}}(i\Delta_{a}+\gamma)}{(i\Delta_{c}+\kappa)(i\Delta_{a}+\gamma)+g^{2}}
\end{align}
\end{subequations}

The phase of the reflected light is $\arg(r(\omega))$, while the intensity reflection and transmission coefficients are given by $R(\omega)=\left|r(\omega)\right|^{2}$ and $T(\omega)=\left|t(\omega)\right|^{2}$, respectively.

Measuring the spectrum of the empty cavity transmission and reflection allows for a quantification of the rates $\kappa_{l}$, $\kappa_{r}$ and $\kappa_{\text{loss}}$. However, the influence of the mode-matching efficiency $\xi$ between the spatial mode profile of the impinging light and that of the cavity mode has to be considered. In Fabry-Perot resonators, any light field that is not coupled to the cavity will be reflected with $R\simeq1$. Thus, $R_{\text{exp}}(\Delta_{c})=(1-\xi)+\xi R(\Delta_{c})$.

Fig. \ref{fig:Transmission_Reflection} shows transmission (a) and reflection (b) spectra, measured in the strong-coupling regime with $(g,\kappa,\gamma)=2\pi\times(7,2.5,3)\,\text{MHz}$ in the coupled (red squares) and uncoupled (black circles) case. The system is overcoupled. For photons impinging from the left side, this means that $\kappa_{l}\gg \kappa_{\text{loss}},\kappa_{r}$. The black lines are fit curves according to the equations above, which yield $\kappa=2\pi\times 2.5(1)\,\text{MHz}$ and $\kappa_{l}=2\pi\times 2.3(1)\,\text{MHz}$. The red curves are calculated using $\Delta_{a}=\Delta_{c}$, the fit values of $\kappa$, a coupling strength $g=2\pi\times 7\,\text{MHz}$, and a mode-matching parameter $\xi=0.9$. The latter has been estimated by measuring the coupling efficiency of light transmitted through the cavity from the right into the antireflection coated optical fiber from which the input mode $a_{\text{in,l}}$ emerges.

Note the good agreement between experiment (red squares) and theory (red line). Compared to the first measurements displayed in Fig. \ref{fig:NormalModesSingleAtom}(b), an improved control over the atomic state, as described below in Sec. \ref{sec:ExperimentalTechniques}, makes the observed spectrum (red squares) approach the ideal expectation (red line). The position of the peaks is centered around $\pm g$. Thus, the spectrum exhibits resonances at the energy eigenstates of the coupled system calculated in Eq. (\ref{eq:EnergyEigenstatesJaynescumming}). The full width at half-maximum (FWHM) of the peaks is in good agreement with the linewidth-averaged value of $\kappa+\gamma$ that is expected from the solution of the master equation [Eq. (\ref{eq:MasterEquation})]. For a derivation, see \cite{agarwal_quantum_2013}. For comparison, the transmission without a coupled atom is also depicted (black circles), showing a Lorentzian resonance with FWHM $2\kappa$.

From the obtained parameters, one can also calculate the phase response of the system, as shown in Fig. \ref{fig:Transmission_Reflection} (c) in transmission and (d) in reflection. Note that in reflection, unlike in transmission, a phase shift of $\pi$ occurs around zero detuning for the case of an empty cavity. With a coupled atom, the phase shift is zero, as observed in \cite{reiserer_nondestructive_2013} in the strong-coupling regime and in \cite{volz_nonlinear_2014, tiecke_nanophotonic_2014} in the Purcell regime. The resulting conditional phase shift is the basis for the realization of robust quantum gates \cite{reiserer_quantum_2014} following the proposal of \cite{duan_scalable_2004}. This will be the topic of Sec. \ref{sub:quantumInfoProcess}.

\subsection{Coupling to a third level} \label{sub:Raman_coupling}

\begin{figure}
\includegraphics[width=86mm]{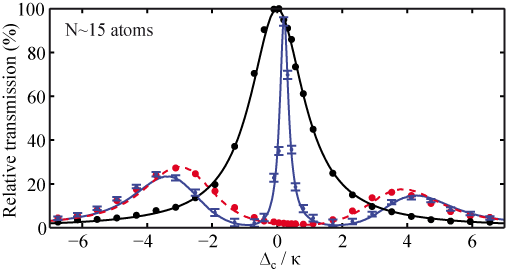}
\caption{ \label{fig:CavityEIT}
(color online) Electromagnetically induced transparency, observed with about 15 atoms trapped in a cavity. When the cavity is empty or the atoms are prepared in the uncoupled state (black), a Lorentzian transmission peak is obtained. When the atoms are prepared in the coupled state, the vacuum-Rabi splitting is observed (red data and red dashed theory curve). When an additional control laser is switched on (blue data with error bars and blue theory curve), the transmission peaks are shifted outwards, and the spectrum exhibits a narrow transparency window, which indicates the presence of a coherent dark state. Adapted from \cite{mucke_electromagnetically_2010}.
}
\end{figure}

The above sections described the coupling of a two-level atom (with states $\ket{c}$ and $\ket{e}$) to the cavity field at a rate $g$. This section introduces an additional uncoupled ground state level, $\ket{u}$, detuned by $\Delta_{u}$ from the resonator frequency, as illustrated in Fig. \ref{fig:AtomInCavity}(b). Assuming $\Delta_u \gg g$, the direct coupling of the third level to the resonator mode is negligible. In addition, there is no decay between the two ground states, $\ket{c}\nleftrightarrow\ket{u}$. Such three-level system is often termed to display a $\Lambda$-configuration.

Now apply an external laser field is applied at a frequency $\omega_L=\omega_a + \Delta_u + \Delta_{ac}$, i.e. in two-photon resonance with the resonator mode and detuned from the atomic level $\ket{e}$ by $\Delta_{ac}$. This control laser is irradiated from the side of the cavity and leads to a coupling $\ket{u}\leftrightarrow\ket{e}$ with a Rabi frequency of $\Omega_{L}$. Setting the energy origin at $\ket{c,0}$ results in the following Hamiltonian:

\begin{equation} \label{eq:ThreeLevel_Hamiltonian}
\begin{split}
\mathcal{H}_{\Lambda} =
\underbrace{\hbar \omega_a \sigma_{ee} + \hbar \Delta_u \sigma_{uu} }_{\text{atom}} + \underbrace{\hbar \omega_c a^\dagger a}_{\text{cavity}} + \\
 + \underbrace{\hbar g(a^\dagger\sigma_{ce} + \sigma_{ec}a)}_{\text{Jaynes-Cummings coupling}} +
\underbrace{\hbar \frac{\Omega_L}{2}(\sigma_{ue}+\sigma_{eu})}_{\text{external laser}}
\end{split}
\end{equation}

The analysis is restricted to the experimentally most relevant case where only a single excitation is present in the system. Here, the energy eigenstates of the coupled system are:

\begin{subequations} \label{eq:EIT_EnergyEigenstates}
\begin{align}
E_0&=\hbar \omega_c \\
E_\pm&=\hbar \omega_c+\frac{\hbar}{2}(\Delta_{ac}\pm\sqrt{4g^2+\Delta_{ac}^2+\Omega_L^2})
\end{align}
\end{subequations}

In analogy to the Jaynes-Cummings doublets introduced in Sec. \ref{sub:JaynesCummings_Model}, there are now state triplets. One of the states is located on the cavity resonance. The splitting of the other two states increases with increasing laser power. The width of the resonances can again be calculated by solving the master equation of the system, compare Eq. (\ref{eq:MasterEquation}) in Sec. \ref{sub:damped_AtomCavitySystems}. Potential decay or dephasing of the ground states can be included as additional jump operators.

Similar to the Jaynes-Cummings results [see Eq. (\ref{eq:EigenvectorsJaynesCummings})], the eigenstates of the coupled three-level system are linear combinations of the uncoupled states $\ket{u,0}$, $\ket{c,1}$ and $\ket{e,0}$. Note that one of these new eigenstates is a dark state that does not contain a contribution from the excited state and, hence, is not subject to spontaneous emission. This state reads:

\begin{equation}
\ket{\Lambda_0} = \cos\theta\ket{u,0}-\sin\theta\ket{c,1}\\
\end{equation}

Here, the mixing angle $\theta$ is given by $\tan\theta=\frac{\Omega_L}{2g}$. The coherent dark state allows one to control the interaction between the atom and a photon in the cavity via the applied laser intensity. This is employed in particular in single-photon generation and storage experiments, as will be discussed in detail in Sec. \ref{sec:QuantumLightSources} and Sec. \ref{sec:QuantumNetworks}. In addition to providing such control, the use of a Raman coupling scheme allows one to trade effective atom-cavity coupling strength for a smaller effective spontaneous emission rate via a proper choice of the detuning $\Delta_{ac}$. For this reason, most processes considered in this review exhibit a scaling with $C$, rather than with the worse of $\frac{g}{\kappa}$ and $frac{g}{\gamma}$.

The first spectroscopic observation of the coherent dark state with a single trapped atom was reported in \cite{mucke_electromagnetically_2010, kampschulte_optical_2010}. Fig. \ref{fig:CavityEIT} shows an experimental transmission spectrum, here recorded with $\sim 15$ atoms to increase the cooperativity and thus the visibility of the observed features. The black curve shows the transmission of the empty cavity, which is a Lorentzian of FWHM $2\kappa$. The red dashed curve shows the vacuum-Rabi splitting obtained in the coupled case without the control laser, i.e. $\Omega_L=0$. Finally, the blue solid curve shows the transmission resonances when $\Omega_L\neq0$. One can clearly resolve the three peaks predicted from Eq. (\ref{eq:EIT_EnergyEigenstates}). As expected, for nonzero control-laser power the vacuum-Rabi peaks are shifted to larger detunings and an additional resonance appears at the frequency of the empty cavity. The width of this transmission resonance is much smaller than both the atomic and cavity linewidths, proving the coherence of the dark state.

\section{Experimental techniques} \label{sec:ExperimentalTechniques}

As mentioned in the introduction, the focus of this article lies on quantum interfaces between optical photons and single atoms trapped in vacuum. Using natural atoms has the advantage that it is easily possible to have several emitters with identical properties that do not change over time. In combination with long atomic ground-state coherence times, this allows one to realize a very clean model system. This, however, comes at the price of three experimental challenges:

First, the small atom-photon interaction strength requires excellent optical cavities. The different resonator designs employed in current experiments are explained in Sec. \ref{sub:resonatorDesigns}.

Second, the requirement to trap single atoms in the vacuum results in a technically demanding experimental setup, including an ultrahigh-vacuum system and well-stabilized lasers. Extensive use can be made of the powerful techniques of laser cooling and trapping \cite{metcalf_laser_1999} that have been developed over the past decades. This will be the topic of Sec. \ref{sub:ObservingAndTrapping} and \ref{sub:Cooling}.

Finally, trapped atoms often exhibit a level structure that is more complicated than the three-level system discussed above. Therefore, techniques to initialize and read out the atomic state are required. These will be presented in Sec. \ref{sub:InternalStateControl}.

\subsection{Optical resonator designs} \label{sub:resonatorDesigns}

\begin{figure*}
\includegraphics[width=172mm]{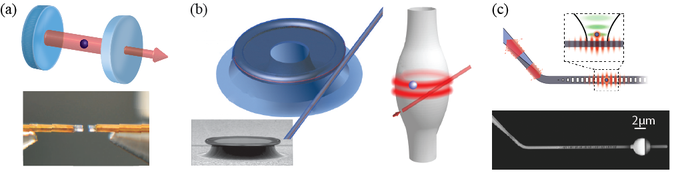}
\caption{ \label{fig:ResonatorTypes}
(color online) Optical resonator designs. (a) Top: Schematic of a Fabry-Perot cavity. The resonator is formed by two facing mirrors with high reflectivity. A small residual mirror transmission allows for in- and outcoupling of light. Bottom: Photograph of a fiber-based Fabry-Perot resonator. The mirrors are the coated end facets of two optical fibers, which allows for improved optical access. The fibers are surrounded by a copper sleeve to facilitate ion trapping in the $\sim 200\text{\textmu m}$ small gap between the mirrors. Adapted from \cite{brandstatter_integrated_2013}. (b) Microtoroid (left) and bottle (right) resonators. Atoms are coupled to the evanescent field of a whispering gallery mode that circulates in the glass structure. The mode diameter is typically on the order of $100\text{\textmu m}$. The resonator is coupled to a nanofiber with a rate that can be adjusted via the fiber distance. Bottom: Scanning electron microscope image of a microtoroid with a diameter of $120\text{\textmu m}$. Adapted from \cite{vahala_optical_2003} and \cite{junge_strong_2013}. (c) Photonic crystal cavity. Top: Sketch of the experiment. Atoms are trapped in a standing-wave optical tweezer, $\sim 0.2\,\text{\textmu m}$ from the surface of the resonator. The electric field of the resonator mode is confined to a volume of less than $\lambda^3$ using a periodic modulation of the refractive index. The cavity is coupled to a propagating mode via a fiber taper visible at the left side. Bottom: Scanning electron microscope image. Adapted from \cite{tiecke_nanophotonic_2014}.
}
\end{figure*}

An optical resonator that allows for the implementation of an efficient quantum interface should fulfill the following two conditions:

\begin{subequations}
\begin{align}
C &\gg 1 \\ 
\kappa_{\text{out}} &\gg (\gamma, \kappa_{\text{loss}}) \label{eq:KappaOut_gg_Loss}
\end{align}
\end{subequations}

These conditions clearly indicate the need for high-quality resonators (i.e. small $\kappa_{\text{loss}}$) with a small mode volume (i.e. large $g$). Several different approaches exist towards realizing such a resonator, the most prominent being the use of Fabry-Perot cavities, whispering-gallery-mode resonators, and photonic-crystal cavities. These will be briefly discussed in the following. For a more detailed review, see \cite{vahala_optical_2003} and \cite{vahala_optical_2004}.

The first approach that has reached $C \gg 1$ in the optical domain is the use of Fabry-Perot resonators. Such a cavity consists of two curved mirrors, facing each other at a short distance, as illustrated in Fig. \ref{fig:ResonatorTypes}(a). In order to achieve the condition expressed by Eq. (\eqref{eq:KappaOut_gg_Loss}), excellent dielectric mirrors with small surface roughness are required. This can be achieved with commercially available superpolished mirrors, typically coated with a stack of dielectric interference layers which are deposited by means of an ion-beam sputtering technique. Transmission and scattering losses both below $1\,\text{ppm}$ per mirror have been reported \cite{rempe_measurement_1992}, corresponding to a finesse of $1.9 \times 10^6$. Details about the characterization of such mirrors can be found in \cite{hood_characterization_2001}. In addition to superpolished mirrors, the fabrication of low-roughness depressions with smaller radii of curvature, either by etching \cite{trupke_microfabricated_2005} or by laser machining \cite{hunger_laser_2012}, is a recent promising direction. The latter has enabled the assembly of cavities with mode volumes on the order of $10^3\,\lambda^3$ and a cooperativity of $C\simeq145$ \cite{colombe_strong_2007}. Optimization of the fabrication process leads to high finesses exceeding $10^5$ \cite{hunger_fiber_2010, muller_ultrahigh-finesse_2010, uphoff_frequency_2015} and frequency-degenerate polarization eigenmodes \cite{uphoff_frequency_2015}.

The use of Fabry-Perot resonators has been widely explored experimentally, with typical cavity linewidths in the range of a few MHz, comparable to the spectral width of an alkaline-atom resonance line. Such resonators have several advantages compared to the other approaches: First, an atom in the resonator can be optically accessed from the side, independent from the resonator mode. This is favorable for Raman coupling, see Sec. \ref{sub:Raman_coupling}, and moreover allows for trapping and cooling in cavity-independent potentials, see Sec. \ref{sub:ObservingAndTrapping} and \ref{sub:Cooling}. Second, the energy of a photon in the resonator is almost completely stored in the vacuum, which makes material absorption less critical and facilitates trapping of several individually addressable atoms in the same resonator mode. Third, the atom is far away from any dielectric surfaces, which prevents the undesired influence of surface forces and charges. Finally, the cavity length can be easily tuned and actively stabilized when the mirrors are mounted on a piezo-electrical crystal.

Unfortunately, these advantages come at the price of an atom-cavity coupling rate that is bounded to a few $100\,\text{MHz}$ since the standing-wave light field of the cavity mode significantly penetrates into the mirror coatings, an effect which sets a lower bound to the achievable mode volume $V$ [see Eqn. (\ref{eq:couplingStrength})] \cite{hood_characterization_2001}. An early approach to improve the coupling strength was the use of so called whispering-gallery modes \cite{vahala_optical_2003} in microspheres, microtoroids, and bottle resonators, see Fig. \ref{fig:ResonatorTypes}(b). In contrast to Fabry-Perot cavities, the field is predominantly guided in the material, typically silicon dioxide. Nevertheless, atoms can be coupled to the cavity mode when they are brought into the evanescent field of the mode, i.e. within a distance of $\sim 100 \,\text{nm}$ to the surface. Typical cavity linewidth values are tens of MHz, which enables to reach the strong-coupling regime in experiments with falling atoms \cite{aoki_observation_2006} with $C\simeq20$, a value comparable to those achieved in Fabry-Perot resonators. Unfortunately, the realization of a larger $C$ seems difficult in these experiments as the atom would need to fly closer to the material where the attractive Van der Waals force would quickly pull the atom onto the surface \cite{alton_strong_2011}. For the same reason, atom trapping has not been achieved with these whispering-gallery resonators so far, which hinders experiments that require long atomic coherence times or the simultaneous presence of several atoms. Beyond these challenges, it has recently been observed that whispering gallery mode resonators open up new possibilities stemming from the nontransversal character of the evanescent light field \cite{junge_strong_2013}.

In contrast to whispering gallery mode resonators, atom trapping has been demonstrated near photonic-crystal resonators, where the photon field is confined by the photonic bandgap that a suitably patterned microstructure can exhibit, see Fig. \ref{fig:ResonatorTypes}(c). The achievable mode volumes can be extremely small, on the order of $\lambda^3$. This results in coupling strengths in the range of GHz, values that have allowed reaching the strong-coupling regime with quantum dots \cite{yoshie_vacuum_2004, reithmaier_strong_2004}. The combination of photonic-crystal cavities and neutral atoms has been early proposed \cite{vuckovic_design_2001} and recently achieved \cite{thompson_coupling_2013}. In this experiment, deterministic loading and stable trapping of a single atom in the cavity mode has been demonstrated using an optical tweezers trap. Further improvements of the resonator design have reduced the resonator linewidth to $\frac{\kappa}{2 \pi} \simeq 13\,\text{GHz}$. With this, the Purcell regime has been reached with a cooperativity of $C\simeq4$ \cite{tiecke_nanophotonic_2014}. The high rates of $g$ and $\kappa$ allow for an investigation of atom-photon interactions on short timescales and make this approach promising for the implementation of quantum interfaces with a large bandwidth.

\subsection{Observing, trapping and cooling single atoms} \label{sub:ObservingAndTrapping}

\begin{figure}
\includegraphics[width=86mm]{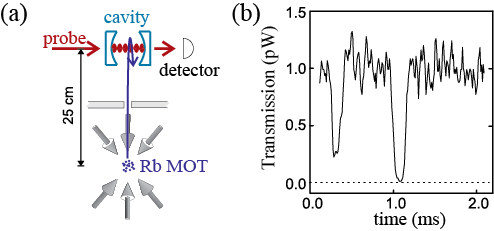}
\caption{ \label{fig:AtomObservation}
(color online) Observation of single atoms launched from an atomic fountain on their pass through a Fabry-Perot cavity. (a) Sketch of the used experimental setup. (b) Cavity transmission signal. When an atom arrives in the mode, the transmission of a weak probe laser drops. The depth and duration of the dip is determined by the cavity parameters and the specific atomic trajectory. Adapted from \cite{munstermann_dynamics_1999}.
}
\end{figure}

CQED with single atoms has a long history. The first experiments in the optical domain employed thermal atomic beams, with a decreasing number of atoms in the cavity, which eventually led to the observation of the vacuum-Rabi splitting with an average atom number of only one \cite{thompson_observation_1992}. However, atom-light interaction was limited to very short times, typically less than a few hundred nanoseconds in these experiments, hampering the unambiguous detection of single atoms.

To improve, the techniques of laser cooling and trapping \cite{metcalf_laser_1999} were introduced to CQED. First steps along these lines were taken by releasing cold atoms from a magneto-optical trap (MOT) such that they fall through a high-finesse optical cavity. This allowed to reach longer interaction times of some ten $\text{\textmu s}$, sufficient to detect single atoms on their transit through the cavity mode by measuring the resonator output when driving the system with a faint laser \cite{mabuchi_real-time_1996, hood_real-time_1998, hennrich_vacuum-stimulated_2000, trupke_atom_2007, terraciano_photon_2009}. This approach has even allowed to continuously observe atoms in real time and to reconstruct individual atomic trajectories \cite{hood_atom-cavity_2000, pinkse_trapping_2000, puppe_single-atom_2004}. Later, it has also been applied to atoms passing toroidal resonators \cite{aoki_observation_2006}, where the transit time through the small resonator mode is even shorter, $\sim 1\,\text{\textmu s}$. Interaction times can be further increased when using an atomic fountain with the cavity at the turning point of the atoms \cite{munstermann_dynamics_1999, munstermann_observation_2000}. Figure \ref{fig:AtomObservation} shows an experimental trace recorded in this configuration.

Albeit many of the basic functionalities of a quantum interface can be demonstrated with flying atoms, it is obvious that the connection of several systems to a quantum network requires atom trapping. The major concern with this respect is to achieve long storage times and the maximum possible coupling strength. The latter requires that the atom is tightly localized, such that the extent $x_0$ of the atomic wavepacket in the ground state of the potential is small compared to the mode structure of the cavity. This usually implies $x_0 \ll \lambda$, favoring the use of tight potentials with trap frequencies on the order of $100\text{kHz}$. Such trap frequencies are larger than the recoil frequency that corresponds to the momentum of a single resonant photon. Therefore, the motional state of the atom rarely changes in absorption and emission events. Such a situation is called the Lamb-Dicke regime \cite{leibfried_quantum_2003} and is very favorable for efficient laser cooling, even to the ground state of the potential.

Trapping in the Lamb-Dicke regime has been achieved in two different approaches: First, by electrical trapping of charged atoms --- usually cations \cite{leibfried_quantum_2003}, and second, by optically trapping neutral atoms in far-detuned laser fields \cite{grimm_optical_2000, ye_quantum_2008}. These two approaches will be discussed in the following.

\subsubsection{Ion traps}

While the first CQED experiments with trapped atoms have used neutral atoms \cite{ye_trapping_1999}, also the implementation of an ion trap in a Fabry-Perot resonator has early been realized \cite{guthohrlein_single_2001, mundt_coupling_2002}. This has allowed for very long trapping times of more than $90\,\text{min}$ \cite{keller_continuous_2004}. In addition, excellent control over the position of trapped ions within the resonator field has quickly become experimentally feasible. This can be seen in Fig. \ref{fig:Trapping}(a), where the position of a single ion has been shifted in a controlled way, thus mapping out the mode structure of a Fabry-Perot resonator \cite{guthohrlein_single_2001}. Recently, position control has also been demonstrated individually for two ions trapped in the same cavity \cite{casabone_heralded_2013}.

Unfortunately, experiments with ion traps have long been limited to rather long cavities because of the technical difficulty that surface charges on the dielectric mirrors of optical resonators tend to disturb the electric trapping potential. Therefore, neither the Purcell nor the strong-coupling regime has been reached with trapped ions until today, albeit there is recent progress due to the development of high-finesse fiber-based Fabry-Perot cavities \cite{hunger_fiber_2010}. Because of their smaller size, the size of the required dielectric surfaces is reduced and the resulting fields can be efficiently shielded using metal sleeves. Thus, such resonators allow for ion trapping in a sufficiently small mode volume to fulfill the condition $g>\gamma$ \cite{steiner_single_2013}. With improved mirror finesse, operation in the Purcell regime might become feasible. Other experiments even try to reach the strong-coupling regime \cite{brandstatter_integrated_2013}.

\subsubsection{Optical dipole traps}

In contrast to trapped ions, a cooperativity larger than unity has been achieved in many experiments trapping neutral atoms. Remarkably, it has early been demonstrated in two independent experiments \cite{hood_atom-cavity_2000, pinkse_trapping_2000} that even the force of a laser beam that contains only one photon on average can suffice to trap an atom on a ms timescale when operating on a strongly coupled transition. In all experiments that employ several atomic levels, however, a state-insensitive trap is required. In this respect, the first milestone was the achievement of a $28\,\text{ms}$ atom storage time \cite{ye_trapping_1999} in a far-off-resonance dipole trap (FORT) (see e.g. \cite{metcalf_laser_1999, grimm_optical_2000}), implemented in a standing-wave configuration along the cavity axis.

Optical traps are typically only a few mK deep. Therefore, it is important to load the trap with precooled atoms. Most experiments use a magneto-optical trap as a starting point. Then, the atoms are transferred to the cavity field by dropping them from above, using an atomic fountain from below, or employing a running-wave transfer trap \cite{nusmann_submicron_2005}. In combination with single-atom imaging, even deterministic atom loading has been demonstrated, either with an optical conveyor belt \cite{fortier_deterministic_2007, khudaverdyan_controlled_2008} or by shifting an optical tweezers trap \cite{thompson_coupling_2013} to the resonator mode. An alternative method uses single-atom extraction from a small Bose-Einstein condensate, transferred to the cavity mode by shifting a confining radio-frequency potential on an atom chip \cite{gehr_cavity-based_2010, volz_measurement_2011}.

\begin{figure}
\includegraphics[width=86mm]{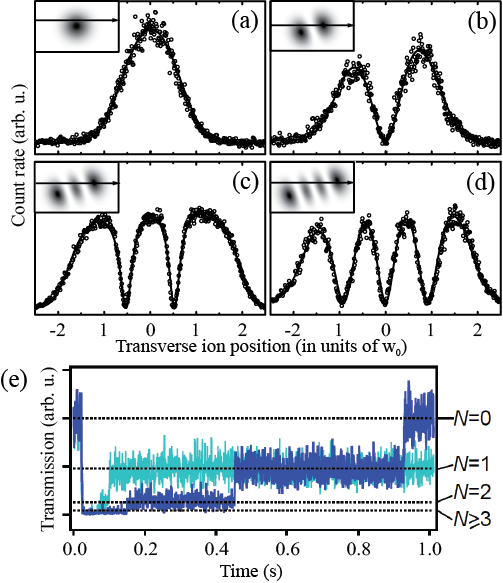}
\caption{ \label{fig:Trapping}
(color online) (a-d) Measurement of the spatial structure of the electromagnetic field within a high-finesse Fabry-Perot cavity. An ion is trapped in a radio-frequency potential. When the trap is shifted, the rate of photons scattered into free space out of a laser beam that drives the resonator mode reveals the spatial structure of the cavity field. The solid curve is a fit that includes saturation of the atomic transition. By changing the incoupling, different resonator modes can be mapped (a to d). The inset shows the calculated mode structure, and an arrow indicates the scan path. Adapted from \cite{guthohrlein_single_2001}.
(e) State-insensitive trapping of single atoms in a Fabry-Perot cavity in the strong-coupling regime. The depth of transmission suppression of a faint resonant probe laser allows for continuous monitoring of the number of atoms $N$ trapped in the resonator (when $N<3$). In both depicted traces, the initial drop is caused by several atoms that are cooled into the cavity mode. Subsequently, the transmission increases in steps whenever one atom is lost from the trap. Adapted from \cite{mckeever_state-insensitive_2003}.
}
\end{figure}

To observe long trapping times even when the atom is constantly illuminated requires efficient cooling, which has first been demonstrated in \cite{mckeever_state-insensitive_2003}. Fig. \ref{fig:Trapping}(b) shows an experimental trace obtained in this study, where the transmission of a faint laser through the resonant cavity is shown. When one or several atoms are present in the trap, the transmission is reduced with increasing atom number. This has allowed to demonstrate single-atom trapping under constant observation for a period of about $500\,\text{ms}$. The techniques developed to further increase the trapping times and to reduce the temperature will be discussed in more detail in Sec. \ref{sub:Cooling}.

As mentioned above, it is highly desirable to achieve tight confinement and high trap frequencies. With optical traps, this typically requires a standing-wave configuration and high intensities. Unfortunately, the latter comes at the price of potentially large scattering rates that can cause heating and decoherence. The common solution is to choose a large detuning $\Delta$ of the trap light with respect to the atomic transition frequency, as the trap depth scales as $\frac{1}{\Delta}$, the scattering rate however as $\frac{1}{\Delta^{2}}$ \cite{grimm_optical_2000}. This is the reason why most CQED experiments with optical dipole traps operate at a detuning of tens or even hundreds of nanometers, often limited by the power the resonator can withstand.

Still, achieving long trapping times with neutral atoms can be hampered by large parametric heating rates caused by fluctuations of the optical trap. This is especially problematic when using an intracavity dipole trap whose intensity is typically very sensitive to acoustic vibrations of the cavity mirrors. In \cite{nusmann_vacuum-stimulated_2005}, this problem was avoided using a cavity-independent FORT. With a combination of vacuum-stimulated and Sisyphus cooling, trapping times of $17\,\text{s}$ have been demonstrated, later increased to more than one minute on average \cite{specht_single-atom_2011}, which is most likely limited by collisions with the room-temperature background gas.

In addition to increased storage times, using a cavity-independent trap has another advantage: it facilitates changing the position of the atom along the beam axis with sub-micron precision by shifting the standing-wave pattern of the trapping laser \cite{nusmann_submicron_2005}. In combination with single-atom imaging, full control over the atomic position with respect to the cavity field has thus also been achieved with neutral atoms \cite{reiserer_ground-state_2013}.

\subsubsection{Cooling} \label{sub:Cooling}

Numerous approaches have been explored to improve the localization of atoms in CQED and to extend the storage times. An early approach was to gain information about the atomic position and momentum from the temporal modulation of a resonant laser beam transmitted through the cavity \cite{hood_atom-cavity_2000, pinkse_trapping_2000}. Subsequent feedback onto the trap allowed to reduce the scattering-induced heating \cite{fischer_feedback_2002}. With the implementation of an optical dipole trap and fast electronics, feedback-increased storage times have been observed \cite{kubanek_photon-by-photon_2009}. With a higher output-coupling efficiency of the cavity, the available rate of information about the atomic motion has been increased. This facilitated cooling of the atomic motion and average trapping times exceeding $1\,\text{s}$ \cite{koch_feedback_2010}.

The first CQED experiments with second-long neutral-atom trapping times \cite{mckeever_state-insensitive_2003}, however, used a different cooling method, namely a combination of red-detuned Doppler cooling (on a closed transition) and blue-detuned Sisyphus cooling (on another transition). In these experiments, Cs atoms were trapped in a magic-wavelength \cite{ye_quantum_2008} FORT in a standing-wave configuration along the cavity axis. With a cavity-independent trap, trapping times have been increased above one minute on average \cite{specht_single-atom_2011} using Rb atoms and intracavity-Sisyphus cooling \cite{nusmann_vacuum-stimulated_2005}.

The above mentioned cooling mechanisms --- feedback, Doppler, and Sisyphus cooling --- all lead to the observation of long storage times. However, it is also important to achieve the highest possible cooling rate, as many experimental protocols require photon scattering that causes heating. Similarly, it is also important to cool the atoms to a low temperature to avoid motion-induced dephasing, fluctuations in the atom-cavity coupling strength and a time-varying AC Stark shift of the atomic levels in case the trap is not a magic one \cite{ye_quantum_2008}. Therefore, the quest for fast and efficient cooling mechanisms has not come to an end yet.

Along these lines, a mechanism has been proposed \cite{horak_cavity-induced_1997, vuletic_laser_2000} that directly makes use of the coupling to a cavity. The first implementation \cite{maunz_cavity_2004} of this cavity-cooling method used neutral Rb atoms in an intracavity dipole trap, while cooling of a $\text{Sr}^+$ ion is reported in \cite{leibrandt_cavity_2009}. Meanwhile, even cooling below the recoil limit has been demonstrated with an atomic ensemble trapped in a long cavity of high finesse \cite{wolke_cavity_2012}. Unfortunately, cavity cooling requires the operation at a specific detuning, which often conflicts with experiments that operate at a different detuning or on resonance. The same holds for the recently proposed \cite{bienert_cavity_2012} and observed \cite{kampschulte_electromagnetically-induced-transparency_2014} cooling based on electromagnetically induced transparency \cite{roos_experimental_2000} in a cavity QED setting. Thus, the use of another technique is often desirable.

To this end, Raman sideband cooling has been investigated, following many successful experiments in free-space with single trapped ions \cite{leibfried_quantum_2003}. The first experiment in CQED has used this mechanism in one dimension only, achieving a ground state probability of $95\%$ along the axis of an intracavity dipole trap \cite{boozer_cooling_2006}. Later, cooling to the three-dimensional ground state could be demonstrated in a three-dimensional optical lattice with high trap frequencies in all directions \cite{reiserer_ground-state_2013}.

\begin{figure}
\includegraphics[width=86mm]{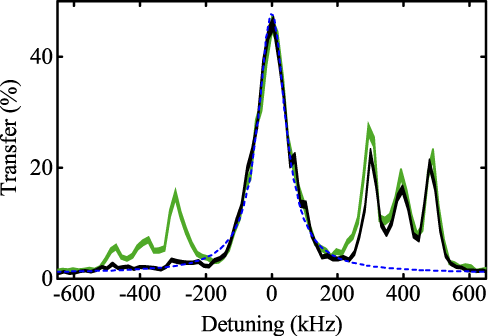}
\caption{ \label{fig:GroundStateCooling}
(color online) Full control over the atomic motion. Atoms are trapped in a three-dimensional optical lattice with different trap frequencies along the three orthogonal axes. When a Raman sideband spectrum is recorded after intracavity Sisyphus cooling (green), three peaks are resolved on either side of the carrier (dashed blue Lorentzian fit), at the positions of the three red (left side) and the three blue (right side) sidebands. After Raman sideband cooling (black), the atom is prepared in the three-dimensional ground state of the trapping potential and the red sideband peaks vanish. From \cite{reiserer_ground-state_2013}.
}
\end{figure}

Fig. \ref{fig:GroundStateCooling} shows the experimental results obtained in this study. The green curve shows a sideband spectrum after intracavity Sisyphus cooling. The mean vibrational quantum number, obtained from the ratio of the red and blue sidebands \cite{leibfried_quantum_2003}, is about or even below one quantum along all three spatial axes. After $5\,\text{ms}$ of Raman sideband cooling, the red sideband peaks vanish completely, proving cooling to the three-dimensional motional ground state with a probability of $89(2)\,\%$.

In addition to its motion, also the position of the atom was fully controlled in the above experiment, thus realizing the ideal CQED situation: A point-like single atom, trapped at a fixed position and strongly coupled to an optical resonator.

\subsection{Measurement and control of the internal atomic state} \label{sub:InternalStateControl}

The implementation of quantum interfaces not only requires control over the position and momentum of the trapped atoms, but also techniques to initialize, manipulate and read out the internal state of the atoms. Many of the required tools have been pioneered in free-space experiments with trapped ions \cite{leibfried_quantum_2003} and neutral atoms \cite{meschede_manipulating_2006}. In the following section, the most common techniques realized in CQED setups are summarized.

\subsubsection{Detection of the atomic state} \label{subsub:StateDetection}

In any quantum network, the projective readout of the stationary nodes is of central relevance. In the following, the properties that an ideal readout process should exhibit are listed: First, it is nondestructive. This trivially means that after the readout procedure, the atom is still trapped. Second, the readout gives an unambiguous result in every attempt. Third, after the measurement the atom is projected to an eigenstate within the qubit subspace with a probability that fully reflects the initial atomic state. Finally, the readout procedure is fast enough to allow for feedback onto the quantum state, which is a prerequisite for measurement-based quantum information processing \cite{briegel_measurement-based_2009}, entanglement purification \cite{bennett_purification_1996, dur_entanglement_2007} and quantum error correction \cite{lidar_quantum_2013}.

In principle, atomic state readout is possible using atom-photon quantum-state transfer, followed by detection of the photon. This technique will be described in detail in Sec. \ref{sub:AtomPhotonQuantumStateTransfer}. Unfortunately, due to limited transfer and detection efficiencies, the readout procedure is not deterministic.

Instead, to achieve the ideal atomic state readout described above, two other techniques can be employed in CQED systems. Both are based on measuring the photons emitted from the cavity while optically exciting the system. The difference is that in the first approach the cavity field is excited while in the second approach the atom is driven by a laser impinging from the side of the resonator.

We start with describing the second approach, fluorescence state detection, which is similar to the electron-shelving technique pioneered in ion-trap experiments \cite{leibfried_quantum_2003}. The basic idea is to drive the atom on a cycling transition, say the transition between the coupled state $\ket{c}$ and the excited state $\ket{e}$, and to detect the fluorescence photons with single photon counters. When the atom is in the 'dark' level $\ket{u}$, which is detuned much further than the linewidth of the transition, then no photons will be detected. Clearly, the achievable fidelity is limited by the probability to lose the atom from the trap or to pump it from the dark to the bright level (or vice versa) before detecting enough photons to discriminate the two.

\begin{figure}
\includegraphics[width=86mm]{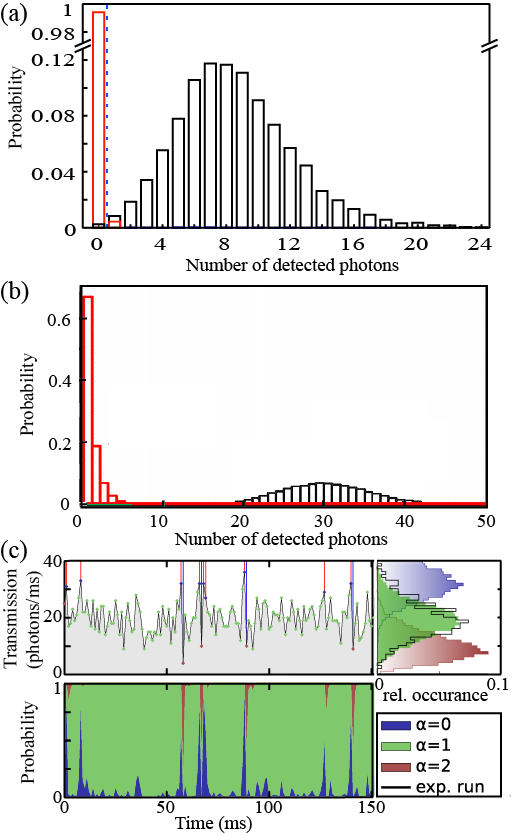}
\caption{ \label{fig:StateDetection}
(color online) Atomic state detection
(a) Fluorescence state detection. The cavity output photons are measured when the atom is driven on a closed transition with a laser beam applied from the side of the cavity. By setting a suitable threshold (blue dashed line), one can unambiguously discriminate the fluorescent 'bright' state (black), where several photons are detected within a certain time interval, from the detuned 'dark' state (red), where no photons are detected. Adapted from \cite{bochmann_lossless_2010}.
(b) Transmission state detection. A weak resonant laser drives the cavity along its axis. When the atom is in a detuned state $\ket{u}$, it does not alter the transmission of the resonator. Within a certain time interval, one observes a Poissonian distribution of detected photons (black). If the atom is in the coupled state ($\ket{c}$, red), the transmission is strongly suppressed. The two obtained distributions are clearly separated, which allows for unambiguous discrimination of the two states. Adapted from \cite{boozer_cooling_2006}.
(c) Continuous detection of and feedback onto the combined state of two atoms trapped in the same resonator. Top: Experimentally obtained cavity transmission signal. The vertical lines indicate the instants when a feedback pulse is applied to transfer one of the atoms to the coupled or uncoupled ground state. Bottom: number of atoms $\alpha$ in the coupled state according to the Bayesian state estimation. Because of the feedback, the probability for $\alpha=1$ is maximized. Right: Histograms of the resonator transmission with exactly $(0,1,2)$ atoms prepared in the coupled state. When two atoms are loaded into the resonator and the feedback is applied, the histogram (black envelope) matches that observed with exactly one atom in the coupled state. Adapted from \cite{brakhane_bayesian_2012}.
}
\end{figure}

When the atom is trapped in a cavity, one can take advantage of the Purcell effect, see Sec. \ref{subsub:PurcellRegime}, which guides the emitted photons into the cavity mode. Thus, very high photon collection efficiencies can be realized, thereby drastically reducing the number of scattered photons that are required to detect the atomic state. This procedure has first been implemented to detect the hyperfine state of a $\Rb$ atom trapped in a cavity in the intermediate coupling regime \cite{bochmann_lossless_2010}. Fig. \ref{fig:StateDetection}(a) shows a histogram of the detected photon number per shot when the atom is in the bright (black) or dark (red) state, respectively. The two states can be discriminated within $85\,\text{\textmu s}$ with a fidelity of $99.4\,\%$.

The fluorescence scheme has two major advantages. First, it is not very sensitive to the atom-cavity detuning. This can facilitate a state measurement for each of several atoms that are trapped in the same cavity and detuned from its resonance. Second, it provides a high contrast between the bright and dark state, even if $C \lesssim 1$. This allows for very fast state detection irrespective of the detector darktime. Thus, a fidelity of $99.65\,\%$ within $3\,\text{\textmu s}$ has recently been achieved in a system with $C \simeq 3$ \cite{reiserer_quantum_2014}. In comparison to \cite{bochmann_lossless_2010}, less time is required to detect the atomic state, which is the result of improved control over the atomic position and momentum \cite{reiserer_ground-state_2013}.

In contrast to the fluorescence scheme, state detection via pumping the cavity typically requires operation at $C > 1$. Here, the detection mechanism is a change of the cavity transmission and reflection when an atom is coupled to the resonator, as explained in more detail in Sec. \ref{sub:CouplingToPropagatingFields}. The achievable fidelity is again limited by the probability of atom loss and bright-to-dark state pumping. Fig. \ref{fig:StateDetection}(b) shows the results obtained in the first experiment that successfully demonstrated state detection using this approach \cite{boozer_cooling_2006}. The transmission is high (black) when the atom is in a detuned state $\ket{u}$, while it is strongly suppressed (red) if the atom is in the coupled state $\ket{c}$. This has allowed for detecting the hyperfine state of trapped Cesium atoms with a fidelity of $>98\,\%$ within $100\,\text{\textmu s}$.

After the first demonstration, state detection via transmission measurements has found applications in several experiments. Examples include the observation and characterization of quantum jumps of single atoms in a cavity \cite{khudaverdyan_quantum_2009} and the application of Bayesian feedback onto the joint spin state of two atoms \cite{brakhane_bayesian_2012}. In this experiment, the parameters are chosen such that the cavity transmission is reduced by ${0, 30\,\%, 60\,\%}$ when the number of atoms $\alpha$ in the coupled hyperfine ground state is equal to ${0, 1, 2}$. A Bayesian update formalism is used to estimate state occupation probabilities as well as transition rates. Based on this analysis, a digital feedback loop switches the intensity of a depumping and a repumping laser beam, aiming to stabilize the system in a mixed state with exactly one atom in each of the two hyperfine ground states. Fig. \ref{fig:StateDetection}(c) shows the major result obtained in this study. With feedback, a system state with one out of two atoms coupled to the cavity mode ($\alpha=1$) can be stabilized to a high degree, as can be seen in the calculated Bayesian probability (bottom) and in the transmission histogram on the right side. Here, the measured curve (black) closely matches the transmission in case a single coupled atom is present, which is clearly distinct from those cases where the two atoms occupy the same ground state.

Note that in the above experiment, the stabilized quantum state is a statistical mixture of $\ket{c_1,u_2}$ and $\ket{u_1,c_2}$, where the subscript denotes the individual atom. To facilitate feedback that stabilizes superposition or even entangled quantum states, two requirements have to be fulfilled: First, the atomic phase must remain unchanged during the time required to detect the atomic population. Second, the atomic state has to be detected without photon scattering into free space as that would reveal information about which atom is in the fluorescing state.

Decisive steps in this direction have been made in a system with a higher cooperativity. First, a state detection fidelity of $99.4\,\%$ has been achieved in a readout interval of $2\,\text{\textmu s}$ \cite{gehr_cavity-based_2010}, which is much shorter than typical coherence times of Zeeman qubits. Second, it has been shown that transmission and reflection measurements allow one to detect the atomic state without energy exchange, i.e. with less than one scattered photon on average \cite{volz_measurement_2011}. This has facilitated the measurement-based creation of entangled Dicke states of an atomic ensemble consisting of up to 41 atoms on average \cite{haas_entangled_2014}.

With respect to single emitters, state detection without energy exchange is very encouraging towards the implementation of quantum feedback and quantum error correction. Most notably, it should facilitate the readout of individual qubits in a quantum register in which the problem of crosstalk, i.e. the scattering-induced decoherence of adjacent unread qubits, has to be avoided. In addition, the technique might allow one to detect the state of quantum systems that do not exhibit a sufficiently closed transition, such as molecules or nitrogen-vacancy centers at high strain splitting. In this respect, the only hurdle is achieving a sufficient cooperativity.

\subsubsection{Initialization of the atomic state}

Most experiments require preparation of the atom in a well-defined internal state, which includes the principal and orbital quantum numbers of the electronic state and the spin of both the electron and the nucleus. While in principle, measurement and feedback can be used to initialize the atomic state, most experiments to date use a procedure known as optical pumping, which simply means the irradiation of laser light that drives transitions from all but one ground state level. Thus, the atomic population will accumulate in this level after a certain time interval.

Most experiments to date are performed with heavy single-electron atoms at low magnetic fields. In this case, the atomic state $\ket{F,m_F}$ is described by two quantum numbers representing the total angular momentum ($F$) and its projection onto the quantization axis ($m_F$) \cite{metcalf_laser_1999}. When this axis coincides with that of an external magnetic field, the $m_F$ number refers to the Zeeman state.

When the used laser light is circularly polarized and irradiated along the quantization axis, it drives $\sigma^+$ or $\sigma^-$ transitions with $\Delta m_F= \pm1$. Thus, one of the two 'stretched' states, $\ket{F,F}$ or $\ket{F,-F}$, can be prepared. When the laser is applied orthogonal to the quantization axis with parallel linear polarization, it drives $\pi$ transitions with $\Delta m_F=0$. The dipole selection rules \cite{metcalf_laser_1999} then allow for the preparation of $\ket{F,m_F=0}$ by irradiating $\pi$-polarized light on the transition $F \leftrightarrow F'=F$. Other Zeeman states can be prepared by applying additional Raman lasers \cite{boozer_optical_2007} or microwave radiation \cite{tiecke_nanophotonic_2014} that spectrally resolve individual Zeeman states. This will be further discussed in Sec. \ref{subsub:Rotations}.

State preparation via optical pumping is a stochastic process. The achievable fidelity of state preparation in a specific Zeeman state is therefore limited by the ratio of the average desired pumping rate, compared to the rate of undesired transfer to other states. The latter can either be caused by qubit decoherence or by technical imperfections such as misaligned lasers or magnetic field fluctuations. Trivially, state preparation in a specific Zeeman level gets more difficult the more levels an atom exhibits. Fortunately, the preparation fidelity can be substantially improved, see e.g. \cite{reiserer_nondestructive_2013}, when an initial stage of optical pumping is combined with fast transmission state detection, explained in Sec. \ref{subsub:StateDetection}, which allows for detecting whether the atom has been pumped to the desired state or not.

While the above procedures can be used to prepare the atom in a specific energy eigenstate, many advanced experiments require to prepare a specific superposition state, e.g. between a coupled and an uncoupled state, $\frac{1}{\sqrt2}(\ket{c}+\ket{u})$. Such states can be prepared with atomic state rotations, which will be the topic of the following section.

\subsubsection{Coherent control of the atomic state} \label{subsub:Rotations}

Exploiting the full potential of quantum networks requires coherent control of the quantum state of each node. The basic operations to achieve such control are often called single-qubit gates, unitary transformations, or coherent rotations. The basic principles of such rotations have been pioneered in nuclear magnetic resonance experiments and later adapted to quantum information processing \cite{vandersypen_nmr_2005} in several different physical systems, including trapped ions \cite{leibfried_quantum_2003} and neutral atoms \cite{meschede_manipulating_2006}.

In atomic systems, coherent control can be implemented by driving the transition between the qubit states $\ket{u}$ and $\ket{c}$ with an external control field. Depending on the energy difference between the states, this field can be at optical or microwave frequencies. Alternatively, it is also possible to couple the levels with a two-photon Raman transition, see e.g. \cite{leibfried_quantum_2003} and Sec. \ref{sub:Raman_coupling}. In both cases, the dipole selection rules of the transitions have to be considered.

In most atomic systems, the qubit states are part of a larger hyperfine manifold. To remain in the qubit subspace, the degeneracy between the Zeeman states is often lifted by applying an external magnetic field. This allows for spectrally selective addressing of individual transitions.

\begin{figure}
\includegraphics[width=86mm]{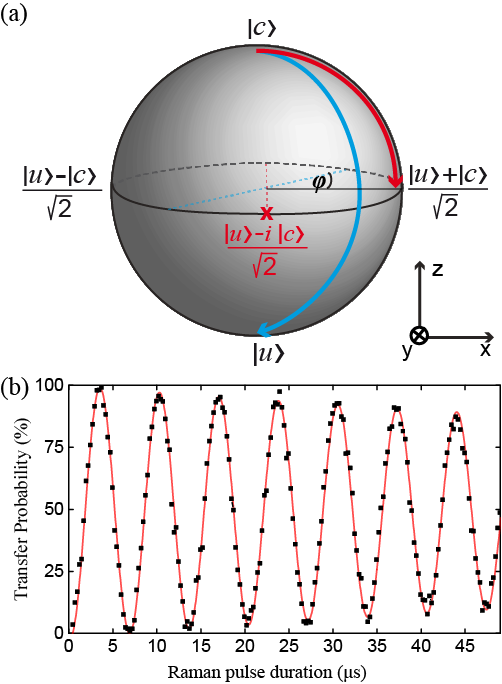}
\caption{ \label{fig:RabiWithBlochSphere}
(color online)
(a) Bloch sphere representation of a two-level system. Any pure atomic superposition state $\alpha \ket{u}+\beta \exp(i\varphi) \ket{c}$ is represented by a point on the surface of a sphere. The latitude represents the ratio of $\alpha$ and $\beta$, while the longitude gives $\varphi$. Mixed quantum states can be represented by points within the sphere.
(b) Rabi oscillations demonstrating controlled atomic state rotations. The atom is prepared in the state $\ket{c}$. Subsequently, a pair of Raman lasers drives transitions to the state $\ket{u}$. The atomic population in $\ket{c}$ is measured by fluorescence state detection. The visibility of the resulting Rabi Oscillations (here: $98\,\%$) is a good measure for the combined quality of the state preparation, rotation and readout process. The observed decay of the oscillation can be caused by fluctuating laser beam intensities and fluctuating ground-state energies, e.g. due to changes in the magnetic field. Adapted from \cite{reiserer_controlled_2014}.
}
\end{figure}

For a visualization of coherent ground state control, the Bloch-sphere picture is very helpful. Here, any superposition $\alpha \ket{c}+\beta \exp(i\varphi) \ket{u}$ (with $\alpha^2+\beta^2=1$) of the two atomic ground states is represented by a point on the surface of a sphere, as can be seen in Fig. \ref{fig:RabiWithBlochSphere} (a).

Assume that the control field is applied for a time $\tau$ with a Rabi frequency of $\Omega_r(t)$ and a detuning $\Delta_r$. In the Bloch sphere picture, this implements the rotation:

\begin{equation}
R_{\vec{n}}(\theta)=\exp(i\theta \vec{n}\vec{\sigma}/2)
\end{equation}

Here, $\theta=\int_{0}^\tau \Omega_r (t) dt$ is the angle of rotation and $\vec{\sigma}$ is a formal vector of the Pauli matrices $\sigma_x$, $\sigma_y$ and $\sigma_z$. The rotation axis $\vec{n}$ is given by:

\begin{equation}
\vec{n}= \cos(\varphi)\textbf{x} + \sin(\varphi)\textbf{y} + \frac{\Delta_r}{\Omega_r} \textbf{z}
\end{equation}

In this equation, $\textbf{x}$, $\textbf{y}$ and $\textbf{z}$ are the coordinate vectors of the Bloch sphere and $\varphi$ is the phase difference between the atomic dipole and the external control field. Clearly, any rotation can be implemented by a suitable choice of control field parameters. In Fig. \ref{fig:RabiWithBlochSphere}(a), two such rotations are indicated. One is the rotation $R_\textbf{y}(\pi/2)$ (red), the other is $R_\varphi(\pi)$ (light blue).

According to Bloch's theorem, any rotation can be decomposed into a series of rotations around two different axes. This reduces the control parameters and thus simplifies the experimental sequence. In many cases, one chooses $\Delta_r=0$, $\varphi=0$ or $\varphi=\pi/2$, and $\Omega_r(t)=\Omega_0$, which leads to rotations at constant speed $\Omega_0$ around the axes $\textbf{x}$ or $\textbf{y}$, respectively. Then, the only remaining parameter that has to be controlled is the pulse duration $\tau$.

Fig. \ref{fig:RabiWithBlochSphere}(b) shows an experimental implementation of such a qubit rotation around the $\textbf{x}$ axis. The control field is applied with constant intensity for a varying duration. This results in the observation of Rabi oscillations with a visibility of $98\,\%$, demonstrating the high quality of the combined state preparation, rotation and readout process achieved in this simple scheme. The fidelity of the rotations can be further increased using more advanced techniques such as shaped or composite pulses, and optimal control theory. For a review, see \cite{vandersypen_nmr_2005}.

\section{Quantum light sources} \label{sec:QuantumLightSources}

Two different approaches are typically considered towards the generation of entanglement in a quantum network and the transfer of quantum states from one network node to another. Both rely on encoding quantum information in nonclassical light fields. While the first approach is based on the transmission of single photons, see e.g. \cite{cirac_quantum_1997, bose_proposal_1999}, the second employs continuous quantum light fields \cite{braunstein_quantum_2005} such as squeezed states, see e.g. \cite{kraus_discrete_2004}.

The latter approach has been very successful in experiments with atomic ensembles \cite{hammerer_quantum_2010}. However, it is still unclear whether it can be used with single-particle nodes, because the existing protocols are typically hampered by errors arising from photon loss. Nevertheless, we emphasize that numerous experiments have investigated the generation of continuous quantum light fields with atom-cavity systems, see e.g. \cite{rempe_optical_1991, birnbaum_photon_2005, schuster_nonlinear_2008, kubanek_two-photon_2008, dayan_photon_2008, armen_spontaneous_2009, ourjoumtsev_observation_2011, koch_three-photon_2011, tiecke_nanophotonic_2014, volz_nonlinear_2014}.

In contrast, quantum networks based on the transmission of single photons have been realized in CQED and will therefore be discussed in detail in this article. Towards this end, qubits $\alpha \ket{\uparrow} + \beta \ket{\downarrow}$ can be encoded in various degrees of freedom of the electromagnetic field. The most fundamental encoding is the simple presence $\ket{\uparrow} \equiv \ket{1}$ or absence $\ket{\downarrow} \equiv \ket{0}$ of a photon. Clearly, this encoding is also prone to errors caused by photon loss, which transfers $\ket{1}$ to $\ket{0}$. Therefore, one typically encodes the qubits in the polarization, the frequency, the arrival time, or the path of propagation of a photon.

While these encodings all have specific advantages, they share the property that photon loss \emph{destroys} the qubit, but does not \emph{rotate} it from one state to the other. One can thus use a sensitive photo detector to select only those cases where the photon has not been lost. This facilitates the implementation of protocols that are insensitive to photon loss or can correct for it in an efficient way. This is the prerequisite for the transfer of quantum states both over large distances and using imperfect devices.

Before turning to the experimental generation of single-photon fields, a few remarks about their characterization seem appropriate. It is well known that the properties of a quantum state cannot be determined with a single measurement. Instead, analysis of a large ensemble of identically prepared states is required. Unfortunately, this condition is often hard to achieve in an experiment. Typically, one will rather prepare a statistical mixture of quantum states, described by a density operator.

The generated mixed state of the light field can be fully reconstructed when it is interfered with a classical laser field with well-known properties, a technique known as homo- or heterodyning (see e.g. \cite{lvovsky_continuous-variable_2009, agarwal_quantum_2013}. Unfortunately, this approach becomes more difficult with decreasing field amplitude due to technical noise. Therefore, in many cases the properties of quantum light fields are characterized using single-photon counting modules (SPCMs), which allow both for heterodyning \cite{thompson_high-brightness_2006} and for the direct detection of single-photon events.

Today's commercial SPCMs typically exhibit high quantum efficiencies ($>50\,\%$) and low dark count rates ($<10\,\text{Hz}$) \cite{eisaman_invited_2011}, providing an excellent signal-to-noise ratio even for weak photon streams. Based on correlation measurements, very powerful techniques have been developed in quantum optics to characterize nonclassical light fields using such detectors. For a detailed explanation, see e.g. \cite{agarwal_quantum_2013}. Similarly, many techniques have been developed to reconstruct single- and multi-qubit states. To reconstruct the density operator, one typically employs a scheme known as quantum state tomography, see e.g. \cite{altepeter_photonic_2005, nielsen_quantum_2010}.

\subsection{Generation of single photons} \label{sub:PhotonGeneration}

From the very beginning of research in CQED, many schemes have been proposed to use the strong atom-photon interaction in a cavity to generate nonclassical light fields. Pioneering work in the microwave domain \cite{rempe_observation_1990, raimond_manipulating_2001} concentrated on the generation and detection of photons trapped within the cavity. Along these lines, early proposals to generate Fock states in the optical domain \cite{parkins_synthesis_1993} also treated the decay of the cavity field as a loss channel.

A paradigm shift emerged from the seminal work of \cite{cirac_quantum_1997}, where it was realized that the cavity decay, when caused by mirror transmission, efficiently couples the atomic state to propagating photonic modes. Photons in this mode can then be sent to and absorbed by another atom-cavity system which thus forms the elementary building block of a quantum network. The first experimental step in this direction is the generation of single photons from an atom-cavity system, which will be described in this section.

To understand the basic mechanism of photon generation, consider a two-level atom trapped in a cavity with $C \gg 1$. The transition from the ground state $\ket{c}$ to the excited state $\ket{e}$ is driven using an external laser field applied from the side of the resonator. From the excited state, the atom can decay under emission of a photon, either into free space or into the cavity mode. The dynamics of this decay have been discussed in Sec. \ref{sub:damped_AtomCavitySystems}. When $\kappa_{\text{out}}\gg\kappa_{\text{loss}}$, intra-cavity photons are emitted into a propagating mode, which facilitates efficient single-photon sources.

In the Purcell regime, the single-photon character of the cavity output field has the same origin as for an atom in free space \cite{kimble_photon_1977}. In fact, upon photon emission, the atom is projected to the ground state $\ket{c,0}$. Before it can emit a second photon, the atom has to be excited again, which takes a finite amount of time. This leads to the observation of antibunching, $g^{(2)}(0)=0$, when only one atom is trapped in the resonator and the (normally-ordered) intensity correlation function $g^{(2)}(\tau)=\frac{\langle a^\dagger (0) a^\dagger (\tau) a(\tau) a (0) \rangle}{\langle a^\dagger a \rangle^2}$ of the light field emitted from the system is measured in a Hanbury Brown and Twiss setup [see e.g. \cite{agarwal_quantum_2013}].

In the strong-coupling regime, the emission of a photon from the cavity does not necessarily project the atom to its ground state $\ket{c}$. Still, when the system, i.e. either the atom or the cavity, is driven on resonance with one of the states $\ket{1,+}$ or $\ket{1,-}$, the output field is a stream of single photons because the nonlinearity of the Jaynes-Cummings ladder makes the coupled system resemble a two-level atom. This topic will be revisited in Sec. \ref{subsub:PhotonPhoton_Gates}.

When the atom is constantly driven, it steadily emits photons at random times. In contrast, most quantum network experiments require the emission of only one photon in a well-defined temporal mode. This can be achieved using two different strategies. First, the atom can be excited to the state $\ket{e}$ using a laser pulse that is much shorter than the decay time of the system \cite{bochmann_fast_2008}. This is the standard technique to generate photons from a system that exhibits only two levels, see \cite{eisaman_invited_2011} for a review.

Alternatively, one can employ a Raman coupling scheme to generate single photons. Such scheme can be implemented both in the Purcell regime \cite{law_deterministic_1997} and in the strong-coupling regime \cite{kuhn_controlled_1999}. It makes use of the presence of a third level $\ket{u}$, as described in Sec. \ref{sub:Raman_coupling}. Here, the transition from the uncoupled ground state $\ket{u}$ to the excited state $\ket{e}$ is driven by an external control laser. Upon emission of a photon, the atom ends up in $\ket{c}$ and is decoupled from the excitation laser due to the large detuning $\Delta_u$. Thus, in every attempt only a single photon is emitted.

Atomic systems typically use the second approach, Raman coupling. When the intensity of the external control laser is varied only on a slow timescale, the system can be kept in a coherent Raman dark state, as described in Sec. \ref{sub:Raman_coupling}. This technique is called vacuum-stimulated Raman adiabatic passage (vSTIRAP) \cite{kuhn_controlled_1999, kuhn_optical_2002}. When the intensity of the control laser is ramped up, the atom will end up in the coupled ground state $\ket{c}$ and emit exactly one photon. We emphasize that the emission process is not necessarily a stepwise process in which the atom first deposits the photon into the cavity from where the photon subsequently escapes. Instead, typically photon generation into and photon emission from the cavity occur simultaneously, and the average number of photons inside the cavity is much smaller than one.

Compared to other photon generation schemes, the vSTIRAP technique has several advantages with respect to quantum networks: First, the excited state is never populated, which leads to reduced spontaneous emission into free space and thus increases the photon collection efficiency. Second, it is reversible, which is the prerequisite for the implementation of the protocol proposed in \cite{cirac_quantum_1997}. This topic will be revisited in Sec. \ref{sub:deterministicNetworks}. Finally, the photon properties, e.g. the frequency and wavepacket envelope, are tunable using the external laser. This allows for the generation of indistinguishable photons from dissimilar emitters, which is required for quantum networks based on two-photon interference and subsequent measurement, as discussed in Sec. \ref{sub:Two-photon_Interference} and \ref{sub:probabilisticNetworks}.

\begin{figure} [h!]
\includegraphics[width=86mm]{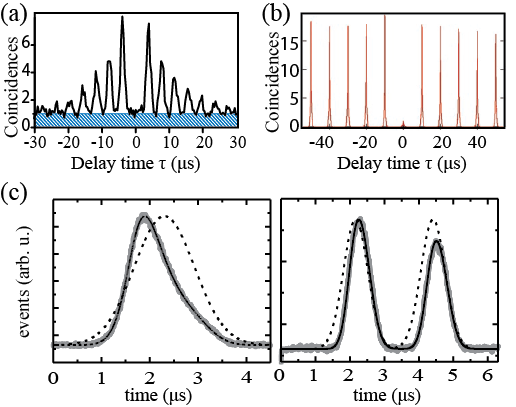}
\caption{ \label{fig:Antibunching_Shaping}
(color online) Cavity-based single-photon sources. Using a vacuum-stimulated Raman transfer from one of the atomic ground states to the other, single photons are emitted into the cavity mode. With alternating repump pulses, a single-photon stream is generated.
(a) Photons are generated with atoms falling through a Fabry-Perot Resonator. The cavity output is analyzed in a Hanbury Brown and Twiss setup to measure the correlation function. With respect to the independently measured background caused by detector dark counts (blue hatched area), the absence of coincidence counts at $\tau=0$ demonstrates the single-photon character of the emitted field. The temporal distance of the peaks is given by the repetition rate. The envelope of the correlation function is determined by the atom transit time. Adapted from \cite{kuhn_deterministic_2002}.
(b) With trapped rather than falling atoms, a continuous stream of single photons can be generated, and the peaks in the correlation function remain at a constant height. Adapted from \cite{mckeever_deterministic_2004}.
(c) Control of the photonic wavepacket envelope. The intensity of the control laser (dashed black line) that drives the Raman transfer is changed and the arrival time of the generated photons is measured. The data points (grey) are in good agreement with a numerical model (solid black line) of the photon generation process. Adapted from \cite{keller_continuous_2004}.
}
\end{figure}

The first experimental steps towards on-demand single-photon generation have been made with atoms falling through a Fabry-Perot cavity. In \cite{hennrich_vacuum-stimulated_2000}, the dark state of the vSTIRAP technique has been observed in the spectral domain. In \cite{kuhn_deterministic_2002}, emission of short single-photon streams was demonstrated. Fig. \ref{fig:Antibunching_Shaping}(a) shows the correlation function recorded in this experiment. The absence of coincidence counts at $\tau=0$ demonstrates the single-photon character of the emitted field. The temporal distance of the peaks is given by the repetition rate of the pulsed experiment. The duration of the photon stream was limited by the atomic transit time through the resonator mode, which can be seen in the decreasing envelope of the correlation function at larger absolute values of the delay time.

This effect can be avoided when using trapped rather than falling atoms. Fig. \ref{fig:Antibunching_Shaping}(b) shows the results of the first realization of such experiment \cite{mckeever_deterministic_2004}. The efficiency to generate a single propagating photon (albeit distributed over two spatial modes due to the use of a symmetric cavity) reached a value of $69\,\%$ in this experiment, thus realizing an almost deterministic on-demand source of single photons.

In the same year, single-photon emission has been demonstrated with a trapped ion with $8\,\%$ efficiency \cite{keller_continuous_2004}. This experiment has also demonstrated that the photonic waveform can be controlled by the external laser field, see Fig. \ref{fig:Antibunching_Shaping}(c). To this end, the intensity of the control laser (dotted line) that drives the Raman transfer is changed and the arrival time of the generated photons is measured. The data points (grey) are in good agreement with a numerical model (solid line) of the photon generation process. The wavepacket envelope can be varied over a large range, with a minimum duration that is set by the cavity decay rate of $2\kappa$.

Following these early demonstrations, single-photon generation has been studied in several other experimental setups with flying \cite{wilk_polarization-controlled_2007, nisbet-jones_highly_2011, nisbet-jones_photonic_2013} and trapped atoms \cite{hijlkema_single-photon_2007, barros_deterministic_2009, mucke_generation_2013}, investigating the achievable wavepacket control and efficiency over a large range of experimental parameters.

\subsection{Time-resolved two-photon interference} \label{sub:Two-photon_Interference}

The previous section has described the generation of single photons in a well-defined spatial mode using atom-cavity systems. This section will focus on the temporal and spectral properties of such photons. In particular, it will be shown that photons can be generated in a pure quantum state. This lays the ground for the implementation of quantum networks using probabilistic protocols that are based on two-photon interference.

A first step in the characterization of a triggered single-photon source is to measure the photon statistics in a Hanbury Brown and Twiss setup as presented in Sec. \ref{sub:PhotonGeneration}. Simultaneously, the temporal envelope of the photons in the ensemble can be measured when the photon duration is longer than the temporal resolution of the detectors, which is typically below $1\,\text{ns}$.

Information about the purity of the quantum state can then be obtained in two-photon interference experiments. The first such experiment was performed with photon pairs from a down-conversion source \cite{hong_measurement_1987}. The photons were interfering with a variable delay on a non-polarizing beam splitter (NPBS). Due to the quantum nature of the light fields, one observes nonclassical interference: Two photons that impinge simultaneously on a NPBS and that are indistinguishable in all of their properties are always found in the same output port, and the coincidence signal between the two ports drops to zero. When the delay between the pulses is increased, the coincidence signal rises to a constant maximum value when the two photonic wavepackets have no temporal overlap.

The characteristic drop in the coincidence signal is known as the Hong-Ou-Mandel dip. Its depth, also called contrast $\mathcal{C}$, is a direct measure for the indistinguishability of the two photons. The first such experiment was performed with simultaneously generated photons. However, with an appropriate delay line the technique can also be applied to photons that are emitted by the same source one after another. This was first demonstrated with a single quantum dot \cite{santori_indistinguishable_2002} and has allowed to analyze shot-to-shot fluctuations in the generated photonic quantum state. Fluctuations on a longer time scale, however, are not detected. Therefore, demonstrating that a pure quantum state is generated requires the interference of photons from independent sources. This has first been realized in \cite{beugnon_quantum_2006} with two atoms in the same free-space trap, and in \cite{nolleke_efficient_2013} with two atoms trapped in separate optical resonators in independent laboratories.

Recording the interference signal with temporal resolution \cite{legero_characterization_2006} allows one to obtain more information about temporal and spectral characteristics of the photons. To this end, consider two photons in the same temporal mode whose frequencies differ by $\Delta_p$. The photons impinge simultaneously in the two input ports $A$ and $B$ of a NPBS. The two output ports $C$ and $D$ are related to the inputs via:

\begin{equation}
a_{C,D}=\frac{1}{\sqrt{2}}(a_B \pm a_A)
\end{equation}

Here, $a$ denotes the annihilation operator of a photon in the respective mode. Depending on which detector clicks, $a_C$ or $a_D$ is applied to the input state $\ket{1_A,1_B}$. Thus, the system is projected onto the superposition state:

\begin{equation}
\ket{\Psi_\pm(\tau)}=\frac{1}{\sqrt{2}}(\ket{1_A,0_B}\pm \text{e}^{i \Delta_p \tau}\ket{0_A,1_B})
\end{equation}

In this equation, the phase difference between the fields, depending on the detection time difference $\tau$, is given by $\Delta_p \tau$. The contrast, i.e. the probability to detect the second photon in the same port as the first one, is then given by \cite{legero_characterization_2006}:

\begin{equation}
\mathcal{C}(\tau)=\frac{1}{2}(1+\cos(\Delta_p \tau))
\end{equation}

\begin{figure} [h!]
\includegraphics[width=86mm]{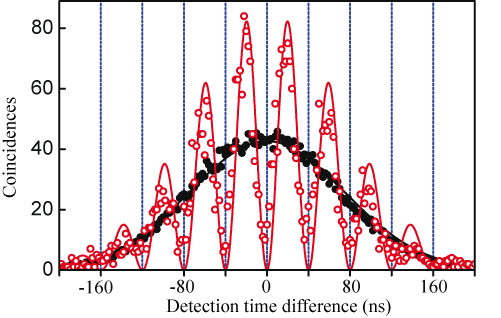}
\caption{ \label{fig:Two-Photon_Interference}
(color online) Time-resolved two-photon interference. Two photons are generated one after the other from an atom-cavity system. One of the photons is sent to an optical delay line, such that both arrive at the same time in the two input ports of a beam splitter. When the photons have the same frequencies, no coincidence counts are observed (not shown). When a frequency offset is externally applied to the photons, a quantum beat is observed (red), which proves the coherence of the photon generation process. The black data show a reference run with distinguishable photons of orthogonal polarization. Adapted from \cite{specht_phase_2009}.
}
\end{figure}

From this equation, one expects to observe an oscillation in the coincidence signal when the two photons exhibit a frequency detuning. This was first demonstrated with falling atoms in \cite{legero_quantum_2004}. Fig. \ref{fig:Two-Photon_Interference} shows the results obtained with trapped atoms in \cite{specht_phase_2009}. In this study, two photons at the same frequency were generated one after the other. One of them was shifted in frequency by sending it through an electro-optical modulator with an increasing voltage ramp applied to the electrodes. With a suitable optical delay line, subsequently generated photons were impinging simultaneously onto a NPBS. The black reference data are recorded with non-interfering photons that had orthogonal polarization, while the red data are obtained with parallel polarization. The predicted beat signal is clearly observed. When $\Delta_p$ is increased (not shown), the oscillation period is reduced as expected.

It must be emphasized that the coincidence signal is always zero around $\tau=0$, even for photons with different wavepacket shapes and arrival times, as has been derived by \cite{legero_time-resolved_2003}. This has important consequences for the implementation of probabilistic quantum-network protocols that are based on two-photon interference: When conditioning on a short detection time difference, a perfect interference contrast can be obtained even with imperfect photon sources! This topic will be revisited in Sec. \ref{sub:probabilisticNetworks}.

In the same section, also two-photon interference data obtained with two independent atom-cavity setups will be presented. The achieved contrast reached $99\,\%$ in this experiment when conditioning on $\tau < 20 \text{ns}$. When integrating over all values of $\tau$, a contrast of $64\,\%$ was observed \cite{nolleke_efficient_2013}. This demonstrates that the photons generated by each of the atom-cavity systems are in a quite pure quantum state. The deviation from unity contrast is attributed to inhomogeneous broadening of the atomic transition frequencies, caused by the uncorrelated motion of the atoms in both traps. A dramatic improvement is therefore expected when cooling to lower temperatures, as later demonstrated in one of the two setups \cite{reiserer_ground-state_2013}.

\subsection{Atom-photon quantum state transfer} \label{sub:AtomPhotonQuantumStateTransfer}

The previous sections have demonstrated the generation of photons in a pure quantum state. This section will go one step further and present the implementation of a \emph{unidirectional} quantum interface that allows for the transfer of the quantum state of a single trapped atom onto that of a single propagating photon. While in principle any photonic qubit implementation can be used towards this end, all experimental implementations to date are based on polarization qubits. This has the advantage that it is easily possible to manipulate the qubit states using polarization optics, such that tomography of the photonic state is straightforward [see e.g. \cite{altepeter_photonic_2005}].

The first experiment that has demonstrated a unidirectional quantum interface has used $\Rb$ atoms falling through a Fabry-Perot cavity \cite{wilk_single-atom_2007}. The atomic level scheme exhibits five levels in a double-$\Lambda$ configuration, as illustrated schematically in Fig. \ref{fig:Photon_Photon_entanglement}(a). The atomic qubit is encoded in two Zeeman-states, such that $m_F= +1$ is represented by $\ket{\uparrow}$ and $m_F= -1$ by $\ket{\downarrow}$. Here, the atomic quantization axis is chosen such that it coincides with the cavity axis. Note that for flying photons, the assignment of polarization states to $\Delta m_F= +1$ or $\Delta m_F= -1$ transitions depends on the propagation direction. In the following, a single-sided cavity is considered, and impinging (emitted) photons exhibit a positive (negative) direction of propagation, respectively.

Assume that the atom is initially prepared in $m_F = -1$ and a photon is generated with the vSTIRAP technique, as explained in Sec. \ref{sub:PhotonGeneration}. To this end, the intensity of a control laser beam, irradiated from the side of the cavity and polarized along its axis, is adiabatically increased from zero to a value larger than $g$. Because of polarization selection rules, the laser couples only $\pi$-transitions with $\Delta m_{F}=0$, indicated by the blue arrows in Fig. \ref{fig:Photon_Photon_entanglement}(a). The cavity supports two frequency-degenerate modes with orthogonal polarization, indicated by the red curly arrows.

As there is no state with $m_F=-2$, the cavity only couples the atom in the excited $m_F=-1$ state to the ground state with $m_F=0$, which is a $\sigma^-$ transition. As the angular momentum of the combined atom-photon state must be conserved, the emitted photon will be right-circularly polarized $\ket{\circlearrowright}$. Similarly, when the atom is in the $m_F=+1$ state, it is only coupled to the cavity on a $\sigma^+$ transition, which results in a left-circularly polarized photon $\ket{\circlearrowleft}$.

The photon generation process is fully coherent. Therefore, when the atom is prepared in a superposition state $\alpha \ket{\uparrow} + \beta \ket{\downarrow}$, the emitted photon will exhibit the polarization $\alpha \ket{\circlearrowleft} + \beta \ket{\circlearrowright}$, and the atom will always end up in the ground state with $m_F=0$. In this way, the atomic qubit state is fully transferred to the photonic qubit state.

The above procedure can also be employed to transfer the state of one atom to that of a remote one, which will be further discussed in Sec. \ref{sec:QuantumNetworks}. In addition, it can be used for a reconstruction of the atomic state. To this end, the photonic polarization is determined via quantum-state tomography. An important feature is that in contrast to the methods presented in Sec. \ref{subsub:StateDetection}, state readout is possible even if the atomic qubit levels are degenerate in energy. However, the readout procedure is probabilistic as both the state-transfer and photon-detection process will exhibit a limited efficiency.

The first demonstration of atom-photon quantum-state transfer achieved an efficiency of $9\,\%$ \cite{wilk_single-atom_2007}, identical to the first experiment with a trapped atom \cite{weber_photon-photon_2009}. Improvement of the cavity mirrors has increased this value to $56\,\%$ in the same experimental setting of \cite{specht_single-atom_2011}. The achieved fidelity was larger than $93\,\%$ in this experiment, limited by imperfect state preparation and laser beam alignment. Optimization of these parameters has meanwhile allowed for a value as high as $98(2)\,\%$.

Atom-photon state transfer has also been achieved with a single trapped ion, with an efficiency of $1\,\%$ (including detector inefficiency) and a fidelity of $66(1)\,\%$ \cite{stute_quantum-state_2013}. To improve, a higher cooperativity would be required in this experiment. Nevertheless, a higher fidelity of $92\,\%$ has been observed for a small subensemble of the data, because conditioning on early photon detection events excludes attempts where the atom has undergone spontaneous scattering before emitting a photon into the cavity. Thus, it is possible to increase the fidelity at the price of a reduced efficiency.

\subsection{Generation of atom-photon entanglement} \label{sub:AtomPhotonEntanglement}

\begin{figure} [h!]
\includegraphics[width=86mm]{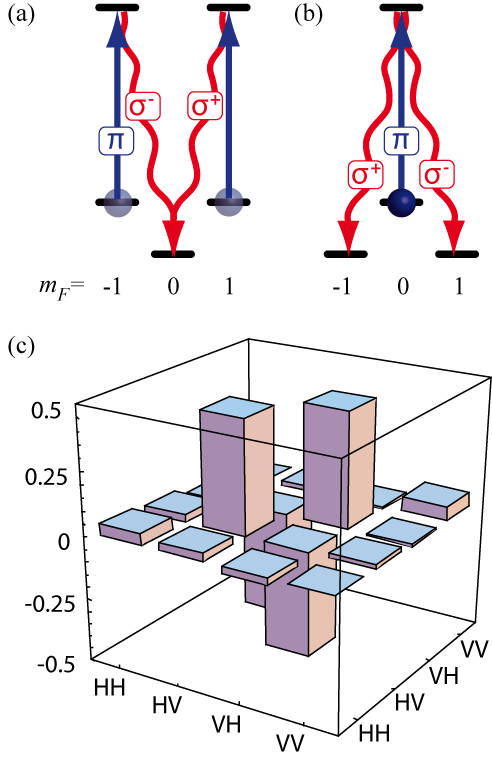}
\caption{ \label{fig:Photon_Photon_entanglement}
(color online)
(a) Atom-photon state transfer. The atomic qubit is encoded in the states $m_F= +1 \equiv \ket{\uparrow}$ and $m_F= -1 \equiv \ket{\downarrow}$. The atom is initially in the superposition state $\alpha \ket{\uparrow}+\beta \ket{\downarrow}$. A photon is generated using a $\pi$-polarized control laser (blue arrows), which transfers the atom to $m_F=0$, independent of its initial state. The polarization of the emitted photons (red curly arrows), however, depends on the initial state, such that $\alpha \ket{\uparrow}+\beta \ket{\downarrow} \rightarrow \alpha \ket{\circlearrowleft}+\beta \ket{\circlearrowright}$.
(b) Generation of atom-photon entanglement. The atom is prepared in the $m_F=0$ state. Using a $\pi$-polarized control laser (blue arrow), a photon can be generated on two different transitions $\sigma^\pm$ with $\Delta m_{F}=\mp1$. Depending on the polarization of the emitted photon (red curly arrows), the atom ends up in a different state. Thus, the process generates atom-photon entanglement.
(c) Density matrix of an atom-photon entangled state, reconstructed by mapping the atomic state onto a second photon and performing quantum state tomography on the two-photon polarization state in the horizontal (H) --- vertical (V) qubit basis. The resulting state has a fidelity of $\mathcal{F}=86\,\%$ with the $\ket{\Psi^-}$ state. Adapted from \cite{wilk_single-atom_2007}.
}
\end{figure}

In the last years, it has been realized that the power of quantum communication and quantum information processing stems from the use of entangled states as a resource for information processing tasks \cite{horodecki_quantum_2009}. There are two canonical models in this respect \cite{nielsen_quantum_2010}. In the first approach, known as the circuit model, entanglement is generated and manipulated between existing qubits based on quantum gate operations. This approach can also be implemented with atoms and photons \cite{reiserer_quantum_2014}, which will be the topic of Sec. \ref{sub:quantumInfoProcess}. The second approach, known as measurement based quantum computing \cite{briegel_measurement-based_2009}, instead relies on the generation of a highly entangled state, and any processing task can be implemented with local measurements and feedback in form of unitary operations. Similarly, the distribution of entanglement in a quantum network and many quantum communication protocols \cite{gisin_quantum_2007, duan_colloquium:_2010} are based on the generation of entangled states between light and matter. This emphasizes the high relevance of efficient atom-photon entangled-state generation, which will be described in this section.

In principle, there are many protocols that allow for the creation of atom-photon entangled states. However, all CQED experiments to date have employed photonic polarization qubits, most of them using a scheme that is described in the following. Consider an atomic four-level-system in a double-$\Lambda$-configuration, as illustrated in Fig. \ref{fig:Photon_Photon_entanglement}(b).
The atom is prepared in the ground state with $m_F=0$. Subsequently, a photon is generated using the vSTIRAP technique, as described in Sec. \ref{sub:PhotonGeneration}. To this end, the intensity of a $\pi$-polarized control laser (blue arrow) is increased, thus coupling the atom to the excited state with $m_F=0$. This state is equally coupled via the cavity mode to the two ground states with $m_F=\pm 1$.

Depending on the polarization of the emitted photon (red curly arrows), the atom will end up in a different Zeeman state. If the photon is right(left-)-circularly polarized, it will drive a $\sigma^-$($\sigma^+$) transition to $m_F=+1(-1)$, respectively. Therefore, the following entangled state is generated:

\begin{equation}
\ket{m_F=0}\xrightarrow{\pi \text{-pol drive}} \frac{1}{\sqrt{2}} (\ket{\downarrow,\circlearrowleft} \pm \ket{\uparrow,\circlearrowright})
\end{equation}

Which of the two signs is realized experimentally depends on the Clebsch-Gordan coefficients of the atomic transitions.

The first experimental realization of this entanglement procedure \cite{wilk_single-atom_2007} again uses $\Rb$ atoms falling through a cavity. Here, the $\ket{\Psi^-}$-Bell state is generated, corresponding to the minus sign in the equation above. To verify entanglement, the atomic state is transferred to the state of a second photon, as explained in Sec. \ref{sub:AtomPhotonQuantumStateTransfer}, and the photon-photon state is reconstructed using quantum state tomography \cite{altepeter_photonic_2005}. Fig. \ref{fig:Photon_Photon_entanglement}(c) shows the obtained density matrix. The combined entanglement and readout fidelity, defined as the overlap with the ideal photonic Bell state, $\mathcal{F}=\bra{\Psi^-}\rho\ket{\Psi^-}$, is $86.0(4)\,\%$. The entanglement-generation efficiency is $15\,\%$, a drastic improvement compared to free-space experiments with single atoms \cite{blinov_observation_2004, volz_observation_2006}, single spins in diamond \cite{togan_quantum_2010, bernien_heralded_2013} or quantum dots \cite{gao_observation_2012, de_greve_quantum-dot_2012} that have not exceeded $0.1\,\%$ yet.

Following this early realization, atom-photon entanglement has been demonstrated in a number of experiments. The first realization with trapped neutral atoms is reported in \cite{weber_photon-photon_2009}. Later, the efficiency has been increased above $50\,\%$ (on the $D_2$ line of $\Rb$), with a fidelity of $94.1(1.5)\,\%$ (on the $D_1$ line) \cite{lettner_remote_2011}. The first CQED implementation of the above protocol with trapped ions has reached an efficiency of $16\,\%$ \cite{stute_tunable_2012}. Here, a bichromatic Raman driving has been used because of non-degenerate atomic ground state levels. To measure the atomic state, fluorescence state detection has been employed instead of atom-photon state transfer. A fidelity of $97.4(2)\,\%$ with a maximally entangled atom-photon entangled state has thus been demonstrated.

There are several proposals how to extend the approach of entanglement generation with atom-cavity systems to cluster states that consist of many subsequently emitted photons, see e.g. \cite{schon_sequential_2005}. Typically, this requires coherent control of the atomic state, which has meanwhile been achieved in several experiments. However, entanglement generation and detection rates decrease exponentially with the number of involved photons. Thus, the largest atom-photon entangled state generated with an optical CQED system to date still consists of only three constituents: an atom and two photons \cite{reiserer_quantum_2014}. Increasing the number of entangled photons by an order of magnitude will require both resonators with very low loss and photodetectors with very high efficiency.

\section{Quantum communication and quantum computation in coherent quantum networks} \label{sec:QuantumNetworks}

The coupling of the atomic state to propagating photons, as presented in Sec. \ref{sec:QuantumLightSources}, facilitates the implementation of a quantum network. To this end, the photons are coupled into optical fibers and thus exchanged between the nodes. In the following, experiments that demonstrate the basic concepts to distribute and process quantum information in elementary quantum networks of single atoms in optical cavities will be described.

We start the discussion with atom-atom quantum state transfer and the generation of remote entanglement based on photon storage at the second node. Although in practical applications photon losses are always present, we call this approach \emph{deterministic}, as the achievable photon generation, transmission and absorption efficiencies can in principle be arbitrarily close to unity in CQED setups. The situation is different for the second approach which is based on two-photon interference. Albeit this technique provides an intrinsic herald, it has an upper bound of $50\,\%$ efficiency, even with perfect photon detectors \cite{calsamiglia_maximum_2001}; we therefore call it \emph{probabilistic}. The first experimental implementations will be presented in Sec. \ref{sub:probabilisticNetworks}.

Subsequently, the implementation of nondestructive photon detection will be described in Sec. \ref{sub:PhotonDetection}. This technique opens up interesting perspectives for the transmission of states over large distances, as it can be used to herald successful photon transmission without affecting the encoded quantum information. Finally, first experiments towards the processing of quantum information with a hybrid system of atoms and photons will be discussed in Sec. \ref{sub:quantumInfoProcess}.

\subsection{Deterministic distribution of quantum information} \label{sub:deterministicNetworks}

\begin{figure*}
\includegraphics[width=178mm]{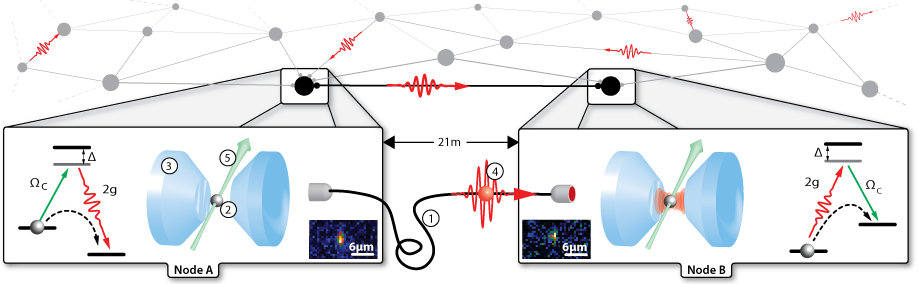}
\caption{ \label{fig:Network}
(color online) An elementary cell of a quantum network that consists of single atoms in optical cavities. An optical fiber (1) connects two independent setups that are separated by $21\,\text{m}$. In both setups, a single atom (2) is trapped in a Fabry-Perot resonator (3). The insets show typical fluorescence images. Quantum states are exchanged between the atoms in form of a single photon (4). To this end, an atomic coherent dark state is controlled with two external laser fields (5). Increasing the field intensity at node A transfers the atomic state to the state of the photon, while decreasing it performs the reverse operation at node B. Adapted from \cite{ritter_elementary_2012}
}
\end{figure*}

The \emph{deterministic} approach towards the implementation of a quantum network with atoms and photons \cite{cirac_quantum_1997} is based on the transmission of a single photon between the two connected nodes $A$ and $B$, as illustrated in Fig. \ref{fig:Network}. The first experimental realization of this scheme \cite{ritter_elementary_2012} used two $\Rb$ atoms, trapped at a distance of $21\,\text{m}$ in Fabry-Perot resonators in the intermediate coupling regime with $C\simeq 1$.

In the following, the different networking experiments performed in this study will be described: First, the transfer of the atomic state from one node to the other is explained in Sec. \ref{subsub:AtomAtomStateTransfer}. Then, the generation of entanglement between the nodes is discussed in Sec. \ref{subsub:AtomAtom_entanglement}.

The used scheme requires that the nodes operate as \emph{bidirectional} quantum interfaces that can both send and receive quantum information. In addition, the nodes must store quantum states for a time that is sufficient to exchange a photon between them. These requirements basically mean that the nodes should operate as a quantum memory for single photons [see e.g. \cite{lvovsky_optical_2009}]. Thus, we will start the discussion with the implementation of such a quantum memory with a single atom.

\subsubsection{Quantum memory} \label{subsub:QuantumMemory}

The first step towards the implementation of a \emph{bidirectional} quantum interface has been to transfer the state of a single atom onto that of a single photon \cite{wilk_single-atom_2007}, as discussed in Sec. \ref{sub:AtomPhotonQuantumStateTransfer}. Let us briefly recall that towards this goal, adiabatic control of a coherent Raman dark state is employed. The system is prepared in the state $\ket{u,0}$ and the intensity of an external control laser that couples the transition from the ground state $\ket{u}$ to the excited state $\ket{e}$ is increased. This transfers the system to $\ket{c,1}$ and thus generates a photon that leaves the cavity, preferentially through the outcoupling mirror. When the coherent coupling $g$ is larger than the dissipative processes $\gamma$ and $\kappa_{\text{loss}}$, the process is fully coherent and thus reversible.

The realization of the reverse process --- coherent photon absorption --- has first been demonstrated with single Cs atoms trapped in a Fabry-Perot resonator in the strong-coupling regime \cite{boozer_reversible_2007}. To mimic a single photon, a faint laser pulse with an average photon number of $\bar{n}=1.1$ is impinging onto the cavity. At the same time, a control laser is irradiated from the side of the cavity, resonant to the transition from $\ket{u}$ to $\ket{e}$. This results in a Raman coupling as explained in Sec. \ref{sub:Raman_coupling}. Thus, the system is initially in a coherent dark state, where the cavity is transparent to the impinging photon, compare the spectrum in Fig. \ref{fig:CavityEIT} \cite{mucke_electromagnetically_2010}.

To store the photon, the intensity of the external control laser field is now ramped down adiabatically. In this way, the state $\ket{c,1}$ is transferred to $\ket{u,0}$ while remaining in the Raman dark state, see also Sec. \ref{sub:Raman_coupling}. To this end, it is important that the control laser ramp exactly matches the temporal envelope of the impinging photon pulse. When this condition is not perfectly achieved, the efficiency of the process will be reduced.

In the experiment of \cite{boozer_reversible_2007}, the achieved efficiency of $5.7\,\%$ is mainly limited by mirror losses and the fact that a symmetric cavity with two output channels is used instead of a single-sided one. Nevertheless, the coherence of the photon absorption process is verified by mapping the atomic state back onto a photon. Using interference with a reference field, a fringe visibility of $46\,\%$ is observed. The achieved value is limited by incoherent transfer processes in which the atom has undergone spontaneous scattering.

In the first experiment that demonstrates an optical quantum memory using a single atom, this problem is avoided by operating at a detuning of $4 \gamma$ from the atomic excited state $\ket{e}$ \cite{specht_single-atom_2011}. Again, a faint laser pulse with average photon number $\bar{n}<1$ is used to mimic impinging single photons. The system employs $\Rb$ atoms in a single-sided cavity in the intermediate coupling regime with $C \simeq 1$. A storage efficiency of $17\,\%$ is achieved, which is substantially lower than the retrieval efficiency of $56\,\%$.

The reason for this reduction is that in contrast to the readout, the storage process relies on perfect spatial and temporal mode matching between the impinging photon, the cavity mode and the control laser field, respectively. The deviation of both efficiency values from unity is caused by the limited cooperativity and by imperfect atomic localization. Improving the latter, a three-fold increase in the combined storage and retrieval efficiency has been observed.

\begin{figure}
\includegraphics[width=86mm]{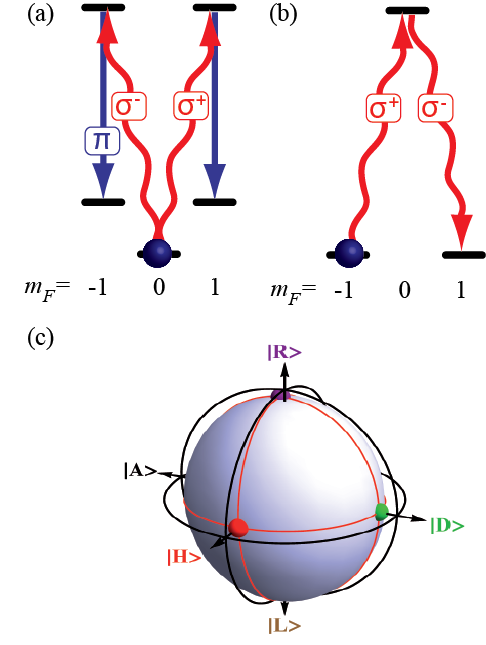}
\caption{ \label{fig:QuantumMemory}
(color online) Single-atom quantum memory.
(a) Atomic level scheme for photon storage using the STIRAP technique. The atom is prepared in the ground state with $m_F=0$ (purple ball). An impinging photon (red curly arrows) drives $\sigma^\pm$-transitions, depending on its polarization. A $\pi$-polarized control laser (blue arrows) thus transfers the photonic polarization onto the spin state of the atom.
(b) Atomic level scheme for an alternative proposal, termed deterministic single-photon Raman passage, that uses an atom-photon SWAP gate to achieve photon storage in the Purcell regime. When the atom is initially in the state $m_F=-1\equiv\ket{\downarrow}$, a left-circularly polarized photon is not coupled to it and thus reflected with unchanged polarization. In contrast, a right-circularly polarized photon will be reflected with left-circular polarization (red curly arrows) due to interference with the atomic emission, and in this process the atom is transferred to the state $m_F=+1\equiv \ket{\uparrow}$. In this way, the photonic polarization is transferred to the atomic state.
(c) Quantum memory results, achieved with the STIRAP scheme. The qubit state after storage and readout is reconstructed for six unbiased input states using quantum state tomography. The result is visualized as colored circles in the Bloch sphere representation. Via quantum process tomography, the memory performance for any input state can be predicted, as visualized by the bluish sphere. The average fidelity of the depicted measurement is $\bar{\mathcal{F}}=93\,\%$. From \cite{specht_single-atom_2011}.
}
\end{figure}

To demonstrate an optical quantum memory, a photonic qubit is encoded in the polarization of the impinging laser pulse. Consider the atomic level scheme of $\Rb$ depicted in Fig. \ref{fig:QuantumMemory}(a). The atom is prepared in the ground state with $m_F=0$. The quantization axis is chosen parallel to the cavity axis, such that the two frequency degenerate polarization modes of the cavity couple $\sigma^+$ and $\sigma^-$ transitions (red curly arrows), depending on the photonic polarization. The external control laser is linearly polarized along the quantization axis, thus driving $\pi$-transitions (blue arrows). In this way, the polarization of an impinging photon in the state $\ket{\Psi_\text{in}}=\alpha \ket{\circlearrowleft} + \beta \ket{\circlearrowright}$ is transferred to the spin state of the atom: $\ket{\Psi_a}= \alpha \ket{\downarrow} + \beta \ket{\uparrow}$, where $m_F=-1\equiv\ket{\downarrow}$ and $ m_F=+1\equiv\ket{\uparrow}$.

To derive a lower bound on the fidelity of the storage process, the atomic state is transferred back to a photon, i.e. the storage process is time-reversed as explained in Sec. \ref{sub:AtomPhotonQuantumStateTransfer}. Then, the photonic polarization state is reconstructed using quantum state tomography. This procedure is repeated for six unbiased input states, which allows one to determine the storage fidelity for any input qubit using quantum process tomography \cite{nielsen_quantum_2010}. The result of this procedure is shown in the Bloch sphere representation in Fig. \ref{fig:QuantumMemory}(c). The colored circles show the measured position of three of the six input qubit states on the Bloch sphere. The colors on the axes indicate the ideally expected states, showing excellent agreement. The fidelity, averaged over all six input state, is:

 \begin{equation}
\overline{\mathcal{F}}_\text{memory} = \overline{ \bra{\Psi_\text{in}} \rho_\text{out} \ket{\Psi_\text{in}} }=93\,\%
\end{equation}

This value is limited by imperfect preparation of the atomic state \cite{specht_single-atom_2011}. The first demonstration of the storage of single photons (rather than faint laser pulses) in a single atom has achieved comparable values of fidelity and efficiency \cite{ritter_elementary_2012}. With improved laser beam alignment and atom localization, the fidelity has later been improved to $98\,\%$ (for faint laser pulses). All quoted fidelities are clearly above the classical threshold. This benchmark is $2/3$ when the quantum memory is characterized with single photons, and increases with the average photon number when coherent input states are used \cite{specht_single-atom_2011}.

A recent alternative approach to the implementation of a heralded quantum memory is the use of a reflection-based atom-photon quantum gate, which will be presented in Sec. \ref{subsub:atom-photon_QuantumGates}, followed by feedback onto the atomic state \cite{kalb_heralded_2015}. Finally, a similar scheme that does not require feedback has been proposed and further investigated in \cite{pinotsi_single_2008, lin_heralded_2009, koshino_deterministic_2010}. It is based on photon reflection from a single sided cavity, too. Consider an atom with two degenerate ground-state levels, e.g. Zeeman states with quantum numbers $m_F=-1$ and $m_F=+1$ as depicted in Fig. \ref{fig:QuantumMemory}(b). Both ground states are equally coupled to a common excited state via two circularly polarized polarization modes (red curly arrows) of the cavity. First consider the situation when the atom is in $m_F=-1\equiv\ket{\downarrow}$. A left-circularly polarized photon that would drive $\sigma^-$ transitions is not coupled to the atom and will therefore simply be reflected from the cavity. Due to the change in propagation direction, it will then exhibit right-circular polarization.

The situation is different for a right-circularly polarized impinging photon. When operating in the Purcell regime ($C\gg1$), there are two possible paths that could lead to a left-circularly polarized photon after the reflection process: Either direct reflection from the cavity, or absorption and subsequent re-emission on the atomic $\sigma^+$ transition. As has been pointed out in \cite{pinotsi_single_2008}, these two paths interfere destructively, and the impinging photons will deterministically be absorbed by the atom and then re-emitted on the $\sigma^+$ transition, i.e. with right-circular polarization. Thus, the atom ends up in the state $m_F=+1\equiv\ket{\uparrow}$. Applying the same arguments to the situation where the atom is initially in $\ket{\uparrow}$, one finds that the effect of the reflection process is a SWAP operation, i.e. an interchange between atomic and photonic state: $(\alpha_p \ket{\circlearrowleft} + \beta_p \ket{\circlearrowright})(\alpha_a \ket{\downarrow} + \beta_a \ket{\uparrow}) \rightarrow (\alpha_a \ket{\circlearrowleft} + \beta_a \ket{\circlearrowright})(\alpha_p \ket{\downarrow} + \beta_p \ket{\uparrow})$, where the subscripts denote the two particles, atom and photon, respectively.

The above mechanism, sometimes termed deterministic one-photon Raman interaction \cite{rosenblum_analysis_2014}, might facilitate the implementation of an atomic quantum memory without additional Raman control fields. A first step in this direction has been made in \cite{shomroni_all-optical_2014}. The experiment uses atoms flying through the evanescent field of a whispering-gallery mode resonator, where the two orthogonal polarization modes in the above description are accompanied by a different direction of propagation along the coupling nano fiber. The presence of an atom is verified and its state is prepared by detecting reflected photons from several short 'control' pulses. Instead of reading out the atomic state after the last control pulse, the transmission (reflection) of another faint 'target' laser pulse is measured, achieving the intended switching with $90\,\%$ ($65\,\%$) probability, respectively. Towards the implementation of a quantum memory, these values would have to be improved, and equal losses and phase stability of the $\sigma^+$ and $\sigma^-$ paths would have to be ensured.

\subsubsection{Atom-atom quantum state transfer} \label{subsub:AtomAtomStateTransfer}

The combination of photon generation and photon storage facilitates the distribution of quantum information in an elementary quantum network \cite{ritter_elementary_2012}. In this context, we start by describing the transfer of the atomic state from one network node to the other. To this end, the atom at node A is prepared in a superposition of Zeeman states, $\ket{\Psi_A}$. This state is first transferred to the polarization state of a photon using the procedure described in Sec. \ref{sub:AtomPhotonQuantumStateTransfer}:

\begin{equation}
\ket{\Psi_A}=\alpha \ket{\downarrow} + \beta \ket{\uparrow} \rightarrow \ket{\Psi_p}=\alpha \ket{\circlearrowright} + \beta \ket{\circlearrowleft}
\end{equation}

Subsequently, a STIRAP pulse is used to transfer the polarization of the photon to the spin state of the atom at node B, as explained in Sec. \ref{subsub:QuantumMemory}. This completes the state transfer $\ket{\Psi_A} \rightarrow \ket{\Psi_B}$. To characterize the process, it is performed for six unbiased input states, and the state transfer fidelity is evaluated using quantum process tomography. This gives an average fidelity of:

\begin{equation}
\overline{\mathcal{F}}_\text{state transfer}= \overline{ \bra{\Psi_A} \rho_B \ket{\Psi_A}} = 84(1)\,\%.
\end{equation}

The achieved value proves the quantum character of the state transfer, as the highest fidelity achievable with classical information exchange between the nodes is again $2/3$. The overall success probability of the protocol is $0.2\,\%$. This is the product of the atom-photon state transfer efficiency at node A ($3\,\%$), the photon transmission through the fiber ($34\,\%$) and the storage efficiency at node B ($20\,\%$). Optimization of the parameters to the values given above should allow for an increase by a factor of $\sim 20$. Further improvement is possible with better atomic localization and with higher cooperativity.

\subsubsection{Remote atom-atom entanglement} \label{subsub:AtomAtom_entanglement}

We now summarize an experiment that demonstrates entanglement of two atoms via the transmission of a single photon \cite{ritter_elementary_2012}. The first step in the used protocol is the creation of atom-photon entanglement at node A using the procedure described in Sec. \ref{sub:AtomPhotonEntanglement}. The generated state thus reads:

\begin{equation}
\ket{\Psi^-}_{AP}= \frac{1}{\sqrt{2}}(\ket{\downarrow}_A \ket{\circlearrowleft} - \ket{\uparrow}_A \ket{\circlearrowright})
\end{equation}

Again, the generated photon is coupled into a glass fiber and transmitted to node B, where it is coherently absorbed using the STIRAP scheme explained in Sec. \ref{subsub:QuantumMemory}. This results in the atom-atom entangled state:

\begin{equation}
\ket{\Psi^-}_{AB}=\frac{1}{\sqrt{2}}(\ket{\downarrow}_A \ket{\uparrow}_B - \ket{\uparrow}_A \ket{\downarrow}_B)
\end{equation}

\begin{figure}
\includegraphics[width=86mm]{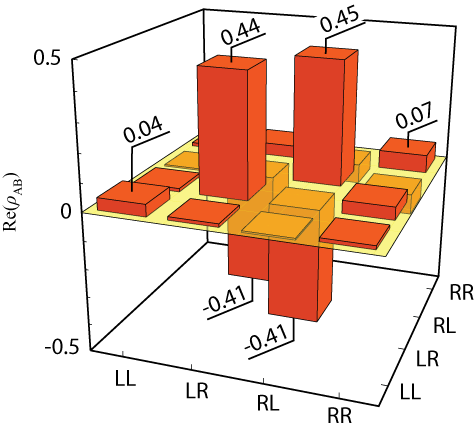}
\caption{ \label{fig:AtomAtomEntanglement}
(color online) Remote atom-atom entanglement. In the experiment, atom-photon entanglement is created at node A and the photon is sent via an optical fiber to node B, where it is coherently absorbed. The resulting atom-atom state is reconstructed using quantum state tomography, which results in the depicted density matrix. The fidelity with the $\ket{\Psi^-}$-Bell state is $85\,\%$. From \cite{ritter_elementary_2012}.
}
\end{figure}

To read out the atomic state at both nodes, it is transferred to a photonic polarization qubit which is subsequently detected. Quantum state tomography then allows for a reconstruction of the atomic density matrix of those experimental runs where the protocol has worked as intended. Fig. \ref{fig:AtomAtomEntanglement} shows the obtained result. The achieved fidelity is:

\begin{equation}
\mathcal{F}_{\text{entanglement}}=\bra{\Psi^-} \rho_{AB} \ket{\Psi^-}=85(1)\,\%.
\end{equation}

This by far exceeds the classical limit of 50\,\%, thus proving the existence of entanglement between the two remote atoms. Fidelities as high as $98.7(2.2)\,\%$ can be achieved by further post-selection of photon detection events, as explained in the discussion in Sec. \ref{subsub:DeterministicApproach_Discussion}.

The achieved success probability of remote atom-atom entanglement is $2\,\%$. Again, this number is the product of the photon generation efficiency ($40\,\%$) at node A, the probability with which the photon is delivered to node B ($34\,\%$) and its storage efficiency at node B ($14\,\%$).

\subsubsection{Discussion} \label{subsub:DeterministicApproach_Discussion}

The experiments described above realize the seminal proposal of \cite{cirac_quantum_1997}. While this scheme is in principle deterministic, the experimentally achieved efficiencies are in the percent regime, predominantly limited by fiber and optics losses, atomic localization, and the finite cooperativity of $C\simeq 1$. None of these limitations is fundamental, and efficiencies above $50\,\%$ should be feasible with current technology, outperforming free-space approaches by orders of magnitude. The only experimental requirements are a suitable atomic level scheme and a large enough cooperativity.

Still, with the achieved limited efficiency values, the successful transfer of quantum information relies on the use of loss-resistant polarization qubits, and is detected only a-posteriori. This means that the detection of read-out photons indicates that the protocol has worked as intended. For scaling to larger quantum networks, a heralded scheme is clearly required, such that finite efficiencies will not necessarily lead to finite fidelities. Such herald can be implemented in a straightforward way: The successful storage of a photon at node B leads to a change of the atomic hyperfine ground state. Thus, detecting whether the atom is still in its original state (using one of the techniques described in Sec. \ref{subsub:StateDetection}) could herald successful realizations of the protocol \cite{lloyd_long_2001}.

Comparing to the probabilistic protocols presented in Sec. \ref{sub:probabilisticNetworks}, higher efficiencies are achievable, and the fidelity of the deterministic approach is robust with respect to the properties of the generated photons. To be more specific, fluctuating photon arrival time, wavepacket envelope or frequency will only reduce the efficiency of the protocol, without affecting its fidelity. The achieved value is thus limited by other experimental imperfections, such as the preparation of the initial atomic state, imperfect laser polarization, and incoherent scattering. In the experimental setting of \cite{ritter_elementary_2012}, all of these imperfections lead to a delayed emission of read-out photons. Thus, post-selecting the data evaluation on those experimental runs where a photon has been detected at the very beginning of the read-out interval leads to increased fidelities. For the atom-atom entanglement experiment, values as high as $98.7(2.2)\,\%$ are reported, which is promising towards scaling remote entanglement to larger distances in a repeater-type architecture.

Towards this goal, also other deterministic approaches may be considered, including entanglement distribution via squeezed light \cite{kraus_discrete_2004}, the preparation of entangled states using dissipative dynamics \cite{kastoryano_dissipative_2011, nikoghosyan_generation_2012}, and dispersive interaction of remote atoms while only virtually populating the photonic modes \cite{pellizzari_quantum_1997, serafini_distributed_2006}. However, achieving high fidelities using one of these schemes would require a strong improvement with respect to the parameters of current experimental systems. Another approach is to combine the generation of atom-photon entanglement at node A with an atom-photon quantum gate, which will be described in Sec. \ref{subsub:atom-photon_QuantumGates}. Subsequent measurement of the photon reflected from node B, followed by feedback onto the atomic state, constitutes a robust, quasi-deterministic, and --- most important --- heralded quantum memory \cite{kalb_heralded_2015} that is highly promising for remote entanglement generation \cite{duan_robust_2005}, which is the prerequisite for quantum repeater schemes \cite{briegel_quantum_1998} and more general for deterministic interactions between network nodes mediated by probabilistic photonic channels \cite{van_enk_photonic_1998}.

\subsection{Probabilistic distribution of quantum information} \label{sub:probabilisticNetworks}

In contrast to the deterministic approach described in Sec. \ref{sub:deterministicNetworks}, the following sections will present the distribution of quantum information between two atoms based on the interference of single photons that are entangled with the atomic state. When the emitted photons are indistinguishable, one can implement probabilistic photon-photon gates using the techniques of linear optical quantum computing \cite{obrien_optical_2007, pan_multiphoton_2012}. Detection of a specific two-photon correlation then heralds successful events. While the employed techniques have been pioneered in free-space experiments \cite{duan_colloquium:_2010}, the use of optical cavities can increase the success probability by several orders of magnitude.

We will start the discussion in Sec. \ref{subsub:OpticalBSM} by explaining how two-photon interference allows one to determine the Bell-State of two photons. This facilitates the basic operations required in quantum networks: transfer of the atomic state via teleportation, described in Sec. \ref{subsub:Teleportation}, and atom-atom entanglement via entanglement swapping, presented in Sec. \ref{subsub:AtomAtomEntanglement_Ions}.

\subsubsection{Optical Bell-state measurement} \label{subsub:OpticalBSM}

\begin{figure} [h!]
\includegraphics[width=86mm]{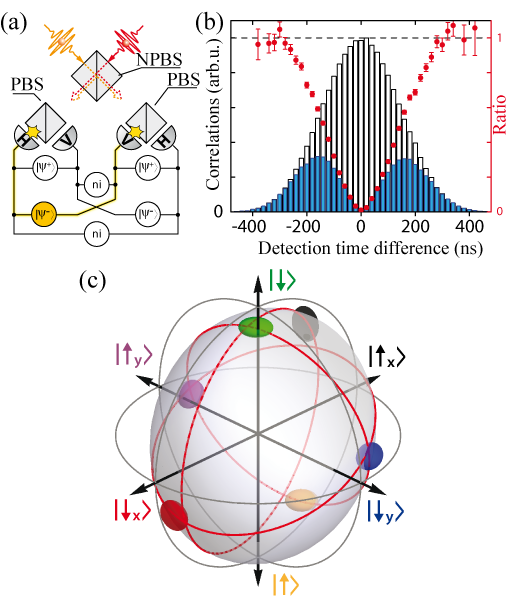}
\caption{ \label{fig:Teleportation}
(color online) (a) Photonic Bell-state measurement. Two photons arrive simultaneously in the two input ports of a non-polarizing beam splitter (NPBS), which is followed by two polarizing beam splitters (PBS). Single photon counters in the individual output ports allow to independently detect horizontal (H) and vertical (V) polarizations. Coincident  detection events project the photonic state onto the $\ket{\Psi^+}$ or the $\ket{\Psi^-}$ Bell state. (b) Time-resolved two-photon quantum interference from two atoms trapped in independent setups. Photons of parallel (blue bars) or orthogonal (open bars) polarization impinge onto a non-polarizing beam splitter. For indistinguishable photons, the ratio of parallel and orthogonal coincidence events (red circles) is expected to be zero. The observed dip indicates photon indistinguishability with a contrast of $64\,\%$. Selecting a shorter coincidence time window allows to obtain a better interference contrast at the price of a reduced number of events. (c) Teleportation between remote atoms. The sender atom is initialized in six unbiased input states. After the teleportation process, the resulting state of the receiver atom is reconstructed, giving the colored circles in the figure. The grey sphere is the result of quantum process tomography. The average fidelity in the depicted experiment was $88.0(1.5)\,\%$. Adapted from \cite{nolleke_efficient_2013}.
}
\end{figure}

Photons are bosons, a fact which implies that their overall wavefunction must be symmetric under particle exchange. To understand how this can be used to detect the Bell state of two photons, consider as an example the setup depicted in Fig. \ref{fig:Teleportation}(a). Two photons that are completely indistinguishable in their temporal and spectral mode impinge onto a NPBS, one in each port. When their combined polarization state is antisymmetric, then their spatial wave function after the NPBS must be antisymmetric as well, which means that the photons leave in different output ports. Thus, when the $H$ and $V$ detectors in different ports register a coincidence event, then the photon-photon state is projected onto the antisymmetric Bell-state $\ket{\Psi^-}_{pp}=\frac{1}{\sqrt{2}}(\ket{H}\ket{V}-\ket{V}\ket{H})$.

The other Bell states exhibit a symmetric polarization state. Thus, they will lead to the detection of photon pairs in the same output port of the NPBS. Still, the $\ket{\Psi^+}_{pp}=\frac{1}{\sqrt{2}}(\ket{H}\ket{V}+\ket{V}\ket{H})$ Bell state can be unambiguously detected because it exhibits orthogonal polarizations. The other two Bell states cannot be distinguished using the depicted setup. In general, the efficiency that is achievable using only linear optical elements exhibits an upper bound of $50\,\%$, irrespective of the specific setup \cite{calsamiglia_maximum_2001, pan_multiphoton_2012}.

The first realization of a Bell-state measurement in optical CQED \cite{nolleke_efficient_2013} used the same two experimental setups that have previously implemented the deterministic approach, see Sec. \ref{sub:deterministicNetworks}, which enables a direct comparison. Two $\Rb$ atoms are trapped in remote cavities in the intermediate coupling regime with $C \simeq 1$. Each atom generates a photon using the vSTIRAP technique (see Sec. \ref{sub:PhotonGeneration}). Spatial mode matching is guaranteed by overlapping the photons on a fiber-based beam splitter. Spectral indistinguishability is provided by locking the control laser that drives the vSTIRAP to the same reference, provided by an optical frequency comb. Thus, the frequency of the photons (determined by the difference between the laser frequency $\omega_L$ and the atomic ground state hyperfine splitting $\Delta_{u}$) is identical, independent of the atomic transition frequency. The remaining critical parameter, the temporal mode, is adjusted via the intensity profile of the control laser.

To characterize the achieved indistinguishability, time-resolved two-photon interference is employed, as explained in Sec. \ref{sub:Two-photon_Interference}. The experimental results are depicted in Fig. \ref{fig:Teleportation}(b). When the photons have orthogonal polarization (open bars) they do not interfere, and one therefore observes correlations in the two output ports of the NPBS. When the photons exhibit parallel polarization (blue bars), they are supposed to be indistinguishable and thus one expects to observe no coincidence counts. While this is indeed the case for short detection time differences $\tau$, the integrated contrast, and thus the fidelity of an optical Bell-state measurement, decreases from $99\,\%$ to $64\,\%$ for larger values of $\tau$. This observation is explained by the fact that the atoms are trapped in deep optical potentials. The uncorrelated motion of both atoms thus leads to inhomogeneous broadening of the two atomic transition frequencies and the two couplings to the cavity modes, which in turn causes a jitter in the temporal mode of the photons.

\subsubsection{Quantum teleportation} \label{subsub:Teleportation}

In the following, the optical Bell-state measurement is employed to transfer the atomic qubit state via teleportation from one network node to another one. To this end, the proposal of \cite{bose_proposal_1999} is extended to polarization qubits. The atom at the sender node A is initialized in the state to be teleported:

\begin{equation}
\ket{\varphi}_A=\alpha\ket{\uparrow}_A+\beta\ket{\downarrow}_A.
 \end{equation}

Then, entanglement is created locally between the spin-state of atom B and the polarization of a photon C using a vSTIRAP, as explained in Sec. \ref{sub:AtomPhotonEntanglement}. The atom-photon state reads:

\begin{equation} \label{eq:Psim_exp}
\ket{\Psi^-}_{BC}=\frac{1}{\sqrt{2}}\left(\ket{\downarrow}_{B}\ket{\circlearrowleft}_{C} - \ket{\uparrow}_{B}\ket{\circlearrowright}_{C}\right).
\end{equation}

Here, $\ket{\circlearrowright}$ and $\ket{\circlearrowleft}$ denote right- and left-circularly polarized photon states, and the subscripts label the individual particles. Photon C is sent to node A via an optical fiber. The three-particle state of the qubits A, B and C can be written as

\begin{align} \label{eq:Teleportation}
\ket{\varphi}_{A}\ket{\Psi^-}_{BC} = &\frac{1}{2}(\left(\ket{\Phi^+}_{AC}\hat{\sigma}_\text{x}\hat{\sigma}_\text{z}\ket{\varphi}_{B}
+\ket{\Phi^-}_{AC}\hat{\sigma}_\text{x}\ket{\varphi}_{B} \right. \nonumber \\
&\left.-\ket{\Psi^+}_{AC}\hat{\sigma}_\text{z}\ket{\varphi}_{B}-
\ket{\Psi^-}_{AC}\ket{\varphi}_{B}\right).
\end{align}

Here, $\ket{\Phi^\pm}$ and $\ket{\Psi^\pm}$ are the four Bell states and $\hat{\sigma}_i$ are the Pauli operators, with $i= \text{x,y,z}$. To transfer the atomic state from A to B, the Bell state of the qubits A and C is measured. This projects atom B onto the initial state $\ket{\varphi}$ of atom A, except for a rotation that depends on the measurement outcome.

In the experiment \cite{nolleke_efficient_2013}, the Bell-state measurement is performed optically, as explained in Sec. \ref{subsub:OpticalBSM}. To this end, the qubit state of atom A is first mapped onto a photon, as explained in Sec. \ref{sub:AtomPhotonQuantumStateTransfer}, and this photon then interferes with photon C.

The success probability of the teleportation protocol is $0.1\,\%$. It is the product of the photon-production efficiency ($39\,\%$ at node A and $25\,\%$ at node B) and the photon-detection efficiency of $31\,\%$ and $12\,\%$, respectively, multiplied by a factor of $0.25$ as only one Bell state is detected. These numbers include all propagation losses and the quantum efficiency of the detectors. Compared to a realization with atoms in free space \cite{olmschenk_quantum_2009}, the use of cavities boosts the efficiency by almost five orders of magnitude. However, compared to the values achieved using the deterministic networking approach presented in Sec. \ref{sub:deterministicNetworks}, the efficiency is more than an order of magnitude smaller, which is caused by the fundamental factor of $0.25$ and by the limited quantum efficiency of the SPCMs (about $50\,\%$ each).

The achieved average fidelity is $\overline{F} = (78.9\pm 1.1)\,\%$. It is determined by the entanglement fidelity and by the limited two-photon interference contrast of $66\,\%$ already mentioned in Sec. \ref{subsub:OpticalBSM}. Reducing the coincidence time window to $80\,\text{ns}$ increases the contrast to $92.8\,\%$ and thus the average teleportation fidelity to $(88.0\pm 1.5)\,\%$, while reducing the efficiency by a factor of four.

As a side remark, the described teleportation protocol can be directly applied to entanglement swapping, as required for the implementation of a quantum repeater. To this end, both atoms A and B would be entangled with remote atoms A' and B', e.g. using the procedure described in Sec. \ref{subsub:AtomAtom_entanglement}. Then, atom-photon quantum state transfer (as explained in Sec. \ref{sub:AtomPhotonQuantumStateTransfer}) followed by the optical Bell-state measurement would allow for entanglement swapping to the nodes A' and B'. The achievable efficiency and fidelity are expected to be comparable.

\subsubsection{Entanglement of two atoms in one cavity} \label{subsub:AtomAtomEntanglement_Ions}

An entanglement swapping protocol based on an optical Bell-state measurement can also be used for the heralded entanglement of two atoms in the same resonator. The first experimental realization \cite{casabone_heralded_2013} follows the scheme proposed in \cite{duan_efficient_2003}. It employs two $^{40}\text{Ca}^+$ ions trapped in a Fabry-Perot cavity in the intermediate coupling regime with $C \simeq 0.25$. Adjusting the trap frequencies of the employed linear Paul trap facilitates controlled localization of both ions at an antinode of the cavity field. Atom-photon entanglement is generated by driving both ions with a bichromatic Raman laser, as first demonstrated in \cite{stute_tunable_2012}. Ideally, this generates the ion-photon state:

\begin{equation}
\ket{\psi}=\frac{1}{\sqrt{2}}(\ket{\uparrow}\ket{H}+\ket{\downarrow}\ket{V})
\end{equation}

Here, $\ket{\uparrow} \equiv \ket{3^2 D_{5/2}, m_j=-3/2}$ and $\ket{\downarrow}\equiv \ket{3^2 D_{5/2}, m_j=-5/2}$ denote the atomic spin state, and $\ket{H}$ ($\ket{V}$) denotes horizontal (vertical) polarization, respectively. The two photons leave the cavity in the same output mode, and are impinging onto a polarizing beam splitter that allows for a photonic Bell-state measurement, similar to that using a NPBS described in Sec. \ref{subsub:OpticalBSM}. Thus, when one photon is detected in each port, the state of the atoms A and B is projected to the entangled state:

\begin{equation}
\ket{\psi}_{AB}=\frac{1}{\sqrt{2}}(\ket{\uparrow}_A\ket{\downarrow}_B + \ket{\downarrow}_A \ket{\uparrow}_B)
\end{equation}

\begin{figure} [h!]
\includegraphics[width=86mm]{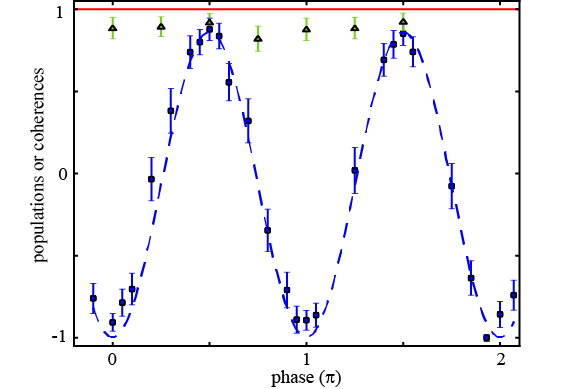}
\caption{ \label{fig:TwoIonEntanglement}
(color online) Entanglement of two ions, heralded by the detection of two photons in the cavity output. The fidelity of the entangled state is bounded by three measurements. First, the sum of population terms $\rho_{\downarrow \uparrow,\downarrow \uparrow}$ and $\rho_{\uparrow \downarrow,\uparrow \downarrow}$, indicated by the red line. Second, the parity after two $\pi/2$ rotations with relative phase $\phi$ on the $\ket{c}\leftrightarrow\ket{u}$ transition (blue circles and dashed sinusoidal fit curve). Third, the parity after one $\pi/2$ pulse with absolute phase $\phi$ (green triangles). From the obtained values, a lower bound to the state fidelity of $92\,\%$ can be derived. Adapted from \cite{casabone_heralded_2013}.
}
\end{figure}

Reconstructing the density matrix of the combined atom-atom state would require individual addressing of both ions. This is not implemented experimentally. Instead, a lower bound to the fidelity of the generated state with the ideally expected one is obtained using common qubit rotations, followed by a measurement of the atomic-state parity given by $p_0+p_2-p_1$. Here, $p_i$ denotes the probability to find $i$ ions in the fluorescent state after the rotation. The experimental results can be seen in Fig. \ref{fig:TwoIonEntanglement}. One clearly observes the expected parity oscillation when varying the phase between the two $\pi/2$ qubit rotations.

The depicted data includes experimental runs in which the two photons have been detected within a time window of $0.5\, \text{\textmu s}$, achieving a fidelity of $91.9(2.5)\%$. The used coincidence time window is short compared to the $40\,\text{\textmu s}$ duration of the Raman laser pulse that generates the photons. This increases the achieved fidelity as it excludes experimental runs in which the atom has undergone several scattering events before emitting a photon into the cavity mode. However, the consequence is a reduction of the success probability by more than an order of magnitude. Nevertheless, the achieved entanglement rate is $0.2$ events per second. To improve, a cavity with a higher cooperativity and reduced scattering losses would be required.

Albeit entanglement of two or more ions within one trap can be generated deterministically with higher fidelity using the direct Coulomb interaction \cite{blatt_entangled_2008}, the used mechanism can also find applications in the generation of entangled states that consist of neutral atoms or other emitters that do not provide direct coupling mechanisms. In addition, the measured efficiency and fidelity allow one to extrapolate the values that should currently be achievable with ions trapped in remote cavities.

\subsubsection{Discussion} \label{subsub:ProbabilisticApproach_Discussion}

The experiments described above \cite{nolleke_efficient_2013, casabone_heralded_2013} are first steps towards larger quantum networks using the probabilistic approach. Compared to free-space approaches \cite{moehring_entanglement_2007, duan_colloquium:_2010, hofmann_heralded_2012, bernien_heralded_2013, pfaff_unconditional_2014}, the use of optical resonators as efficient quantum interfaces allows for high success probabilities. Remarkably, in \cite{nolleke_efficient_2013} the time required for one successful teleportation event, $0.1\,\text{s}$, is about an order of magnitude shorter than the coherence time observed with single trapped atoms \cite{meschede_manipulating_2006}, even at a moderate repetition rate of $10\,\text{kHz}$ that is currently limited by intermittent cooling intervals.

Still, scaling to larger quantum networks is challenging and would require much higher event rates. One way to achieve this is the implementation of faster cooling mechanisms. In addition, the efficiency of the protocol can still be improved by two orders of magnitude compared to the realization in \cite{nolleke_efficient_2013}. A factor of ten can be gained by increasing the photon production efficiency, which is clearly feasible with better atomic localization and higher coupling strength. Gaining another order of magnitude would require an improved optics setup and SPCMs with an efficiency approaching unity, as invented recently \cite{marsili_detecting_2013}.

This might allow for further approaching the fundamental limit of $50\,\%$. Beating this limit would require a deterministic photon-photon gate, which might soon be realized in CQED, see sec. \ref{subsub:PhotonPhoton_Gates}. Alternatively, the use of photon states that are entangled in several degrees of freedom and thus allow for deterministic Bell-state measurement with linear optics can be considered \cite{walborn_hyperentanglement:_2008}. Finally, to further increase the teleportation efficiency, the Bell state can also be measured by applying an atom-photon quantum gate, described in Sec. \ref{subsub:atom-photon_QuantumGates}, followed by detection of the atomic and photonic state \cite{bonato_cnot_2010}.

Apart from the efficiency, also the fidelity values achieved so far using the probabilistic approach pose a problem with respect to up-scaling. Improvement would require higher cooperativity in the ion-cavity experiment \cite{casabone_heralded_2013}. In the neutral-atom experiment \cite{nolleke_efficient_2013}, reducing the atomic temperature \cite{reiserer_ground-state_2013} or using a magic wavelength trap \cite{mckeever_state-insensitive_2003} is expected to improve the interference contrast. Then, the fidelity values might approach the high values achieved with the deterministic networking approach presented in Sec. \ref{sub:deterministicNetworks}.

\subsection{Nondestructive photon detection} \label{sub:PhotonDetection}

Scaling quantum networks to large distances necessitates heralded protocols that combat photon absorption in the used fiber links. The simplest approach to herald that a photon has been transmitted through a fiber is to measure its presence with an SPCM. This clearly destroys the encoded quantum information and is thus not useful for quantum networks. One can do better by using two-photon interference to transfer the state of the photon to that of a remote particle, e.g. a second photon or a trapped atom (as it has been described in the teleportation experiment in Sec. \ref{subsub:Teleportation}). However, this process is intrinsically probabilistic, with $0.1\,\%$ being the highest success probability achieved to date \cite{nolleke_efficient_2013}. This section will present an alternative scheme that facilitates a deterministic detection of the presence of a single photon without absorbing it. In this process, quantum information that is encoded in the photonic polarization or arrival time can be preserved, making the scheme promising for applications in quantum networks.

In order to nondestructively detect a photon, the interaction mechanism proposed in \cite{duan_scalable_2004} is employed. In the experiment \cite{reiserer_nondestructive_2013}, an incoming photon $\ket{1}$, contained in a faint laser pulse (with average photon number $\bar{n}=0.12$), is reflected from an atom-cavity system in the strong-coupling regime (with $C \simeq 3$). A theoretical treatment of this process is given in Sec. \ref{sub:CouplingToPropagatingFields}. Recall from Fig. \ref{fig:Transmission_Reflection} (c) that the phase of the combined atom-photon state remains unchanged when the atom is in $\ket{c}$ (blue), i.e. $\ket{c}\ket{1} \rightarrow \ket{c}\ket{1}$. In contrast, when the atom is in $\ket{u}$, the phase is flipped, i.e. $\ket{u}\ket{1}\rightarrow - \ket{u}\ket{1}$. Here, the arrow denotes the reflection process.

\begin{figure*}
\includegraphics[width=172mm]{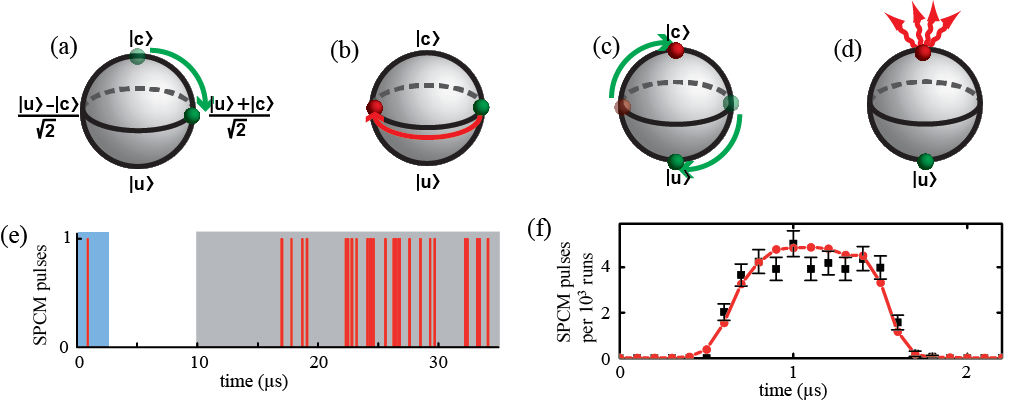}
\caption{ \label{fig:PhotonDetection_Results}
(color online) Nondestructive detection of an optical photon. (a)-(d) Ramsey sequence. The atomic state, visualized on the Bloch sphere, is first prepared in the superposition state $(\ket{u}+\ket{c})/\sqrt{2}$ (a). When a single photon is reflected from the resonator (b), the phase of the superposition state is flipped: $(\ket{u}+\ket{c})/\sqrt{2}\rightarrow (\ket{u}-\ket{c})/\sqrt{2}$. This state is transferred to $\ket{c}$ in a subsequent Raman rotation (c). State detection in fluorescence (d) thus facilitates detecting whether a photon has been reflected or not. (e) Result of a nondestructive detection process. A photon is destructively detected by an SPCM after reflection from an atom-cavity system (red line in the left blue trigger interval). But before being absorbed, the photon has left its trace in the atomic state. Therefore, after $\pi/2$ rotation, many fluorescence photons are observed in the $25\,\text{\textmu s}$ long readout interval (right, gray), unambiguously signaling the atomic state change induced by the reflected photon. (f) Arrival-time histogram of the photons detected with the SPCMs after reflection from the setup. The data taken during the nondestructive photon measurement (black squares) do not show a significant deviation from the reference curve recorded without atom (red points). Adapted from \cite{reiserer_nondestructive_2013}.
}
\end{figure*}

To use this conditional phase shift for photon detection, the atom is prepared in a superposition of its ground states using the techniques described in Sec. \ref{sub:InternalStateControl}, see Fig. \ref{fig:PhotonDetection_Results} (a). If there is no reflected photon, the state remains unchanged, as symbolized by the green filled circle on the right side of (b). Reflection of a photon, symbolized by the red arrow, however, leads to a flip of the superposition phase (red filled circle on the left side):
\begin{equation}
\frac{1}{\sqrt{2}}(\ket{c}+\ket{u})\ket{1}\rightarrow \frac{1}{\sqrt{2}}(\ket{c}-\ket{u})\ket{1}
\end{equation}

To measure this phase flip, a $\pi/2$ rotation (c) maps the atomic state $(\ket{u}+\ket{c})/\sqrt{2}$ onto $\ket{u}$, while $(\ket{u}-\ket{c})/\sqrt{2}$ is rotated to $\ket{c}$. Subsequently, cavity-enhanced fluorescence state detection (d) is used to discriminate between the atomic states $\ket{u}$ and $\ket{c}$. Fig. \ref{fig:PhotonDetection_Results} (e) shows a typical experimental trace. A photon is reflected in the first $2.5\,\text{\textmu s}$. After the reflection, it is detected by a conventional SPCM. Following a rotation of the atomic state, many fluorescence photons are detected by the same SPCM in the readout interval, which means that the photon has indeed been detected twice: Before it was absorbed by the SPCM, it has left its trace in the atom, which is subsequently read out.

The probability that an impinging photon is reflected (rather than absorbed or scattered) is $66\,\%$. The achieved single-photon detection efficiency is $74\,\%$. In contrast to absorbing photon detectors, it can be further improved by attempting more measurements. This yields $87\,\%$ for two concatenated setups, and $89\,\%$ for three or more devices. The performance of the setup is not fundamentally limited. It could be further increased with a higher cooperativity, smaller cavity losses, and reduced technical imperfections in the atomic state preparation, rotation and readout.

To investigate whether the presence of a photon can be heralded without affecting encoded quantum information, the wavepacket envelope of the reflected pulses has been measured, see Fig. \ref{fig:PhotonDetection_Results} (f). Except for a small reduction in amplitude, the data of the nondestructive photon detection experiment (black squares) do not show a significant deviation from the reference curve recorded without atom (red points). This means that the photonic wavepacket shape remains unchanged, and therefore it should be possible to detect a photon without changing the quantum information encoded in its arrival time. Similarly, polarization qubits can be employed when the atom is equally coupled to two degenerate polarization modes of a resonator.

\subsection{Quantum computing} \label{sub:quantumInfoProcess}

While the above sections have focused on the distribution of quantum information, experiments towards its processing in optical CQED setups will be summarized in the following. A first step in this direction is to control the transmission, reflection, and phase of a light field via a single, strongly coupled atom. This will be the topic of Sec. \ref{subsub:atom-controlled_switching}. Combining these techniques with atomic state control facilitates the realization of an atom-photon quantum gate, described in Sec. \ref{subsub:atom-photon_QuantumGates}. When this gate is applied to two photons consecutively, it can mediate a photon-photon quantum gate. This and other approaches towards photon-photon interactions in CQED are discussed in Sec. \ref{subsub:PhotonPhoton_Gates}.

\subsubsection{Atom-controlled amplitude switching and phase shifting} \label{subsub:atom-controlled_switching}

As explained in Sec. \ref{sub:CouplingToPropagatingFields}, the coupling of a single atom to an optical resonator can lead to a large change in the phase and amplitude of the transmission and reflection of light from a cavity when the cooperativity of the system fulfills the condition $C \gg 1$. Since the first observation of this effect \cite{thompson_observation_1992} with an atomic beam (with one intracavity atom on average traversing a Fabry-Perot resonator), many experiments have investigated atom-controlled switching. The first switching of the amplitude transmission with a single trapped atom is reported in \cite{maunz_normal-mode_2005, boca_observation_2004}, using a Fabry-Perot cavity in the strong-coupling regime. Similar experiments have meanwhile been realized in solid-state systems, e.g. using a single quantum dot coupled to a photonic crystal cavity \cite{englund_controlling_2007}. Recent experiments with optical ensembles trapped in a resonator have also demonstrated a hybrid approach \cite{chen_all-optical_2013}. Here, excitations at the single-photon level are trapped and released in a mode that is orthogonal to the cavity using the effect of collective enhancement of the absorption cross section. The coupling of the atoms to the resonator mode then leads to a reduced transmission whenever a photon is stored in the ensemble.

In order to harness atom-induced switching for quantum information processing, a high contrast between transmission and reflection is required, combined with excellent mode matching between the input and resonator modes. Towards this end, whispering-gallery-mode resonators have been identified as a promising system, with the additional advantage of operation in a fiber-coupled configuration with an adjustable outcoupling rate \cite{spillane_ideality_2003}. Several experiments have demonstrated switching by single atoms falling through the mode \cite{dayan_photon_2008, aoki_efficient_2009, oshea_fiber-optical_2013, shomroni_all-optical_2014}, but coherent control of the atomic ground state and thus the demonstration of a quantum gate is hardly possible without trapping.

Quantum information processing can also be implemented with atom-induced phase (rather than amplitude) control. The first experiment in this direction was the observation of a phase shift of several ten degrees, observed in transmission with an atomic beam passing through a Fabry-Perot cavity in the Purcell regime (with $C\simeq 1$) \cite{turchette_measurement_1995}. Higher phase shifts of $140$ degrees in transmission have been observed recently with a single trapped atom in a strongly-coupled system \cite{sames_antiresonance_2014}. Even stronger effects can be observed in the reflection from a single-sided cavity. Here, ideally a $\pi$ phase shift is achievable without a change in the reflected amplitude, see the calculations in Sec. \ref{sub:CouplingToPropagatingFields} depicted in Fig. \ref{fig:Transmission_Reflection}.

Recently, this has been observed in three independent experiments, both in the strong-coupling regime using a Fabry-Perot resonator \cite{reiserer_nondestructive_2013}, and in the Purcell regime using a photonic crystal cavity \cite{tiecke_nanophotonic_2014} or a whispering-gallery mode resonator \cite{volz_nonlinear_2014}. A phase shift in reflection has also been observed with quantum dot-cavity systems \cite{fushman_controlled_2008, kim_quantum_2013}. With neutral atoms, the phase shift has facilitated the implementation of an atom-photon quantum gate, which will be described in Sec. \ref{subsub:atom-photon_QuantumGates}. In addition, the mechanism is promising for the implementation of photon-photon quantum gates, see Sec. \ref{subsub:PhotonPhoton_Gates}.

\subsubsection{An atom-photon quantum gate} \label{subsub:atom-photon_QuantumGates}

For an intuitive explanation of the mechanism that facilitates an atom-photon quantum gate \cite{duan_scalable_2004}, consider a three-level atom in a lossless cavity that consists of a perfectly reflecting mirror and a coupling mirror which has a small transmission, see Fig. \ref{fig:AtomInCavity}. A resonant photon is impinging onto and reflected from the setup.

When the atom is in the uncoupled ground state $\ket{u}$, any transition is far detuned. Therefore, the photon can enter the cavity, as the light-field leaking out of the resonator interferes destructively with the direct reflection at the first mirror. In the reflection process, the combined atom-photon state thus experiences a phase shift of $\pi$, as discussed in Sec. \ref{sub:CouplingToPropagatingFields}. Now consider the case when the atom is in the coupled state $\ket{c}$. When the system is operated in a regime with $C \gg 1$, the photon cannot enter the cavity and is reflected without a phase shift. The effect of the reflection process is thus a conditional phase shift of $\pi$, i.e. a sign change, of the combined atom-photon state.

We emphasize that the conditional phase shift is independent of the exact atom-cavity coupling strength and detuning, making the scheme robust with respect to most experimental imperfections. There are only three prerequisites: First, the photonic bandwidth has to be small compared to the cavity linewidth. Second, the frequency of the impinging photons must be stable with respect to the cavity resonance. Third, the spatial mode of the impinging photons has to match that of the cavity.

When these conditions are fulfilled, the controlled phase shift enables the implementation of a quantum gate. To this end, a loss-tolerant photonic qubit is encoded in the polarization of the impinging photon. The original proposal \cite{duan_scalable_2004} employs a PBS, such that only one polarization component is reflected from the cavity and the other one is reflected from a mirror. Alternatively, one can use a resonator that only supports one resonant polarization mode and directly reflects the other, or employ an atomic level scheme that only couples to one polarization. The latter is used in the first experimental realization of an atom-photon quantum gate \cite{reiserer_quantum_2014}, which is described in this section.

In the experiment, the atomic qubit is defined such that $\ket{u} \equiv \ket{\downarrow}^{a}$ and $\ket{c} \equiv \ket{\uparrow}^{a}$, and the photonic qubit such that $\ket{\circlearrowleft} \equiv \ket{\downarrow}^{p}$ and $\ket{\circlearrowright} \equiv \ket{\uparrow}^{p}$. The photon is only coupled to the atom in the state $\left|\uparrow^{a}\uparrow^{p}\right\rangle$. Thus, the reflection of a photon results in the following conditional sign change of the combined atom-photon state:

\begin{subequations}
\begin{align}
\left|\uparrow^{a}\uparrow^{p}\right\rangle 	&\rightarrow	 \left|\uparrow^{a}\uparrow^{p}\right\rangle \\
\left|\uparrow^{a}\downarrow^{p}\right\rangle 	&\rightarrow	 -\left|\uparrow^{a}\downarrow^{p}\right\rangle \\
\left|\downarrow^{a}\uparrow^{p}\right\rangle 	&\rightarrow	 -\left|\downarrow^{a}\uparrow^{p}\right\rangle \\
\left|\downarrow^{a}\downarrow^{p}\right\rangle &\rightarrow	 -\left|\downarrow^{a}\downarrow^{p}\right\rangle
\end{align}
\end{subequations}

The above operation is a controlled-phase gate that is universal for quantum computation. This means that any two-qubit unitary can be decomposed into a sequence of controlled-phase gates and rotations of the individual qubits. Thus, with respect to the rotated photonic basis states $\ket{\uparrow_{x}^{p}}\equiv\frac{1}{\sqrt{2}}(\ket{\uparrow^{p}}+\ket{\downarrow^{p}})$ and $\ket{\downarrow_{x}^{p}}\equiv\frac{1}{\sqrt{2}}(\ket{\uparrow^{p}}-\ket{\downarrow^{p}})$, the conditional phase shift represents an atom-photon controlled-not (CNOT) gate that performs a flip of the photonic target qubit, controlled by the quantum state of the atom, similar to its classical analogue.

\begin{figure}
\includegraphics[width=86mm]{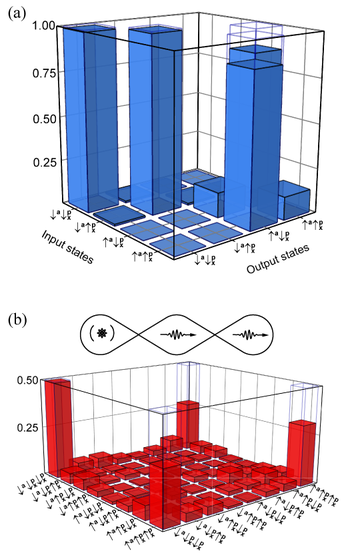}
\caption{ \label{fig:Gate}
(color online) (a) Truth table of the atom-photon CNOT quantum gate, showing the normalized probability of obtaining a specific output state for a complete set of atom-photon input states. The experimentally obtained values (filled bars) are close to the ideally expected ones (open bars). (b) Atom-photon-photon entangled state, generated via subsequent interactions of two photons with the same atom. The depicted density matrix is obtained by quantum state tomography. The fidelity with a maximally-entangled GHZ state (open bars) is $67\,\%$, proving genuine three-particle entanglement. From \cite{reiserer_quantum_2014}.
}
\end{figure}

A first step to characterize the gate is therefore to measure a classical truth table that shows the normalized probabilities to detect a certain output state for each of the orthogonal input states. The measurement result is depicted in Fig. \ref{fig:Gate}(a). The control and target qubits are expected to be unchanged when the control qubit is in the state $\ket{\downarrow^{a}}$, which is accomplished with a probability of $99\,\%$. When the control qubit is in $\ket{\uparrow^{a}}$, the expected flip of the photonic target qubit is observed with a probability of $86\,\%$.

The decisive feature that discriminates a quantum gate from a classical one is its capability to generate entangled output states from separable input states. To demonstrate this, the gate is applied to the photons contained in one or two sequentially impinging laser pulses ($\bar{n}=0.07$, FWHM $0.7\,\text{\textmu s}$). Post-selecting events where one photon was detected in each of the input pulses, a maximally entangled $\ket{\Psi^+}$ Bell-state or Greenberger-Horne-Zeilinger (GHZ) state \cite{greenberger_going_1989} is expected, respectively:

\begin{subequations}
\begin{align}
\ket{\downarrow_{x}^{a}\downarrow_{x}^{p}}\rightarrow\ket{\Psi^+}=\frac{1}{\sqrt{2}}(\ket{\uparrow^{a}\uparrow_{x}^{p}}+e^{-i\varphi}\ket{\downarrow^{a}\downarrow_{x}^{p}}) \\
\ket{\downarrow_{x}^{a}\downarrow_{x}^{p}\downarrow_{x}^{p}}\rightarrow\ket{\text{GHZ}}=\frac{1}{\sqrt{2}}(\ket{\uparrow^{a}\uparrow_{x}^{p}\uparrow_{x}^{p}}-e^{-i\varphi}\ket{\downarrow^{a}\downarrow_{x}^{p}\downarrow_{x}^{p}}) \label{eq:GHZ}
\end{align}
\end{subequations}

The density matrix of the generated quantum state is reconstructed using quantum state tomography. For the $\ket{\Psi^+}$ state with $\varphi=0$, a fidelity of $83\,\%$ is achieved, proving quantum operation of the gate. The result for the GHZ state is shown in Fig. \ref{fig:Gate}(b). With $\varphi=0.21\pi$, the achieved fidelity is $67(2)\,\%$, proving genuine three-particle (atom-photon-photon) entanglement.

In principle, the gate mechanism could be applied to more than two reflected photons in order to generate larger atom-photon cluster states. However, this would require an increase in the fidelity. In the experiment, the by far dominant source of error is imperfect spatial mode matching between the impinging photons and the cavity mode. It should be feasible to strongly improve this in an optimized optics setup, or with additional spatial filters.

Then, the experimentally achievable cluster-state size would depend on the efficiency of the used source of input photons and the efficiency of the gate itself. The experimental value of $67\,\%$ is limited by the moderate cooperativity of $C \simeq 3$ and scatter and absorption losses of the cavity mirrors. However, the achievable efficiency scales quite favorably with the atom-cavity coupling strength, such that gate efficiencies above $90\,\%$ seem feasible with current cavity technology.

The hybrid character of the atom-photon quantum gate makes the scheme very promising for applications in quantum information processing. In combination with photon measurement and feedback, the mechanism allows for the implementation of a heralded quantum memory and quantum gates between two atoms in the same or in remote cavities \cite{xiao_realizing_2004, cho_generation_2005, duan_robust_2005}. In addition, it may mediate a deterministic photon-photon gate, which will be discussed at the end of the following section.

\subsubsection{Towards photon-photon quantum gates} \label{subsub:PhotonPhoton_Gates}

\begin{figure}
\includegraphics[width=86mm]{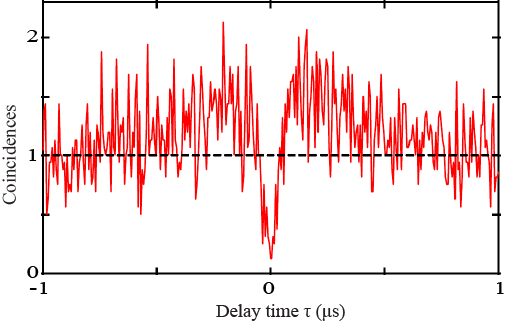}
\caption{ \label{fig:PhotonBlockade}
(color online) A two-mode photon-blockade effect. A faint, linearly polarized probe laser, detuned from the empty cavity resonance by the coupling strength $g$, drives the cavity mode. After passing an orthogonal polarizer, the transmitted light is analyzed in a Hanbury Brown and Twiss setup. The obtained correlation function exhibits clear antibunching. This demonstrates that because of its nonlinear eigenspectrum, photons can pass the atom-cavity system only one-at-a-time. Adapted from \cite{birnbaum_photon_2005}.
}
\end{figure}

The realization of deterministic interactions between single photons is a long-sought goal in quantum optics, and several different approaches are investigated in this context \cite{chang_quantum_2014}. In CQED systems, one can consecutively apply the atom-photon gate described in Sec. \ref{subsub:atom-photon_QuantumGates}. This approach will be revisited at the end of this chapter. Alternatively, one can make use of the nonlinear character of the atom-controlled switching mechanisms described in Sec. \ref{subsub:atom-controlled_switching} to mediate interactions between photons.

Towards this end, consider a single atom that is trapped in a cavity in the strong-coupling regime. A stream of photons impinges onto the system, i.e. either the atom or the cavity, with a frequency that is shifted from the empty cavity resonance frequency by the coupling strength $g$. When the atom is in the coupled state, the photons can resonantly drive transitions to one of the states in the first excited doublet of the Jaynes-Cummings ladder, $\ket{1,+}$ or $\ket{1,-}$, shown in Fig. \ref{fig:JaynesCummingsLadder}. Therefore, a single photon impinging onto the setup has a high chance to be absorbed and re-emitted. If, however, two photons impinge simultaneously, only one of them can be absorbed, as the corresponding doubly-excited state $\ket{2,+}$ or $\ket{2,-}$ is detuned from the light field by $(\sqrt{2}-1)g$.

It follows that, photons can only be emitted one-at-a-time, so that one expects to observe antibunching when measuring, e.g., the cavity output photon stream in a Hanbury Brown and Twiss setup. The first experiment that investigates this effect \cite{birnbaum_photon_2005} used crossed polarizations for excitation of the cavity and photon detection behind the cavity in order to suppress photons transmitted through the cavity in instances when the atom was not strongly coupled to the mode or no atom was present. The resulting normalized number of cross-polarization coincidences is shown in Fig. \ref{fig:PhotonBlockade}. Around zero time delay, sub-Poissonian photon statistics was observed, proving the single-photon character of the output field. The width of the dip in the correlation function is set by the linewidth of the normal mode, $\frac{\gamma+\kappa}{2}$.

The nonlinearity of the Jaynes-Cummings-ladder can also be used to implement a 'two-photon gateway' \cite{kubanek_two-photon_2008}: When operating at a different frequency, namely one with vanishing two-photon detuning, two photons are more likely transmitted simultaneously than at large temporal separation. Note that this situation should not be confused with a source of two-photon Fock states, as the one-photon transmission probability can be comparable or even larger than the joint two-photon transmission probability. In fact, a measurement of three-photon correlations reveals interesting dynamics \cite{koch_three-photon_2011}.

The experiments mentioned above all operate in the strong-coupling regime and exploit the nonlinear eigenspectrum of the Jaynes-Cummings ladder. Alternatively, one can also observe nonlinear amplitude or phase switching when operating in the Purcell regime at $\Delta_a=\Delta_c=0$, making use of the fact that a single two-level atom can absorb and emit only one photon at a time. Following the early work of \cite{turchette_measurement_1995}, such nonlinearities have been investigated in several experiments with atoms near whispering-gallery mode resonators \cite{dayan_photon_2008, aoki_efficient_2009, oshea_fiber-optical_2013, volz_nonlinear_2014}, and with atoms or quantum dots in photonic crystal cavities \cite{reinhard_strongly_2012, tiecke_nanophotonic_2014}.

While in principle, the observation of nonlinear effects in the experiments mentioned above leads to an interaction between photons, the question whether such nonlinearity allows for the implementation of a photon-photon quantum gate has been the subject of a long debate, starting with the work of \cite{shapiro_single-photon_2006}. The basic argument is that such quantum gate requires conflicting properties of the photonic wavepackets: On the one hand, matching the cavity bandwidth requires the photon duration to be long. But with long two-photon pulses, the average population of the atomic excited state becomes small, and thus the nonlinearity vanishes. Therefore, the observation of nonlinear switching explained above is restricted to a subensemble of all events \cite{rosenblum_photon_2011}. Still, it can be used for probabilistic gates and photonic Bell-state measurements \cite{witthaut_photon_2012}.

In contrast, consecutive application of an atom-photon gate in a three-level configuration is expected to facilitate deterministic photon-photon quantum gates. Towards this goal, there are two possible mechanisms, that of \cite{duan_scalable_2004} which implements a CNOT gate as explained in Sec. \ref{subsub:atom-photon_QuantumGates}, and that of \cite{koshino_deterministic_2010} which should allow for a $\sqrt{\text{SWAP}}$-operation.

A first step towards realizing the latter is reported in \cite{shomroni_all-optical_2014}, where it is shown that using the gate mechanism in a SWAP-configuration, photons contained in faint light pulses can toggle the state of an atom, as described in Sec. \ref{subsub:QuantumMemory}. Then, the atom can switch the transmission of another photon, as described in Sec. \ref{subsub:atom-controlled_switching}. Towards photonic quantum gates, a higher switching contrast and faithful operation with quantum superposition states would have to be demonstrated.

Considering the mechanism of \cite{duan_scalable_2004}, the implementation of an atom-photon quantum gate has been demonstrated \cite{reiserer_quantum_2014}. To extend this towards photon-photon gates, a first reflected photon is entangled with the atomic state, which in turn toggles the polarization of a second photon. To realize a quantum gate, the atom has to be disentangled from the photons, which can be done by reflecting the first photon a second time from the setup \cite{duan_scalable_2004}. Alternatively, a quantum eraser protocol can be employed \cite{hu_deterministic_2008}.

The latter approach has been used in \cite{reiserer_quantum_2014} to generate a maximally entangled photon-photon state out of two separable input photons by post-selection. The experiment starts by generating the state $\ket{\text{GHZ}}$, see Eq. (\ref{eq:GHZ}) in Sec. \ref{subsub:atom-photon_QuantumGates}. Then, a $\frac{\pi}{2}$ rotation is applied to the atom. When $\varphi=0$, the state is transformed to:

\begin{equation}
\frac{1}{\sqrt{2}}\left[\ket{\uparrow^{a}}(\ket{\uparrow_{x}^{p}\uparrow_{x}^{p}}-\ket{\downarrow_{x}^{p}\downarrow_{x}^{p}})-\ket{\downarrow^{a}}(\ket{\uparrow_{x}^{p}\uparrow_{x}^{p}}+\ket{\downarrow_{x}^{p}\downarrow_{x}^{p}})\right]
\end{equation}

Subsequent measurement of the atomic state disentangles the atom, which results in an entangled two-photon state: If the atom is found in $\ket{\uparrow^{a}}$ ($\ket{\downarrow^{a}}$), the resulting state is $\ket{\Phi^{-}}$ ($\ket{\Phi^{+}}$), respectively. The experimentally achieved fidelity with a maximally entangled state is $76(2)\,\%$, proving photon-photon entanglement.

The above post-selected measurement demonstrates that the atom-photon quantum-gate mechanism can mediate a photon-photon gate that is in principle deterministic. To this end, fast feedback onto the photonic state would have to be implemented, which requires that the two photons are stored during the time required to rotate and read out the atomic state, which takes about $3\,\text{\textmu s}$ in the setup of \cite{reiserer_quantum_2014}. Thus, the feedback procedure can technically be realized with an optical fiber of less than one kilometer length and an electro-optical modulator. This makes the implementation of a deterministic photon-photon quantum gate via the mechanism proposed in \cite{duan_scalable_2004} a realistic perspective for the near future.

\section{Summary and outlook} \label{sec:Outlook}

The experiments described in this review article have demonstrated that optical resonators with low internal loss provide an efficient interface between stationary material qubits and flying optical qubits. Moreover, steady progress in atom trapping and cooling has enabled full control over individual atoms. Thus, the ideal situation assumed in most CQED proposals can nowadays be realized experimentally: A point-like two- or multi-level atom that is trapped at a fixed position within the field of an overcoupled optical cavity which provides controllable strong coupling to propagating photons.

The achieved experimental control has enabled the realization of first prototype quantum networks, consisting of two nodes connected by one channel. The nodes are universal, meaning that they can perform all operations required in a quantum network: sending, receiving, storing and processing of quantum information. Nevertheless, scaling the demonstrated elementary networks to more nodes and larger distances poses a formidable challenge, both experimentally and theoretically. The reason is that all of the described processes exhibit finite efficiencies and fidelities.

In the following, we first consider the scaling to larger distances, where the main obstacle is the loss in both fiber and free-space photonic quantum channels. In free-space networks, the main source of loss is atmospheric scattering and distortion of the mode. An interesting approach to mitigate this problem is to establish photonic channels via satellites \cite{peng_experimental_2005}. In fiber-based networks, long-distance experiments will likely require operation in the so called telecommunications window, i.e. at a wavelength around $1.55\,\text{\textmu m}$ where the loss is lowest, typically about $0.2\,\frac{\text{dB}}{\text{km}}$. Towards this end, one can use atoms with suitable transition frequencies, employ wavelength-bridging entanglement \cite{radnaev_quantum_2010} or convert the photonic wavelength via non-linear processes \cite{tanzilli_photonic_2005, de_greve_quantum-dot_2012} that can achieve remarkable efficiencies above $30\,\%$ \cite{zaske_visible--telecom_2012}. Still, direct fiber-based links will be limited to internode separations of tens of kilometers at most, as the transmission probability $p_{\text{trans}}(L)=e^{-L/L_\text{loss}}$ of a photon decreases exponentially with distance $L$. Here, $L_\text{loss}\simeq 20\,\text{km}$ denotes the $1/e$ loss length of today's telecommunication fibers.

Many proposals have been put forward to overcome this unfavorable exponential scaling by using a divide-and-conquer strategy. To understand the basic idea, consider a situation where one tries to generate entanglement between two nodes $A$ and $B$ that are separated by a distance $L_\text{0}$. The average success rate is then given by $R_\text{success}=R_\text{rep} \times p_\text{int} \times p_\text{trans}(L_{\text{0}})$, where $p_\text{int}$ is the intrinsic efficiency of the entanglement generation and distribution protocol and $R_\text{rep}$ is the repetition rate. When the distance between the nodes is increased to $n L_0$, the success rate will drop proportional to $p_\text{trans}(n L_{\text{0}}) = p_\text{trans}(L_0)^{n}$, i.e. exponentially with $n$.

Now consider adding $n$ intermediate network nodes, equally distributed between $A$ and $B$, and each equipped with two qubits. Generating entanglement between one of the qubits of adjacent intermediate nodes will still succeed with an average rate $R_\text{success}$. When the entanglement generation is heralded and the quantum state of individual qubits can be preserved until all neighboring nodes are entangled, then entanglement over an arbitrary distance can almost instantaneously be generated (without an exponential reduction of the efficiency) by repeated application of entanglement swapping \cite{zukowski_``event-ready-detectors_1993}. The latter process is equivalent to teleportation \cite{bennett_teleporting_1993} of one of the qubits of an entangled state to the next node (and subsequently all along the chain).

Ideally, perfect qubit readout and both deterministic and error-free two-qubit quantum gates are used for entanglement swapping and teleportation. However, ideal systems are not required as two seminal proposals have shown that remote entanglement and state transfer can also be achieved with imperfect \cite{briegel_quantum_1998} and even probabilistic \cite{duan_long-distance_2001} local operations at the (polynomial) cost of an increased number of network nodes. In order to make this work, additional purification \cite{bennett_purification_1996} steps are employed to distill remote entangled pairs of high fidelity from two or several pairs of lower fidelity. Recent approaches employ the concept of quantum error correction for entanglement distillation \cite{jiang_quantum_2009} and quantum state transmission \cite{munro_quantum_2012, muralidharan_ultrafast_2014}. Devices that implement one of the mentioned or the numerous related proposals are called quantum repeaters. A detailed analysis of the scaling of different protocols with the fidelity and efficiency of the individual operations is presented in \cite{duan_colloquium:_2010, sangouard_quantum_2011}. Still, to date, no experiment has demonstrated an improvement of the remote entanglement rate per optical mode, compared to the fundamental limit encountered for direct photon transmission \cite{takeoka_fundamental_2014}. However, achieving this task over a limited distance seems feasible with state-of-the-art atom-cavity setups.

We now consider the scaling to a larger number of network nodes. Like in the previous scenario of large distances, the success probability of naively trying to generate a maximally entangled state of the network will again drop exponentially with the number of nodes, as any experimental entanglement protocol will exhibit finite efficiencies. And again, there are several proposals how to overcome this problem with the techniques developed for the implementation of universal quantum computation. One approach is to use cluster states \cite{raussendorf_one-way_2001}  as a resource, which are generated using probabilistic gates between remote qubits. The scaling of such networks is analyzed, e.g., in \cite{duan_colloquium:_2010}. A different approach is to combine probabilistic remote operations with local deterministic ones, as in the quantum repeater schemes mentioned above. In such setting a deterministic interaction of remote quantum memories via probabilistic but heralded photonic channels is possible using a repeat-until-success strategy \cite{van_enk_photonic_1998}, where failure after a certain number of trials leads to a reduction in fidelity that can be efficiently corrected \cite{fowler_surface_2010, nickerson_freely_2014}.

In summary, both the scaling towards larger distances and a larger number of network nodes require similar experimental capabilities. The cavity-based network nodes described in this review article have already demonstrated many of these requirements. A realistic avenue towards scalable quantum networks, however, will still require the following developments:

First, the nodes have to be equipped with two or more quantum bits, as has recently been implemented in atom-cavity setups, both with neutral atoms \cite{reimann_cavity-modified_2015} and with ions \cite{casabone_enhanced_2015}. Yet, it still has to be demonstrated that it is possible to optically address a single qubit within a node without inducing decoherence of the others. Possible solutions to this problem include the use of off-resonant Raman schemes, shelving of the qubits in protected internal states, shifting individual atoms to uncoupled locations, and employing different atomic species for either communication or storage of qubits.

Second, the nodes must allow for local entanglement swapping with very high probability and fidelity. Otherwise, even though the success rates scale polynomially with distance, they will be too low for a useful experimental implementation \cite{duan_colloquium:_2010, sangouard_quantum_2011}. Towards the implementation of efficient entanglement swapping, deterministic local two-qubit gates will be beneficial, which can be implemented using the coulomb interaction of trapped ions \cite{blatt_entangled_2008, casabone_enhanced_2015} or the dipolar interaction of neutral atoms excited to a Rydberg state \cite{saffman_quantum_2010}. Alternatively, numerous proposals exist to implement two-qubit gates via the common interaction with the same cavity mode, see e.g. \cite{xiao_realizing_2004, cho_generation_2005, duan_robust_2005}.

Third, the scaling of quantum networks requires that remote nodes can be entangled in a heralded way at a rate that is orders of magnitude faster than the inverse of the qubit coherence time. In this respect, the envisioned use of quantum error correction to increase the memory lifetime \cite{terhal_quantum_2015} will be beneficial. But even without such techniques, the rates achieved with optical cavities, up to $100\,\text{Hz}$ \cite{ritter_elementary_2012} with the deterministic and $10\,\text{Hz}$ \cite{nolleke_efficient_2013} with the probabilistic approach, both at a distance of $21\,\text{m}$, are very promising. Still, they would have to be substantially increased for practical scaling.

Finally, the major challenge towards scaling is to improve the fidelity of both local and remote operations. On the one hand, theoretical progress in quantum error correction and entanglement purification protocols over the last decade has constantly lowered the error thresholds for fault-tolerant quantum operations to a few per cent \cite{fowler_surface_2012, terhal_quantum_2015}, even in a network-type architecture \cite{fowler_surface_2010, nickerson_freely_2014, zwerger_universal_2013}. On the other hand, keeping the number of qubits at realistic levels still requires error rates well below $10^{-3}$. Albeit the fidelity of most experiments presented in this article is not physically bounded, technically achieving such thresholds will require considerable effort. In addition, the application of fast qubit measurements and quantum feedback, which is a prerequisite for quantum error correction \cite{devitt_quantum_2013, lidar_quantum_2013}, entanglement purification \cite{bennett_purification_1996, dur_entanglement_2007} and measurement based quantum computing \cite{briegel_measurement-based_2009}, is still in its infancy. In this respect, the use of optical resonators to efficiently couple stationary qubits to photons, for which efficient detectors and powerful measurement techniques are readily available, provides unique possibilities for current and future experiments in quantum communication and quantum information processing.

To conclude, albeit large-scale quantum networks are still far beyond the current state of the art, the experiments presented in this review have identified the main challenges that have to be addressed towards scaling. Continuing research in this direction will help to find answers to a very fundamental question: How large can a system become without losing its quantum properties? In this question, the word 'large' can again refer to both physical distance and number of involved particles. In the first sense, next steps might be the realization of a loophole-free test of a Bell inequality \cite{brunner_bell_2014} that facilitates exploring the ultimate limits of privacy \cite{ekert_ultimate_2014}, and a vision for the future might be the realization of a quantum network on a global scale \cite{kimble_quantum_2008} using quantum repeater protocols \cite{briegel_quantum_1998}. In the second sense, where 'large' refers to the number of quantum bits, future studies will investigate collective decoherence effects and the network-based simulation of quantum systems \cite{georgescu_quantum_2014}. In this respect, quantum networks might provide an open-system approach that is complementary to investigations with ultracold atoms \cite{bloch_quantum_2012} and superconducting qubits \cite{houck_-chip_2012}. Experimental steps towards this goal will include studies on quantum phase transitions of light \cite{torma_transitions_1998, hartmann_strongly_2006, greentree_quantum_2006, carusotto_quantum_2013}, on the percolation of entanglement \cite{acin_entanglement_2007}, and on different quantum network topologies including quantum random networks \cite{perseguers_quantum_2010}. Apart from quantum simulation, the long-term vision of scaling to more network nodes is the development of a distributed quantum computing architecture, in which small quantum registers are connected by optical photons \cite{monroe_scaling_2013, awschalom_quantum_2013}.

At the current stage, however, realistic practical applications of quantum networks demand for a reduction of the experimental overhead. With this respect, emergent photonic technologies \cite{obrien_photonic_2009} might lead to an on-chip integration of light sources, switches, single photon detectors, and maybe even cavities, thus reducing the complexity and price of experimental setups and improving their stability. Still, atom trapping and cooling necessitates ultrahigh vacuum and several controlled laser beams. Avoiding these difficulties using solid state systems seems to be a prerequisite for a transition that is comparable to that from the drift tube to the transistor. However, because of the many challenges solid state systems still pose today, atomic physics experiments will likely continue to play a leading role in the exploration of fundamental phenomena in distributed quantum networks.

\begin{acknowledgments}
The authors thank R. Blatt, S. Bogdanovic, M. Keller, H. J. Kimble, M. D. Lukin, D. Meschede, T. Northup, A. Rauschenbeutel, T. Tiecke, K. Vahala, and J. Volz for granting permission to reproduce their figures and for discussions. GR would like to thank J. Bochmann, E. Figueroa, C. Hahn, M. Hennrich, M. Hijlkema, A. Kuhn, T. Legero, D. L. Moehring, M. M{\"u}cke, A. Neuzner, C. N{\"o}lleke, S. Nu{\ss}mann, S. Ritter, H. P. Specht, M. Uphoff, C. J. Villas-Boas, B. Weber, S. C. Webster, and T. Wilk for their contributions to the described experiments. This work was supported by the European Union (Collaborative Project SIQS) and by the Bundesministerium f{\"u}r Bildung und Forschung via IKT 2020 (QK\_QuOReP).
\end{acknowledgments}

\bibliography{Bibliothek.bib}

\end{document}